\documentclass[journal]{IEEEtran}

\usepackage[utf8]{inputenc}
\usepackage[pdftex]{graphicx}
\graphicspath{{./figures/}{./Bio/}}
\DeclareGraphicsExtensions{.pdf,.png,.jpg,.jpeg,.eps}

\usepackage{cite,amsmath,amssymb,url,multirow}
\usepackage{algorithmicx}
\usepackage{algpseudocode}
\usepackage{cases}
\usepackage{ulem}
\usepackage{epstopdf}
\usepackage{listings}
\usepackage{array}
\usepackage{makecell}
\usepackage{tabularx}
\usepackage{multirow}

\usepackage[x11names,table]{xcolor}
\usepackage{colortbl}

\usepackage{caption}
\usepackage{booktabs}

\usepackage{rotating}

\usepackage{tikz}
\usetikzlibrary{arrows, calc, decorations.markings, positioning,trees}

\makeatother

\usepackage[edges]{forest}

\usepackage{xcolor}
\newcommand\ytl[2]{
	\parbox[b]{6em}{\hfill{\color{cyan}\bfseries\sffamily #1}~$\cdots\cdots$~}\makebox[0pt][c]{$\bullet$}\vrule\quad \parbox[c]{9cm}{\vspace{7pt}\color{darkgray!40!black!80}\raggedright #2.\\[7pt]}\\[-3pt]}

\usepackage{pifont}% http://ctan.org/pkg/pifont
\newcommand{\cmark}{\ding{51}}%
\newcommand{\xmark}{\ding{55}}%

\usepackage{tikz}

\usetikzlibrary{shapes,arrows,decorations,positioning,calc,shadows,trees}
\usetikzlibrary{mindmap}

\usepackage[nopostdot,nomain,nonumberlist,acronym]{glossaries}
\setglossarystyle{super}

{\centering\tablehead{}\tabletail{}%
	\begin{supertabular}{lp{\glsdescwidth}}}%
{\end{supertabular}}%

\makeglossaries

\usepackage{hyperref}

\newacronym{LQE}{LQE}{Link Quality Estimation}
\newacronym{PRR}{PRR}{Packet Reception Ratio}
\newacronym{WSN}{WSN}{Wireless Sensor Network}
\newacronym{SNR}{SNR}{Signal-to-Noise Ratio}
\newacronym{BER}{BER}{Bit Error Rate}
\newacronym{LQI}{LQI}{Link Quality Indicator}
\newacronym{RSSI}{RSSI}{Received Signal Strength Indicator}
\newacronym{WMEWMA}{WMEWMA}{Window Mean with an Exponentially Weighted Moving Average}
\newacronym{KDP}{KDP}{Knowledge Discovery Process}
\newacronym{CDF}{CDF}{Cumulative Distribution Function}
\newacronym{4B}{4B}{Four-Bit}
\newacronym{4C}{4C}{Foresee}
\newacronym{PSR}{PSR}{Packet Success Ratio}
\newacronym{RSS}{RSS}{Received Signal Strength}
\newacronym{ETX}{ETX}{Expected Transmission count}
\newacronym{RNP}{RNP}{Required Number of Packets}
\newacronym{SGD}{SGD}{Stochastic Gradient Descent}
\newacronym{FLI}{FLI}{Fuzzy-logic Link Indicator}
\newacronym{F-LQE}{F-LQE}{Fuzzy-logic based LQE}
\newacronym{WNN-LQE}{WNN-LQE}{Wavelet Neural Network based LQE}
\newacronym{MSE}{MSE}{Mean Squared Error}
\newacronym{RUS}{RUS}{Random Under-Sample}
\newacronym{ROS}{ROS}{Random Over-Sample}
\newacronym{TCP}{TCP}{Transmission Control Protocol}
\newacronym{KDD}{KDD}{Knowledge Discovery and Data mining}
\newacronym{ML}{ML}{Machine Learning}
\newacronym{AI}{AI}{Artificial Intelligence}

\newacronym{SVM}{SVM}{Support Vector Machine}
\newacronym{SVR}{SVR}{Support Vector Regressor}
\newacronym{LQ}{LQ}{Link Quality}
\newacronym{NLQ}{NLQ}{Neighbor Link Quality}
\newacronym{MAE}{MAE}{Mean Absolute Error}
\newacronym{RMSE}{RMSE}{Root-Mean-Square Error}
\newacronym{PER}{PER}{Packet Error Rate}
\newacronym{ROC}{ROC}{Receiver Operating Characteristic}

\definecolor{light-gray}{gray}{0.95}
\definecolor{gray}{gray}{0.85}
\definecolor{warn}{RGB}{255,127,0}

\newcolumntype{L}[1]{>{\hsize=#1\hsize\raggedright\arraybackslash}X}%
\newcolumntype{R}[1]{>{\hsize=#1\hsize\raggedleft\arraybackslash}X}%
\newcolumntype{C}[1]{>{\hsize=#1\hsize\centering\arraybackslash}X}%
\ifCLASSOPTIONcompsoc
	\usepackage[nocompress]{cite}
	\usepackage[caption=false,font=normalsize,labelfont=sf,textfont=sf]{subfig}
\else
	\usepackage{cite}
	\usepackage[caption=false,font=footnotesize]{subfig}
\fi

\definecolor{awesome}{rgb}{1.0, 0.6, 0.0}

\makeatletter

\def\ps@IEEEtitlepagestyle{%
	\def\@oddfoot{\mycopyrightnotice}%
	\def\@evenfoot{}%
}
\def\mycopyrightnotice{%
	{\begin{minipage}{2\linewidth}\footnotesize\bfseries \copyright 2021 IEEE.  Personal use of this material is permitted.  Permission from IEEE must be obtained for all other uses, in any current or future media, including reprinting/republishing this material for advertising or promotional purposes, creating new collective works, for resale or redistribution to servers or lists, or reuse of any copyrighted component of this work in other works.\hfill\end{minipage}}
	\gdef\mycopyrightnotice{}
	
}

\begin{document}

\title{Machine Learning for Wireless Link Quality Estimation: A Survey}

% \author{\IEEEauthorblockN{%
% 	Gregor Cerar\IEEEauthorrefmark{1}\IEEEauthorrefmark{2}, %
% 	Halil Yetgin\IEEEauthorrefmark{1}\IEEEauthorrefmark{3}, \textit{Member, IEEE}, 
% 	Mihael Mohor\v{c}i\v{c}\IEEEauthorrefmark{1}\IEEEauthorrefmark{2}, \textit{Senior Member, IEEE}, and 
% 	Carolina Fortuna\IEEEauthorrefmark{1}\\%
% }
% \IEEEauthorblockA{%
% 	\IEEEauthorrefmark{1}Department of Communication Systems, Jo\v{z}ef Stefan Institute, Jamova 39, SI-1000, Slovenia.\\%
% 	\IEEEauthorrefmark{2}Jo\v{z}ef Stefan International Postgraduate School, Jamova 39, SI-1000, Slovenia.\\%
% 	\IEEEauthorrefmark{3}Department of Electrical and Electronics Engineering, Bitlis Eren University, 13000 Bitlis, Turkey.\\%
% \{gregor.cerar $\mid$ halil.yetgin $\mid$ miha.mohorcic $\mid$ carolina.fortuna\}@ijs.si%
% }}

\author{Gregor Cerar$^{1,2}$,~\IEEEmembership{Student~Member,~IEEE,}
        Halil Yetgin$^{1,3}$,~\IEEEmembership{Member,~IEEE,}
        Mihael Mohor\v{c}i\v{c}$^{1,2}$,~\IEEEmembership{Senior~Member,~IEEE,} and 
        Carolina Fortuna$^{1}$\\
$^{1}$Department of Communication Systems, Jo{\v z}ef Stefan Institute, SI-1000 Ljubljana, Slovenia.\\
$^{2}$Jo{\v z}ef Stefan International Postgraduate School, Jamova 39, SI-1000 Ljubljana, Slovenia.\\
$^{3}$Department of Electrical and Electronics Engineering, Bitlis Eren University, 13000 Bitlis, Turkey.\\
\{gregor.cerar $\mid$ halil.yetgin $\mid$ miha.mohorcic $\mid$ carolina.fortuna\}@ijs.si    
}

\maketitle

\begin{abstract}
	% What is important?
	Since the emergence of wireless communication networks, a plethora of research papers focus their attention on the quality aspects of wireless links.
	% What is the problem?
	The analysis of the rich body of existing literature on link quality estimation using models developed from data traces indicates that the techniques used for modeling link quality estimation are becoming increasingly sophisticated. A number of recent estimators leverage \gls{ML} techniques that require a sophisticated design and development process, each of which has a great potential to significantly affect the overall model performance.
	% What part of this problem we are solving here?
	In this paper, we provide a comprehensive survey on link quality estimators developed from empirical data and then focus on the subset that use \gls{ML} algorithms. We analyze \gls{ML}-based \gls{LQE} models from two perspectives using performance data. Firstly, we focus on how they address quality requirements that are important from the perspective of the applications they serve. Secondly, we analyze how they approach the standard design steps commonly used in the ML community. Having analyzed the scientific body of the survey, we review existing open source datasets suitable for \gls{LQE} research. Finally, we round up our survey with the lessons learned and design guidelines for \gls{ML}-based \gls{LQE} development and dataset collection.
	
%	We revealed that each step of the machine learning process significantly influences the estimator's overall performance. 
\end{abstract}

% \fi
\IEEEpeerreviewmaketitle


\section{Introduction}
\label{sec:intro}

In wireless networks, the propagation channel conditions for radio signals may vary significantly with time and space, affecting the quality of radio links~\cite{Bai2003}. In order to ensure a reliable and sustainable performance in such networks, an effective link quality estimation (LQE) is required by some protocols and their mechanisms, so that the radio link parameters can be adapted and an alternative or more reliable channel can be selected for wireless data transmission. To put it simply, the better the link quality, the higher the ratio of successful reception and therefore a more reliable communication. However, challenging factors that directly affect the quality of a link, such as channel variations, complex interference patterns and transceiver hardware impairments just to name a few, can unavoidably lead to  unreliable links~\cite{baccour2012radio}. On one hand, incorporating all these factors in an analytical model is infeasible and thus such models cannot be readily adopted in realistic networks due to highly arbitrary and dynamic nature of the propagation environment~\cite{Zanella2016}. On the other hand, effective prediction of link quality can provide great performance returns, such as improved network throughput due to reduced packet drops, prolonged network lifetime due to limited retransmissions~\cite{Yetgin2017}, constrained route rediscovery, limited topology breakdowns and improved reliability, which reveal that the quality of a link influences other design decisions for higher layer protocols. Eventually, variations in link quality can significantly influence the overall network connectivity. Therefore, effectively estimating or predicting the quality of a link can provide the best performing link from a set of candidates to be utilized for data transmission.

\begin{figure}[t]
	\centering
	\begin{tikzpicture}[
	scale=2.0,
	blok/.style={rectangle, minimum size=10mm, minimum width=20mm, rounded corners=3mm, very thick, draw=black!50, top color=white, bottom color=black!20},
	oblak/.style={cloud, cloud puffs=10, cloud puff arc=110, aspect=2, inner ysep=1em, very thick,draw=black!50, top color=white, bottom color=black!20},
	smer/.style={draw, stealth'-stealth', thick},
	marker/.style={draw, dotted, rectangle, red, thick, rounded corners},
	]
	
	\node[oblak] (ch) {channel};
	
	% Stack (left)
	\node[blok, left=2em of ch] (phy1) {physical layer};
	\node[blok, above=2em of phy1] (link1) {link layer};
	\node[blok, opacity=0.4, above=2em of link1] (net1) {network layer};
	\node[above=2em of net1] (dummy1) {};
	
	\draw[smer] (phy1) -- (link1);
	\draw[smer, opacity=0.4] (link1) -- (net1);
	\draw[smer, opacity=0.4,stealth'-] (net1) -- (dummy1);
	
	% Stack (right)
	\node[blok, right=2em of ch] (phy2) {physical layer};
	\node[blok, above=2em of phy2] (link2) {link layer};
	\node[blok, opacity=0.4, above=2em of link2] (net2) {network layer};
	\node[above=2em of net2] (dummy2) {};
	
	\draw[smer] (phy2) -- (link2);
	\draw[smer, opacity=0.4] (link2) -- (net2);
	\draw[smer, opacity=0.4,stealth'-] (net2) -- (dummy2);
	
	% Channel connections
	\draw[smer] (ch) -- (phy1);
	\draw[smer] (ch) -- (phy2);
	
	% Logical connections
	\draw[smer, opacity=0.4, dash pattern=on \pgflinewidth off 4pt] (net1) -- (net2) node[pos=0.5, above]{peer layer comm.}; % logical network
	\draw[smer, dash pattern=on \pgflinewidth off 4pt] (link1) -- (link2) node[pos=0.5, above]{peer layer comm.}; % link layer communication
	
	% Rectangle to mark what LQE actually models!
	\draw [marker]
	($(link1.north west) + (-.3em, .5em)$)
	rectangle
	($(phy2.south east) + (.1em, -0.7em)$);
	
	\end{tikzpicture}
	\caption{The unified model of data-driven LQE comprising of physical layer (layer 1) and link layer (layer 2).}
	\label{fig:link-abstraction}
\end{figure}
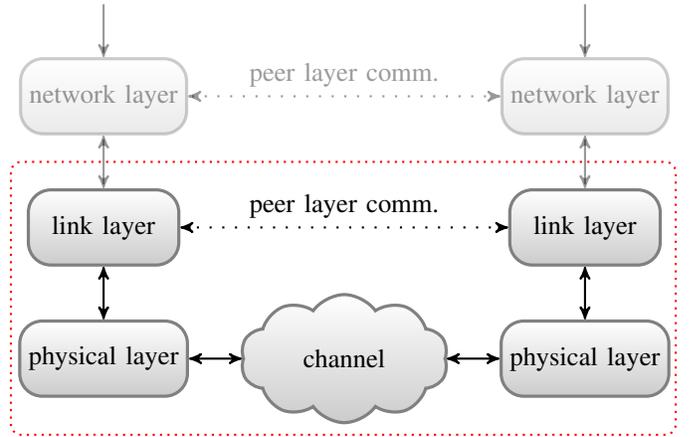

More broadly, the quality of a wireless link is influenced by the design decisions taken for: i) wireless channel, ii) physical layer technology, and iii) link layer, as depicted in Fig.~\ref{fig:link-abstraction}. The channel used for communication can be described by several parameters, such as operating frequency, transmission medium (e.g. air, water), environment (e.g. indoor, outdoor, dense urban, suburban) as well as relative position of the communicating parties (e.g. line-of-sight, non-line-of-sight)~\cite{Bai2003}. The physical layer technology implemented at the transmitter and receiver comprises several complex and well-engineered blocks, such as the antenna (e.g. single, multiple or array), frequency converter, analog to digital converter, synchronization and other baseband operations. The link layer is responsible for successfully delivering the data frame via a single wireless hop from transmitter to receiver, therefore it comprises of frame assembly and disassembly techniques, such as attaching/detaching headers, encoding/decoding payload, as well as mechanisms for error correction and controlling retransmissions~\cite{Zanella2016}. While the quality of a link is ultimately influenced by a sequence of complex, well studied, designed and engineered processing blocks, the performance of the realistic and operational systems is quantified by a relatively limited number of observations~\cite{baccour2012radio}, the so-called \textit{link quality metrics}, which are detailed later in Section~\ref{subsec:input} using Table~\ref{tab:feature-analysis}.

In this paper, we refer to the wireless link abstraction as comprising of link layer and physical layer. More explicitly, \textit{link quality} is referred to the quality of a wireless link that is concerned with the link layer and the physical layer. The \gls{LQE} models reviewed in this survey paper are based on physical and link layer metrics, namely all potential metrics for the evaluation of link quality that lie within the dotted rectangle of Fig.~\ref{fig:link-abstraction}.

To briefly overview, the research on data-driven \gls{LQE} using real measurement data started in the late 90s~\cite{nguyen1996trace} and is still carried on with a plethora of  publications in the last decade~\cite{demetri2019automated,nguyen1996trace, balakrishnan1998explicit, woo2003taming, senel2007kalman, fonseca2007four, srinivasan2008prr, boano2010triangle, baccour2010fuzzy, guo2013fuzzy, rekik2015fli, audeoud2018quick}. Early studies on this particular topic mainly utilized recorded traces and the models were developed manually~\cite{nguyen1996trace, balakrishnan1998explicit, woo2003taming, senel2007kalman, fonseca2007four, srinivasan2008prr, boano2010triangle, baccour2010fuzzy, guo2013fuzzy, rekik2015fli, audeoud2018quick}. Over the past few years, researchers have paid a lot of attention to the development of \gls{LQE} using \gls{ML} algorithms~\cite{liu2011foresee, liu2014temporal, sun2017wnn,demetri2019automated}.

%TODO:, while the \textit{link} ensures that the frames are delivered from a transmitter to the receiver over a wireless channel.

% bottoms up for forest tree, simply set: grow=0,reversed,

\begin{figure*}[!htb]
\vspace{-1cm}
\newcolumntype{H}[1]{>{\centering}p{#1}}
\begin{forest}
	forked edges,
	for tree={
		if level=0{align=center}{%
			align={@{}H{28mm}@{}},
		},
		grow=east,
		text width=2.62cm,
		draw,
		font=\sffamily\scriptsize,
		edge path={
			\noexpand\path [draw, \forestoption{edge}] (!u.parent anchor) -| +(0.5mm,0) -- +(5mm,0) |- (.child anchor)\forestoption{edge label};
		},
		parent anchor=east,
		child anchor=west,
		l sep=10mm,
		tier/.wrap pgfmath arg={tier #1}{level()},
		edge={thin},
		fill=white,
		rounded corners=2pt,
		drop shadow,
	}
[Machine Learning for \\ Wireless Networks
	[Physical Layer
		[Localization
			[Regression
				[Kernel Methods \cite{5871594}]
				[Deep Learning \cite{8320781}]
				[Statistical \cite{8554304}]
			]
		]
		[Channel Equalization
			[Regression
				[Deep Learning \cite{8403666}]
				[Statistical \cite{5290078}]
			]
			[Clustering
			]
		]
		[Modulation and Coding
			[Classification
				[Deep Learning \cite{8267032,8242643}]
				[Kernel Methods \cite{8792186}]
			]
		]
		[Detection Algorithm
			[Regression
				[Deep Learning \cite{8454325}]
			]
		]
		[Channel Modeling
			[Regression
				[Deep Learning \cite{8052521,8272484}]
				[Kernel methods \cite{1315948}]
				[Statistical \cite{6595054}]
			]
			[Clustering
				[Kernel methods \cite{8013075}]
			]
		]
	]
	[Link Layer, bottom color=gray!20
		[Access Control
			[Classification
				[Reinforcement Learning \cite{8453000}]
			]
		]		
		[Rate Adaptation
			[Classification
				[Stochastic \cite{4407695}]
			]
		]
		[Fault Identification
			[Classification
				[Statistical \cite{8676028}]
				[Kernel Methods \cite{8676028}]
			]
		]
		[Frame Size Optimization
			[Regression
				[Neural Networks \cite{5075557}]
			]
		]
		[Link Quality Estimation, bottom color=gray!20
			[Classification, bottom color=gray!20
				[Statistical \cite{guo2013fuzzy, liu2011foresee, liu2012talent, cerar2020}, bottom color=gray!20 ]
				[Deep Learning \cite{luo2019link}, bottom color=gray!20]
			]
			[Regression, bottom color=gray!20
				[Statistical \cite{senel2007kalman}, bottom color=gray!20]
			]
		]
	]
	[Network Layer
		[Traffic Engineering
			[Clustering
				[Statistical \cite{6684161}]
			]
		]
		[Protocol Identification
			[Clustering
				[Statistical \cite{7004871}]
			]
			[Classification
				[Statistical \cite{8002901}]
			]
		]
		[Routing Optimization
			[Regression
				[Reinforcement Learning \cite{1420665}]
			]
		]
	]
	[Application Layer
		[QoE
			[Classification
				[Kernel Methods \cite{7167700}]
				[Statistical \cite{7167700}]
			]
		]
		[Anomaly Detection
			[Classification
				[Kernel Methods \cite{6406630}]
			]
		]
		[Service Optimization
			[Classification
				[Deep Learning \cite{8270639}]
			]
		]
	]
]
\end{forest}
\caption{Layered taxonomy of machine learning solutions for wireless communication networks.}
\label{fig:mlsol}
\end{figure*}
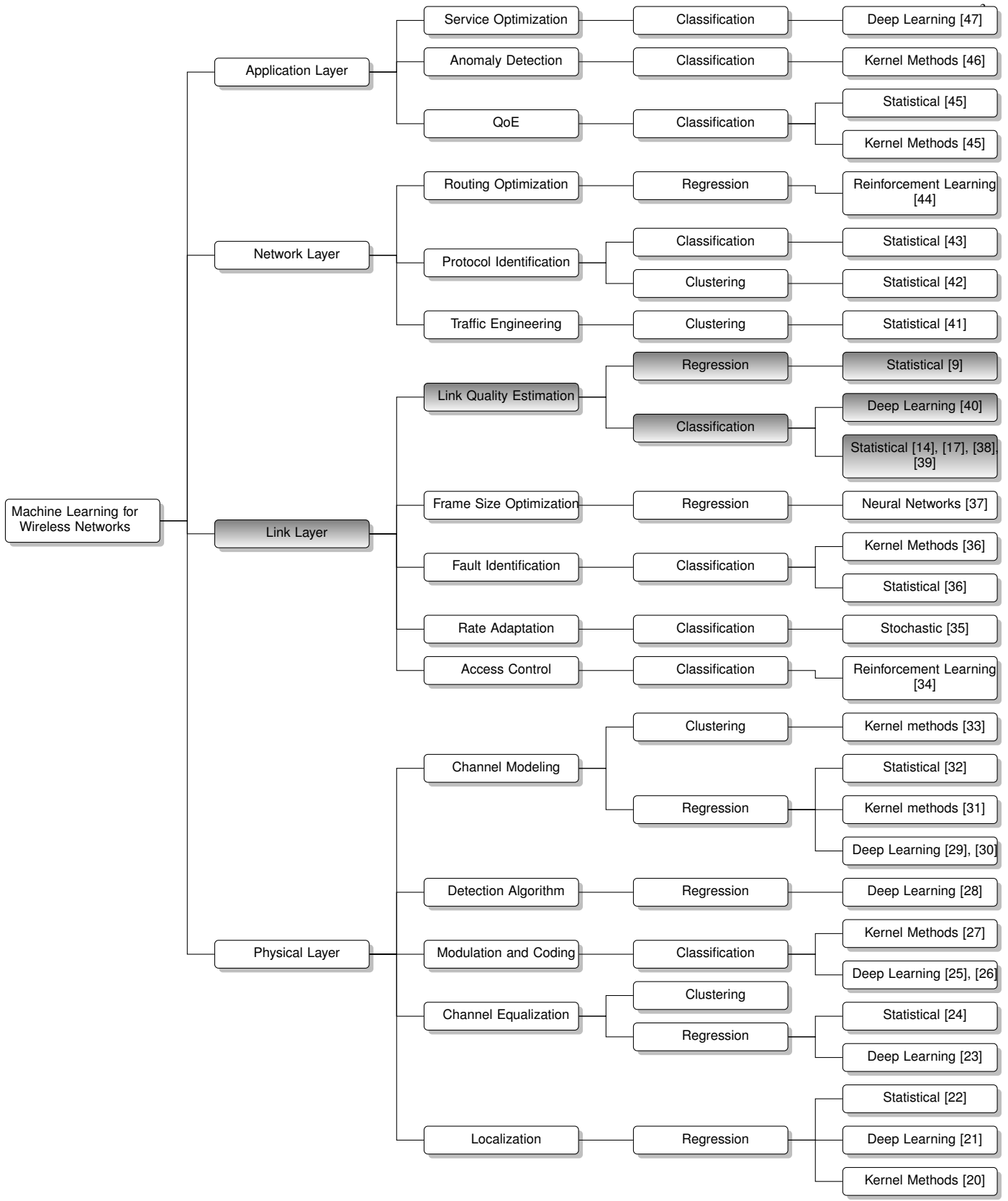

\begin{table*}[!htb]
	\centering
	\scriptsize
	\renewcommand{\arraystretch}{1.2}
	\caption{Existing surveys and tutorials relating to the terms that can define the quality of a link in the state-of-the-art literature.}
	\label{tab:existingsurveys}
	\begin{tabularx}{\linewidth}{| L{0.16} | L{1.04} | L{0.56} | L{0.24} |}
		\hline\rowcolor{gray}
		\bfseries \cellcolor{gray!20}Publication & \bfseries \cellcolor{gray!20}A summary with particular focus & \bfseries \cellcolor{gray!20}Related context in the relevant publication & \bfseries \cellcolor{gray!20}Its related section\\\hline
		
		\cite{baccour2012radio}, 2012 & A survey on empirical studies of low power links in wireless sensor networks as well as on \gls{LQE} without paying any special attention to procedures using ML techniques & Characteristics of low-power links and link quality estimation & Section V\\\hline
		
		\cite{Azarfar2012}, 2012  & A tutorial on improving the reliability of wireless communication links using cognitive radios & Failures in wireless networks & Section II-B\\\hline
		
		\cite{QDong2013}, 2013   & A survey of the techniques and protocols to handle mobility in wireless sensor networks & Prediction of link quality for mobility estimation & Section IV\\\hline
		
		\cite{SHI65170502014}, 2014 & A survey on fair resource sharing/allocation in wireless networks & The impact of link quality on packet delay & Section III-B\\\hline
		
		\cite{Gupta73174902016}, 2016 & A survey of communication related issues in unmanned aerial vehicle communication networks & Dynamic topology changes and time-varying links & Sections I-B/I-C\\\hline
		\cite{Jiang82592702018}, 2018 & A survey on link- and path-level reliable data transfer schemes in underwater acoustic networks & Channel quality control on physical layer as shown in Table II & Section III\\\hline		
		
		\cite{Alimi81134732018}, 2018 & A tutorial on key technologies of cloud access radio network optical fronthaul & Link performances of radio over fiber transport schemes illustrated in Table X & Section VII-E\\\hline
		
		\cite{Mao83821662018}, 2018 & A survey on deep learning applications for different layers of wireless networks & A brief discussion on deep learning for link evaluation & Section IV-C\\\hline
		
		\cite{Zhang86666412019}, 2019 & A survey on deep learning techniques applied to mobile and wireless networking research & Deep learning driven network control and network-level mobile data analysis & Sections I/VI\\\hline
		
		\cite{Amjad87643942019}, 2019 & A survey of effective capacity models used in various wireless networks & A brief discussion on selection of better quality links & Section VII-B\\\hline
		
		\cite{Sharma87125272019}, 2019 & A survey of current issues and machine learning solutions for massive machine type communications in ultra-dense cellular Internet of things networks & Learning link quality and reliability to adapt communication parameters & Section VI-A\\\hline
		
		This survey & A comprehensive survey of data-driven \gls{LQE} models, application quality aspects regarding the development of ML-based LQE models, ML design process for LQE models and publicly available trace-sets suitable for \gls{LQE} research. Additionally, we provide a comprehensive performance data for wireless link quality classification and for design decisions taken throughout the LQE model development. Finally, we also put forward a comprehensive lessons learned section for the development of ML-based LQE model as well as the design guidelines for ML-based LQE development and dataset collection. & Data-driven link quality estimation models & All sections\\\hline	
	\end{tabularx}
\end{table*}

\subsection{Applications of ML in wireless networks}

The use of \gls{ML} techniques in \gls{LQE} is promising to significantly improve the performance of wireless networks due to the ability of the technology to process and learn from large amount of data traces that can be collected across various technologies, topologies and mobility scenarios. These characteristics of \gls{ML} techniques empower \gls{LQE} to become much more agile, robust and adaptive. Additionally, a more generic and high level understanding of wireless links could be acquired with the aid of \gls{ML} techniques. More explicitly, an intelligent and autonomous mechanism for analyzing wireless links of any transceiver and technology can assist in better handling of current operational aspects of increasingly heterogeneous networks. This opens up a new avenue for wireless network design and optimization~\cite{fortuna2016data, jiang2017unleashing} and calls for the \gls{ML} techniques and algorithms to build robust, agile, resilient and flexible networks with minimum or no human intervention. A number of contributions for such mechanisms can be found in the literature, for instance radio spectrum observatory network is designed in~\cite{zheleva2018enabling} and~\cite{rajendran2018electrosense}. 

The diagram provided in Fig.~\ref{fig:mlsol} exhibits a broad picture of what problems are being solved by \gls{ML} in wireless networks and what broad classes of ML methods are being used for solving these particular problems. It can be observed that improvements on all layers of the communication network stack, from physical to application, are being proposed using classification, regression and clustering techniques. For each technique, algorithms having statistical, kernel, reinforcement, deep learning, and stochastic flavors are being used. The scope of the ML works analyzed in this paper is shaded with gray in Fig.~\ref{fig:mlsol} and further detailed later in Fig.~\ref{fig:mllqetax}. For a more comprehensive and intricate analysis,~\cite{Mao83821662018} and ~\cite{Zhang86666412019} survey deep learning in wireless networks, and~\cite{fortuna2009trends} surveys \gls{AI} techniques, including ML and symbolic reasoning in communication networks, but without investing any particular effort on LQE.

\subsection{Existing surveys on LQE}
To contrast our study against existing survey papers on the aspects of link quality estimation, we have identified a comprehensive list of survey and tutorial papers summarized in Table~\ref{tab:existingsurveys}. We have observed that there are existing discussions on the ``link quality'' considering various wireless networks, as outlined in Table~\ref{tab:existingsurveys}. However, only Baccour~\textit{et al.} attempted to address \gls{LQE} in~\cite{baccour2012radio}. They highlighted distinct and sometimes contradictory observations coming from a large amount of research work on \gls{LQE} based on different platforms, approaches and measurement sets. Baccour~\textit{et al.} provide a survey on empirical studies of low power links in wireless sensor networks\footnote{This survey paper is also a more recent contribution on link quality estimation models than~\cite{baccour2012radio} from 2012. Besides, we focus our attention on the data-driven \gls{LQE} models with \gls{ML} techniques.} without paying any special attention to procedures using \gls{ML} techniques. In this survey paper, we complement the aforementioned survey by analyzing the rich body of existing and recent literature on link quality estimation with the focus on model development from data traces using \gls{ML} techniques. We analyze the ML-based LQE from two complementary perspectives: application requirements and employed design process. First, we focus on how they address quality requirements that are important from the perspective of the applications they serve in Section~\ref{sec:lqe-ML-analysis-app}. Second, we analyze how they approach the standard design steps commonly used in the ML community in Section~\ref{sec:lqe-ML-analysis}. Moreover, we also review publicly available data traces that are most suitable for \gls{LQE} research.

\subsection{Contributions}

Considering recent contributions on \gls{LQE} using \gls{ML} techniques, it can be challenging to reveal the relationship between design choices and reported results. This is mainly because each model relying on \gls{ML} assumes a complex development process~\cite{fayyad1996data, witten2016data}. Each step of this process has a great potential to significantly affect the overall performance of the model, and hence these steps and their associated design choices must be well understood and carefully considered. Additionally, to provide the means for fair comparison between existing and future approaches, it is of critical importance to be able to reproduce the \gls{LQE} model development process and results~\cite{gil2007examining, van2016contextual, baker20161}, which indeed also requires open sharing of data traces.

The major contributions of this paper can be summarized as follows.

\begin{itemize}
	\item We provide a comprehensive survey of the existing literature on \gls{LQE} models developed from data traces. We analyze the state of the art from several perspectives including target technology and standards, purpose of LQE, input metrics, models utilized for LQE, output of LQE, evaluation and reproducibility. The survey reveals that the complexity of \gls{LQE} models is increasing and that comparing LQE models against each other is not always feasible.
	
	\item We provide a comprehensive and quantitative analysis of wireless link quality classification by extracting the approximated per class performance from the reported results of the literature in order to enable readers to readily distinguish the performance gaps at a glimpse.
	
	\item We analyze the performance of candidate classification-based LQEs and reveal that autoencoders, tree based methods and SVMs tend to consistently perform better than logistic regression, naive Bayes and artificial neural networks whereas the non-ML TRIANGLE estimator performs considerably well on the two, i.e., \textit{very good} and \textit{good} quality links, of the five classes included in the analysis.
	
	\item We identify five quality aspects regarding the development of an ML-based LQE that are important from the application perspective: reliability, adaptivity/reactivity, stability, computational cost and probing overhead. We provide insightful analyses on how ML-based LQE models address these five quality aspects considering the use of ML methods for a diverse set of specific problems.
	
	\item Starting from the standard \gls{ML} design process, we investigate and quantify the design decisions that the existing ML-based LQE models considered and provide insights for their potential impact on the final performance of the \gls{LQE} using the accuracy as well as the F1 score and precision vs. recall metrics.
	
	\item We survey publicly available datasets that are most suitable for \gls{LQE} research and review their available features with a comparative analysis.
	
	\item We provide an elaborated lessons learned section for the development of \gls{ML}-based \gls{LQE} model. Based on the lessons learned from this survey paper, we derive generic design guidelines recommended for the industry and research community to follow in order to effectively design the development process and collect trace-sets for the sake of LQE research.
\end{itemize}

\begin{figure}[!h]
	\centering
	\includegraphics[scale=0.78]{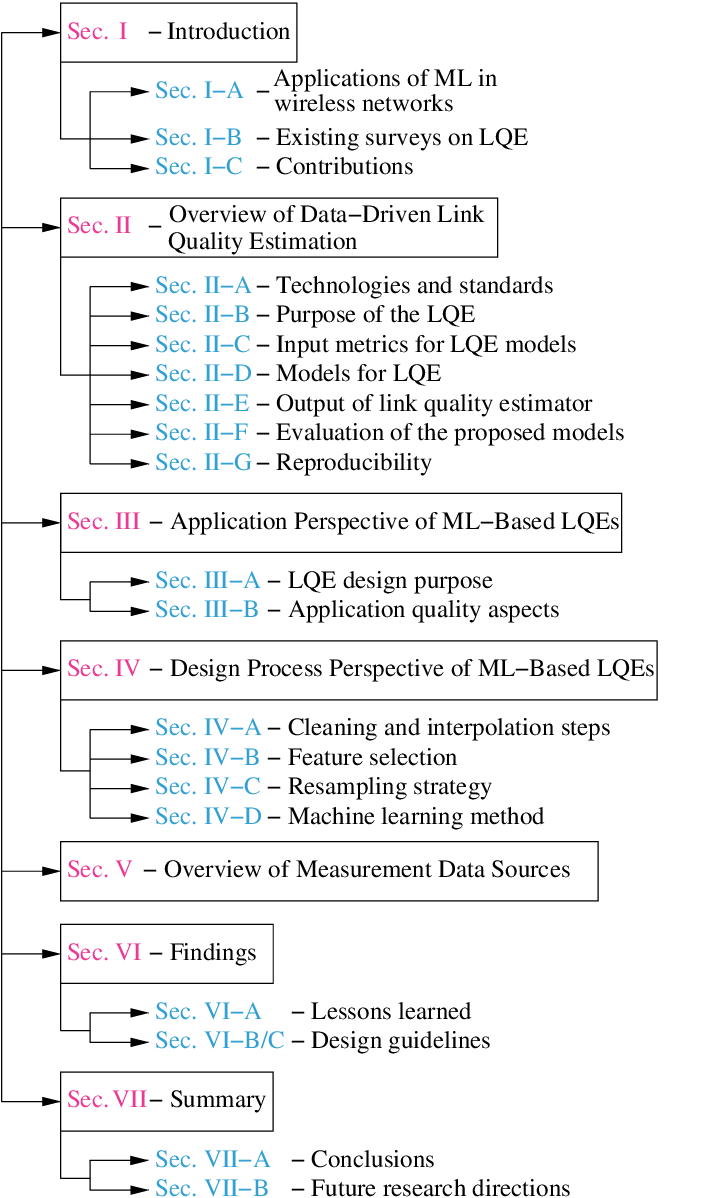}
	\caption{Structure overview of this survey paper.}
	\label{fig:structure}
\end{figure}
\begin{figure*}
	\caption{Timeline of the most prominent models in the evolution of wireless LQE.}
	\centering
	\begin{minipage}[t]{0.1\linewidth}
		\color{gray}
		\rule{\linewidth}{1pt}
		\color{black}
		\rotatebox{90}{\hspace{1cm} Machine learning  \hspace{1.59cm} Traditional approach  \hspace{0.94cm}}
		\color{gray}
		\bigskip
		\rule{\linewidth}{1pt}%
	\end{minipage}
	\begin{minipage}[t]{.7\linewidth}
		\color{gray}
		\rule{\linewidth}{1pt}
		\ytl{1996}{Link loss of pre-WiFi networks using Markov model~\cite{nguyen1996trace}}
		\ytl{1998}{Transport layer protocol that uses link loss notification~\cite{balakrishnan1998explicit}}
		\ytl{2003}{Improved multi-hop routing based on link quality model~\cite{woo2003taming}}
		\ytl{2007}{Four bit cross-layer (phy, link and network) information based LQE~\cite{fonseca2007four}}
		\ytl{2010}{Triangle, a PRR, LQI and SNR based LQE~\cite{boano2010triangle}}
		\ytl{2010}{Fuzzy logic based LQE ~\cite{baccour2010fuzzy}}
		\hrule\hrule
		\ytl{2011}{Bayes, regression and neural network LQE classification~\cite{liu2011foresee,liu2012talent}}
		\ytl{2015}{Topological features + SVM, k-NN, regression trees, gaussian regression LQE~\cite{millan2015time}}
		\ytl{2017}{Reinforcement learning-based LQE~\cite{ancillotti2017reinforcement}}
		\ytl{2017}{LQE for a mmWave base station handover system~\cite{okamoto2017machine}}
		\ytl{2019}{Stacked autoencoder + SVM based link quality estimator~\cite{luo2019link}}
		\ytl{2019}{Satellite image + network metrics LQE using SVM~\cite{demetri2019automated}}
		\bigskip
		\rule{\linewidth}{1pt}%
	\end{minipage}%
	\label{fig:evollqe}
\end{figure*}

The rest of this paper is structured as portrayed in Fig.~\ref{fig:structure}. Section~\ref{sec:overview-of-lqe} provides a comprehensive survey of the state-of-the-art literature on \gls{LQE} models built from data traces. Section~\ref{sec:lqe-ML-analysis-app} and Section~\ref{sec:lqe-ML-analysis} analyze \gls{ML}-based \gls{LQE} models from the perspective of application requirements, and of the design process, respectively. Section~\ref{sec:dataset-overview} then provides a comprehensive analysis of the open datasets suitable for \gls{LQE} research. As a result of our extensive survey, Section~\ref{sec:findings} provides lessons learned and design guidelines, while Section~\ref{sec:summary} finally concludes the paper and elaborates on the future research directions.

\section{Overview of Data-Driven Link Quality Estimation}
\label{sec:overview-of-lqe}
% Motivation
With the emergence and spread of wireless technologies in the early 90s~\cite{van1999new}, it became clear that packet delivery in wireless networks was inferior to that of wired networks~\cite{nguyen1996trace}. At the time of the experiment conducted in~\cite{nguyen1996trace}, wireless transmission medium was observed to be prone to unduly larger packet losses than the wired transmission mediums. Up until today, roughly speaking, numerous sophisticated communication techniques, including modulation and coding schemes, channel access methods, error detection and correction methods, antenna arrays, spectrum management, high frequency communications and so on, have emerged. As part of this combination of revolutionary techniques, a diverse number of estimation models for the assessment of link quality, based on actual data traces in addition to or instead of simulated models, have been proposed in the literature. 

% Pattern: motivation - approach - result
The research of data-driven \gls{LQE} based on measurement data reaches back into late 90s~\cite{nguyen1996trace} and has gained momentum particularly in the last decade~\cite{demetri2019automated}. As summarized in the timeline depicted in  Fig.~\ref{fig:evollqe}, early attempts on \gls{LQE} research mainly hinge on the recorded traces with statistical approaches and the manually developed models~\cite{nguyen1996trace, balakrishnan1998explicit, woo2003taming, senel2007kalman, fonseca2007four, srinivasan2008prr, boano2010triangle, baccour2010fuzzy, guo2013fuzzy, rekik2015fli, audeoud2018quick}. On the other hand, only after 2010, researchers have started paying a great attention to the development of \gls{LQE} model using \gls{ML} algorithms~\cite{liu2011foresee, liu2014temporal, sun2017wnn}.

To date, many analytical and statistical models have been proposed to mitigate losses and improve the performance of wireless communication. These models include channel models, radio propagation models, modulation/demodulation and encoding/decoding schemes, error correction codes, and multi-antenna systems just to name a few. Such models are essentially based on model-driven link quality estimators, where they calculate predetermined variables based on the communication parameters of the associated environment. However, their one significant shortcoming is that they abstract the real environment, and thus consider only a subset of the real phenomena. Data-driven models, on the other hand, rely on actual measured data that capture the real phenomena. The data are then used to fit a model that best approximates the underlying distribution. As it can be readily seen in Fig.~\ref{fig:evollqe}, up until 2010, statistical approaches were the favored tools for \gls{LQE} research. From then on, as in other research areas of wireless communication, portrayed in Fig.~\ref{fig:mlsol}, \gls{ML}-based models replaced the conventional approaches and became the preferred tool for \gls{LQE} research.

% Approach
Empirical observation of wireless link traffic is a crucial part of the data-driven \gls{LQE}. An observation of link quality metrics within a certain estimation window, e.g. time interval or a discrete number of events, allows for constructing different varieties of data-driven link quality estimators. However, there are a few drawbacks of the data-driven approaches that need to be taken into account. Since the ultimate model strictly depends on the recorded data traces, it has to be carefully designed in a way that records adequate information about the underlying distribution of the phenomena. If sufficient measurements of the distribution can be captured, then it is possible to automatically build a model that can approximate that particular distribution. Data-driven \gls{LQE} models are in no way meant to fully replace or supersede model-driven estimators but to complement them. It is certainly possible to incorporate a model-driven estimator into a data-driven one as the input data.

To some extent, different varieties of data-driven metrics and estimators were studied in~\cite{baccour2012radio}, where the authors made three independent distinctions among hardware- and software-based link quality estimators. The software-based estimators are further split into \gls{PRR}-based, \gls{RNP}-based, and score-based subgroups. The first distinction is based on the estimator's origin presenting the way how they were obtained. The second distinction is based on the mode their data collection was done, which can be in passive, active and/or hybrid manner, depending on whether dummy packet exchange was triggered by an estimator. The third distinction is based on which side of the communication link was actively involved. \gls{LQE} metrics can be gathered either on the receiver, transmitter or both sides.

Going beyond~\cite{baccour2012radio}, Tables~\ref{tab:lqe-part1} and~\ref{tab:lqe-part2} provide a comprehensive summary of the most related publications that leverage a data-driven approach for \gls{LQE} research. All the studies summarized in Tables~\ref{tab:lqe-part1} and \ref{tab:lqe-part2} rely on real network data traces recorded from actual devices. The first column in Tables~\ref{tab:lqe-part1} and~\ref{tab:lqe-part2} contains the title, reference and the year of publication. The second column provides the testbed, the hardware and the technology used in each publication, whereas the third column lists the objectives of these publications with respect to \gls{LQE} approach. Columns four, five and six focus on the characteristics of the estimators, particularly on their corresponding input(s), model and output. The last two columns summarize statistical aspects of the data traces and their public availability of the trace-sets for reproducibility, respectively.

\subsection{Technologies and standards}
\label{sec:techstandards}
% Common denominators
As outlined in the second column of Tables~\ref{tab:lqe-part1} and \ref{tab:lqe-part2}, earlier studies on \gls{LQE} were performed on WaveLAN~\cite{nguyen1996trace,balakrishnan1998explicit}, a precursor on the modern Wi-Fi. The study in~\cite{nguyen1996trace} aimed to characterize the loss behavior of proprietary AT\&T WaveLAN. It used packet traces with various configurations for the transmission rate, packet size, distance and the corresponding packet error rate. Then, they built a two-state Markov model of the link behavior. The same model was then utilized in~\cite{balakrishnan1998explicit} to estimate the quality of wireless links in the interest of improving \gls{TCP} congestion performance. More recently,~\cite{okamoto2017machine,bote2018online} used IEEE~802.11 standard in their studies for throughput and online link quality estimators.

Later on, the majority of publications related to \gls{LQE} focused on wireless sensor networks relying on IEEE~802.15.4 standard and only a few targeted other type of wireless networks, such as Wi-Fi (IEEE~802.11) or Bluetooth (IEEE~802.15.1). This can be explained by the fact that IEEE 802.15.4-based wireless sensor networks are relatively cheaper to deploy and maintain. Perhaps, the first such larger testbed was available at the University of Berkley~\cite{woo2003taming} using MicaZ nodes and TinyOS~\cite{levis2005tinyos}, which is an open source operating system for constrained devices. Other hardware platforms, such as TelosB and TMote, and operating systems, e.g. Contiki, have emerged and enabled researchers to further experiment with improving the performance of single and multi-hop communications for wireless networks composed of battery-powered devices. 

Finally, one recent contribution focuses on LoRA technology, a type of Low Power Wide Area Network (LPWAN) for estimating the quality of links, and therefore aiming for the improvement of the coverage for the technology~\cite{demetri2019automated}.

Whereas earlier research on \gls{LQE} leveraged proprietary technologies~\cite{nguyen1996trace}, wireless sensor networks utilized relatively low cost hardware and open source software, therefore enabled a broader effort from the research community. This resulted in a large wave of research focusing on ad-hoc, mesh and multihop communications~\cite{woo2003taming, fonseca2007four, baccour2010fuzzy, liu2011foresee, liu2012talent, guo2013fuzzy, rekik2015fli, sun2017wnn, shu2017research,audeoud2018quick,luo2019link}, all of which rely on the estimation of link quality. The nodes implementing the aforementioned technologies are still being maintained in various university testbeds.

\begin{table*}[!htbp]
	\centering
	\scriptsize
	\renewcommand{\arraystretch}{1.2}
	\caption{Existing work on link quality estimation using real network data traces (Part 1 of 2)}
	\label{tab:lqe-part1}
	\begin{tabularx}{\linewidth}{| C{1.4} | C{1.} | C{1.} | C{1.} | C{1.} | C{1.} | C{1.} | C{.6} |}
		
		\hline\rowcolor{gray}
		\bfseries Title
		& \bfseries Tech.
		& \bfseries Goal
		& \bfseries Input % Metrics~/~Features
		& \bfseries Model
		& \bfseries Output % Actual output
		& \bfseries Data
		& \bfseries Reproduce
		\\\hline
		
		A trace-based approach for modeling wireless channel behavior~\cite{nguyen1996trace}, 1996 % Title
		& WaveLAN, BARWAN testbed, BSD~2.1 % Tech
		& Maximize throughput, channel error model % Goal
		& SNR, signal quality, throughput, PRR % Input
		& Improved two-state Markov model % Model
		& Probability of error to occur and persist % Output
		& Not specified ($<$1500~bytes/packet, 1000~s/trace) % Data
		& No* % Reproduce
		\\\hline 
		
		% Complete article available on: http://nms.lcs.mit.edu/~hari/papers/globecom98/
		Explicit loss notification and wireless web performance~\cite{balakrishnan1998explicit}, 1998 % Title
		& WaveLAN, University of California testbed % Tech
		& Improve TCP Reno on wireless links, maximize throughput % Goal
		& Bitrate, packet size, no. bits, throughput, BER % Input
		& CDF of error and error-free durations % Model
		& Probability of error to occur and persist % Output
		& 800\,000~packets (100\,000~packets/experiment, 8~experiments) % Data
		& No* % Reproduce
		\\\hline
		
		Taming the underlying challenges of reliable multihop routing in sensor networks~\cite{woo2003taming}, 2003 % Title
		& Proprietary, MicaZ mote, TinyOS % Tech
		& Improve routing table management % Goal
		& PRR % Input
		& Shortest path, minimum transmission, broadcast, destination sequenced distance vector % Model
		& Decision on keep/remove routing table entry % Output
		& $\approx$600\,000~packets (8~packets/s, 200~packets/P$_\textrm{Tx}$) % Data
		& No* % Reproduce
		\\\hline
		
		(4B) Four-bit wireless link estimation~\cite{fonseca2007four}, 2007 % Title
		& Intel~Mirage: 85x~MicaZ; USC~TutorNet: 94x~TelosB; IEEE~802.15.4, TinyOS % Tech
		& Improve routing table management % Goal
		& LQI, PRR, broadcast, ACK~count % Input
		& Construct 4-bit score of link state % Model
		& Estimated link quality % Output
		& Mirage: N.A., 40-69~min/experiment; TutorNet: N.A., 3-12h/experiment; % Data
		& No* % Reproduce
		\\\hline
		
		A Kalman filter-based link quality estimation scheme for wireless sensor networks~\cite{senel2007kalman}, 2007 % Title
		& TelosB, IEEE~802.15.4 % Tech
		& PRR estimation % Goal
		& RSSI, noise floor % Input
		& Kalman filter + SNR~to~PRR mapping % Model
		& PRR estimation % Output
		& 25\,200\,000 (500~samples/s, 14~h) % Data
		& No % Reproduce
		\\\hline
		
		PRR is not enough~\cite{srinivasan2008prr}, 2008 % Title
		& IEEE~802.11, IEEE~802.15.4 % Tech
		& Link state estimation % Goal
		& PRR % Input
		& Gilbert-Elliott Model (2-state Markov process); \textit{good} and \textit{bad} state % Model
		& Link quality transition probability % Output
		& Rutgers and Mirage trace-sets % Data
		& Yes % Reproduce
		\\\hline
		
		The triangle metric: fast link quality estimation for mobile wireless sensor networks~\cite{boano2010triangle}, 2010 % Title
		& Tmote Sky, Sentilla JCreate, IEEE~802.15.4, Contiki~OS % Tech
		& New LQE % Goal
		& RSSI, noise floor, LQI % Input
		& Pythagorean equation maps to distance from the origin (hypotenuse) % Model
		& Estimated link quality as \textit{very good}, \textit{good}, \textit{average} or \textit{bad} % Output
		& 30\,000 + N.A., (64~packets/s, all channels, unicast) % Data
		& No % Reproduce
		\\\hline
		
		F-LQE: A fuzzy link quality estimator for wireless sensor networks, \cite{baccour2010fuzzy}~2010, \cite{baccour2011radiale}~2011 % Title
		& RadiaLE testbed, 49x~TelosB, IEEE~802.15.4, TinyOS % Tech
		& Link quality estimation, improve routing % Goal
		& PRR % Input
		& Fuzzy logic maps current to estimated link quality % Model
		& Binary high/low-quality (HQ/LQ) link estimation % Output
		& N.A. (bursts, packet sizes, 20-26 channel) % Data
		& No* % Reproduce
		\\\hline
		
		Foresee (4C): Wireless link prediction using link features~\cite{liu2011foresee}, 2011 % Title
		& 54x~Tmote (local), 180x~Tmote Sky (Motelab), IEEE~802.15.4, % Tech
		& Improve routing % Goal
		& PRR, RSSI, SNR, LQI % Input
		& Logistic regression model % Model
		& Probability of receiving next packet % Output
		& 80\,000 + 80\,000 noise floor ($\approx$10~packets/s) % Data
		& No* % Reproduce
		\\\hline
		
		Fuzzy logic-based multidimensional link quality estimation for multihop wireless sensor networks \cite{guo2013fuzzy}, 2013 % Title
		& (local) 15x~TelosB, TinyOS, IEEE~802.15.4 % Tech
		& Improve routing, minimize topology changes % Goal
		& PRR % Input
		& Fuzzy logic link quality estimator % Model
		& Binary high/low-quality link estimation % Output
		& N.A., (20~min/experiment, 12h) % Data
		& No % Reproduce
		\\\hline
		
		Temporal adaptive link quality prediction with online learning, \cite{liu2012talent}~2012, \cite{liu2014temporal}~2014 % Title
		& Motelab, Indriya and (local) 54x~Tmote testbed, IEEE~802.15.4 % Tech
		& Link quality estimation, improve Routing % Goal
		& PRR, RSSI, SNR, LQI % Input
		& Logistic regression with SGD and s-ALAP adaptive learning rate % Model
		& Binary, estimates if link quality above desired threshold % Output
		& 480\,000, (30 bytes size, 6\,000 per exp., 10/sec.), Rutgers and Colorado trace-sets % Data
		& No~\cite{liu2012talent} Yes~\cite{liu2014temporal} % Reproduce
		\\\hline
		
		Low-Power link quality estimation in smart grid environments~\cite{rekik2015fli}, 2015 % Title
		& IEEE~802.15.4 % Tech
		& Improve routing, LQE reactivity % Goal
		& RNP, SNR, PRR % Input
		& Optimized F-LQE \cite{baccour2010fuzzy} with better reactivity % Model
		& Binary high/low-quality link estimation % Output
		& N.A., 500kV substation env. data, TOSSIM~2 simulator % Data
		& No % Reproduce
		\\\hline
		
		Time series analysis to predict link quality of wireless community networks~\cite{millan2015time}, 2015 % Title
		& Conventional routers, IEEE~802.15.4, IEEE~802.11, AX.25, (FunkFeuer mesh network) % Tech
		& Link quality estimation, regression, clustering, time-series analysis % Goal
		& LQ, NLQ, ETX % Input
		& SVM, k-nearest neighbor, regression trees, Gaussian process for regression % Model
		& Predicted LQ value for different windows sizes % Output
		& N.A., (404 nodes, 2\,095 links, 7 days of data) % Data
		& No* % Reproduce
		\\\hline
		
		Machine-learning based channel quality and stability estimation for stream-based multichannel wireless sensor networks~\cite{rehan2016machine}, 2016 % Title
		& CC2420, IEEE~802.15.4, Matlab simulation % Tech
		& Evaluation of new algorithm with two possible extensions % Goal
		& RSSI, LQI, channel rank, channel % Input
		& Normal equation-based channel quality prediction, weighted input extension, stability extension % Model
		& Channel quality estimation based on 3-class estimator % Output
		& Simulation % Data
		& Yes % Reproduce
		\\\hline
		
		WNN-LQE: Wavelet-neural-network-based link quality estimation for smart grid WSNs~\cite{sun2017wnn}, 2017 % Title
		& 10x~CC2530 WSNs, IEEE~802.15.4 % Tech
		& Improve routing, estimate PRR range % Goal
		& SNR % Input
		& Wavelet-neural-network-based link quality estimator % Model
		& Upper and lower bound of confidence interval for PRR % Output
		& 2\,500 (20 bytes size, 3.33 per second) % Data
		& No % Reproduce
		\\\hline
		
		\multicolumn{8}{l}{
			\footnotesize Note: Asterisk (*) indicates that the experiment was performed on a public testbed, but no data is available.
		}
	\end{tabularx}
\end{table*}
\begin{table*}[!htp]
	\centering
	\scriptsize
	\renewcommand{\arraystretch}{1.2}
	\caption{Existing work on link quality estimation using real network data traces (Part 2 of 2)}
	\label{tab:lqe-part2}
	\begin{tabularx}{\linewidth}{| C{1.4} | C{1.} | C{1.} | C{1.} | C{1.} | C{1.} | C{1.} | C{.6} |}
		
		\hline\rowcolor{gray}
		\bfseries Title
		& \bfseries Tech.
		& \bfseries Goal
		& \bfseries Input % Metrics~/~Features
		& \bfseries Model
		& \bfseries Output % Actual output
		& \bfseries Data
		& \bfseries Reproduce
		\\\hline
		
		A reinforcement learning-based link quality estimation strategy for RPL and its impact on topology management~\cite{ancillotti2017reinforcement}, 2017 % Title
		& \textbf{Sim.:} Cooja simulator (Contiki~3.x); \textbf{Exp.:} 23x~TelosB, CC2420, IEEE~802.15.4 % Tech
		& Improve RPL protocol % Goal
		& PER, RSSI, energy consumption % Input
		& Unsupervised ML % Model
		& PRR estimation % Output
		& \textbf{Sim.:} $\infty$; \textbf{Exp.:} N.A., 178 links, mobile nodes (0.5~m/s), University of Pisa % Data
		& \textbf{Sim.:} Yes; \textbf{Exp.:} No % Reproduce
		\\\hline
		
		Research on Link Quality Estimation Mechanism for Wireless Sensor Networks Based on Support Vector Machine~\cite{shu2017research}, 2017 % Title
		& 2x~TelosB, CC2420, IEEE~802.15.4, TinyOS~2.x % Tech
		& link quality estimation, comparison % Goal
		& RSSI, LQI, PRR % Input
		& SVM classifier % Model
		& Classification, 5 classes % Output
		& 121 datapoints % Data
		& No % Reproduce
		\\\hline
		
		Machine-learning-based throughput estimation using images for mmWave communications~\cite{okamoto2017machine}, 2017 % Title
		& 2x~IEEE~802.11ad @ 60~GHz (mmWave), RGB-D camera (Kinect) % Tech
		& Throughput estimation, obstacle detection, comm. handover w/o control frames % Goal
		& Throughput, depth value (Kinect) % Input
		& Online adaptive regularization of weight vectors (AROW) % Model
		& regression, throughput estimation % Output
		& N.A. % Data
		& No % Reproduce
		\\\hline
		
		Quick and efficient link quality estimation in wireless sensors networks~\cite{audeoud2018quick}, 2018 % Title
		& Grenoble testbed FIT-IoT, 28x~AT86RF231, IEEE~802.15.4 % Tech
		& Analysis of LQI, fast decisions, improve routing % Goal
		& LQI % Input
		& Classification based on arbitrary values % Model
		& Classify link as \textit{good}, \textit{uncertain} or \textit{weak} % Output
		& N.A. (2\,000 per link, 16 channels) % Data
		& No* % Reproduce
		\\\hline
		
		Online ML algorithms to predict link quality in community wireless mesh networks~\cite{bote2018online}, 2018 % Title
		& Conventional routers, IEEE~802.15.4, IEEE~802.11, AX.25, (FunkFeuer mesh network) % Tech
		& Link quality estimation, online regression, compares online ML algorithms % Goal
		& LQ, NLQ, ETX % Input
		& online perceptrons, online regression trees, fast incremental model trees, adaptive model rules % Model
		& Metric estimation, regression % Output
		& N.A. ($\approx$500 nodes, $\approx$2\,000 links, FunkFeuer distributed community network) % Data
		& No* % Reproduce
		\\\hline
		
		Link Quality Estimation Method for Wireless Sensor Networks Based on Stacked Auto-encoder~\cite{luo2019link}, 2019 % Title
		& 8x~TelosB, TinyOS, IEEE~802.15.4 % Tech
		& Link quality estimation, classification % Goal
		& SNR, RSSI, LQI, and PRR from transmitter and receiver % Input
		& Neural network-based classification % Model
		& Estimated link quality as \textit{very bad}, \textit{bad}, \textit{common}, \textit{good}, \textit{very good} % Output
		& N.A., interior corridors, grove, parking lots, road % Data
		& No % Reproduce
		\\\hline
		
		Automated Estimation of Link Quality for LoRa: A Remote Sensing Approach~\cite{demetri2019automated}, 2019 % Title
		& Dragino LoRa 1.3 (RF96 chip), LoRa % Tech
		& Link quality estimation, environment classification % Goal
		& Node/Gateway position, time-stamp, RSSI, SNR, multispectral aerial images % Input
		& SVM classification of LoRa coverage % Model
		& Mapping LoRa coverage onto geographical map % Output
		& 8\,642 samples, 23 sites, 1 packet per 40s, Delft (NL) % Data
		& No % Reproduce
		\\\hline
		
		On Designing a Machine Learning Based Wireless Link Quality Classifier~\cite{cerar2020}, 2020 % Title
		& 29x IEEE~802.11 % Tech
		& Link quality prediction, importance of preprocessing % Goal
		& RSSI % Input
		& logistic, regression, SVM, decision trees, random forest, multi-layer perceptron % Model
		& Classification of future link state as \textit{good}, \textit{intermediate} or \textit{bad}% Output
		& Rutgers dataset % Data
		& Yes % Reproduce
		\\\hline
		
		\multicolumn{8}{l}{
			\footnotesize Note: Asterisk (*) indicates that the experiment was performed on a public testbed, but no data is available.
		}
	\end{tabularx}
\end{table*}

\subsection{Purpose of the LQE}
\label{subsec:purpose}
With respect to the research goal summarized in the third column of Tables~\ref{tab:lqe-part1} and~\ref{tab:lqe-part2}, the surveyed papers can be categorized into two broad groups. The goal of the first group was to improve the performance of a protocol or process. The goal of the second group of papers was to propose a new or improve an existing link quality estimator. For this class of papers, any protocol improvement in the evaluation process was secondary.

\subsubsection{\gls{LQE} for protocol performance improvement} The authors of~\cite{balakrishnan1998explicit, nguyen1996trace} investigated \gls{TCP} performance improvement, whereas others focused on routing protocol performance. This group of papers proposed a novel link quality estimators as an intermediate step towards achieving their goal, e.g. performance improvement of \gls{TCP}, routing optimization and so on.

One of the earliest publications from this group is~\cite{woo2003taming} that aimed for improving the reactivity of routing tables in constrained devices, such as sensor nodes. They collected traces of transmissions for nodes located at various distances with respect to each other. Then, they computed reception probabilities as a function of distances and evaluated a number of existing link estimation metrics. They also proposed a new link estimation metric called \gls{WMEWMA} and showed an improvement in network performance as a result of more appropriate routing table updates. The improvements were shown both in simulations and in experimentation. This study was also among the earliest studies introducing the three different grade regions of wireless links, i.e., \textit{good}, \textit{intermediate} and \textit{bad}. 

Later,~\cite{fonseca2007four} noticed that by considering additional metrics alongside \gls{WMEWMA}, also from higher levels of the protocol stack, the link estimation could be better coupled with data traffic. Therefore, they introduced a new estimator referred to as \gls{4B}, where they combined information from the physical (\gls{PRR}, \gls{LQI}), link (ACK count) and network layers (routing) and demonstrated that it performs better than the baseline they chose for the evaluation. 

In~\cite{baccour2010fuzzy}, the authors developed a new link quality estimator named \gls{F-LQE} that is based on fuzzy logic, which exploits average values, stability and asymmetry properties of \gls{PRR} and \gls{SNR}. As for the output, the model classifies links as high-quality (HQ) or low-quality (LQ). The same authors compared \gls{F-LQE} against \gls{PRR}, \gls{ETX}~\cite{de2005high-ETX}, \gls{RNP}~\cite{cerpa2005temporal-RNP} and \gls{4B}~\cite{fonseca2007four} on the RadiaLE testbed~\cite{baccour2011radiale}. The comparison of the metrics was performed using different scenarios including various data burst lengths, transmission powers, sudden link degradation and short bursts. Among their findings, they showed that \gls{PRR}, \gls{WMEWMA} and \gls{ETX}, which are \gls{PRR}-based link quality estimators, overestimate the link quality, while \gls{RNP} and \gls{4B} underestimate the link quality. The authors of~\cite{baccour2011radiale} demonstrated that \gls{F-LQE} performed better estimation than the other estimators compared.

The authors of~\cite{guo2013fuzzy} used fuzzy logic and proposed a \gls{FLI} for link quality estimation. The \gls{FLI} model uses \gls{PRR}, the coefficient of variance of \gls{PRR} and the quantitative description of packet loss burst, which are gathered independently, while the previous \gls{F-LQE}~\cite{baccour2010fuzzy} requires information sharing of \gls{PRR}. \gls{FLI} was evaluated in a testbed for 12 hours worth of simulation time against \gls{4B}~\cite{fonseca2007four}, and it was reported to perform better. 

\gls{4C}~\cite{liu2011foresee} is the first metric from this group focused on protocol improvement that introduced statistical \gls{ML} techniques. The authors used \gls{RSSI}, \gls{SNR}, \gls{LQI}, \gls{WMEWMA} and smoothed \gls{PRR} as input features into the models. They trained three \gls{ML} models based on na\" ive Bayes, neural networks and logistic regression. TALENT~\cite{liu2012talent} was then improved on \gls{4C} by introducing adaptive learning rate.

More recently, \cite{ancillotti2017reinforcement} proposed enhancement to the RPL protocol, which is used in lossy wireless networks. This backward compatible improvement (mRPL) for mobile scenarios introduces asynchronous transmission of probes, which observe link quality and trigger the appropriate action.

\subsubsection{New or improved link quality estimator} 
Srinivasan~\textit{et al.}~\cite{srinivasan2008prr} proposed a two-state model with \textit{good} and \textit{bad} states, and 4 transition probabilities between the states to improve on the existing \gls{WMEWMA}~\cite{fonseca2007four} and \gls{4B}~\cite{fonseca2007four}. Then, Senel~\textit{et al.}~\cite{senel2007kalman} took a different approach and developed a new estimator by predicting the likelihood of a successful packet reception. Besides, Boano \textit{et al.}~\cite{boano2010triangle} introduced the TRIANGLE metric that uses the Pythagorean equation and computes the distance between the instant \gls{SNR} and \gls{LQI}. This study identifies four different link quality grades including \textit{very good}, \textit{good}, \textit{average} and \textit{bad} links. Some of the classifiers propose a five-class model~\cite{shu2017research,luo2019link} and a three-class model~\cite{audeoud2018quick,cerar2020} for \gls{LQE} research. Other \gls{LQE} models leverage regression rather than classification in order to generate a continuous-valued estimate of the link~\cite{okamoto2017machine,bote2018online,demetri2019automated}.

\begin{table*}[!thb]
	\centering
	\scriptsize
	\renewcommand{\arraystretch}{1.2}
	\caption{Metrics that can be used to measure the quality of a link.}
	\label{tab:feature-analysis}
	\begin{tabularx}{\linewidth}{| C{.8} | c | c | c | c | c | c | c | c | c | c | C{1.2} |}
		\hline%\rowcolor{gray}
		\bfseries Link quality
		& \bfseries Hardware
		& \multicolumn{3}{c|}{\bfseries Software-based}
		& \bfseries Image
		& \bfseries Topological
		& \multicolumn{2}{c|}{\bfseries Sides involved}
		& \multicolumn{2}{c|}{\bfseries Gathering method}
		& % empty
		\\\cline{3-5}\cline{8-11}
		
		%\rowcolor{gray}
		\bfseries  metrics
		& \bfseries based
		& \bfseries PRR-based
		& \bfseries RNP-based
		& \bfseries Score-based
		& \bfseries based
		&
		& \bfseries Rx
		& \bfseries Tx
		& \bfseries Passive
		& \bfseries Active
		& \multirow{-2}{*}{\bfseries Related base-metric(s)}
		\\\hline
		
		\gls{RSSI} % feature
		& \cmark % HW-based?
		& % PRR-based?
		& % RNP-based?
		& % Score-based?
		& % Image-based?
		& % topological
		& \cmark % Rx-side?
		& % Tx-side?
		& \cmark % pasive?
		& % active?
		& \gls{RSS}, SNR % Relation
		\\\hline
		
		\gls{LQI} % feature
		& \cmark % HW-based?
		& % PRR-based?
		& % RNP-based?
		& % Score-based?
		& % Image-based?
		& % topological
		& \cmark % Rx-side?
		& % Tx-side?
		& \cmark % pasive?
		& % active?
		& Vendor-specific % Relation
		\\\hline
		
		SNR % feature
		& \cmark % HW-based?
		& % PRR-based?
		& % RNP-based?
		& % Score-based?
		& % Image-based?
		& % topological
		& \cmark % Rx-side?
		& % Tx-side?
		& \cmark % pasive?
		& % active?
		& \gls{RSS}, noise floor % Relation
		\\\hline
		
		BER % feature
		& \cmark % HW-based?
		& % PRR-based?
		& % RNP-based?
		& % Score-based?
		& % Image-based?
		& % topological
		& \cmark % Rx-side?
		& % Tx-side?
		& \cmark % pasive?
		& % active?
		& -- % Relation
		\\\hline
		
		PRR % feature
		& % HW-based?
		& \cmark % PRR-based?
		& % RNP-based?
		& % Score-based?
		& % Image-based?
		& % topological
		& \cmark % Rx-side?
		& % Tx-side?
		& \cmark % pasive?
		& % active?
		& PER % Relation
		\\\hline
		
		WMEWMA % feature
		& % HW-based?
		& \cmark % PRR-based?
		& % RNP-based?
		& % Score-based?
		& % Image-based?
		& % topological
		& \cmark % Rx-side?
		& % Tx-side?
		& \cmark % pasive?
		& % active?
		& PER, PRR % Relation
		\\\hline
		
		4B % feature
		& % HW-based?
		& % PRR-based?
		& \cmark % RNP-based?
		& % Score-based?
		& % Image-based?
		& % topological
		& \cmark % Rx-side?
		& \cmark % Tx-side?
		& \cmark % pasive?
		& \cmark % active?
		& LQI, PRR, ACK, broadcast % Relation
		\\\hline
		
		LQ, NLQ % feature
		& % HW-based?
		& % PRR-based?
		& \cmark % RNP-based?
		& % Score-based?
		& % Image-based?
		& % topological
		& \cmark % Rx-side?
		& \cmark % Tx-side?
		& % pasive?
		& \cmark % active?
		& --  % Relation
		\\\hline
		
		ETX % feature
		& % HW-based?
		& % PRR-based?
		& \cmark % RNP-based?
		& % Score-based?
		& % Image-based?
		& % topological
		& \cmark % Rx-side?
		& \cmark % Tx-side?
		& % pasive?
		& \cmark % active?
		& LQ, NLQ % Relation
		\\\hline
		
		4C % feature
		& % HW-based?
		& % PRR-based?
		& % RNP-based?
		& \cmark % Score-based?
		& % Image-based?
		& % topological
		& \cmark % Rx-side?
		& % Tx-side?
		& \cmark % pasive?
		& % active?
		& LQI, PRR, SNR, RSSI % Relation
		\\\hline
		
		TRIANGLE % feature
		& % HW-based?
		& % PRR-based?
		& % RNP-based?
		& \cmark % Score-based?
		& % Image-based?
		& % topological
		& \cmark % Rx-side?
		& % Tx-side?
		& \cmark % pasive?
		& % active?
		& SNR, LQI % Relation
		\\\hline
		
		Image-based% feature
		& % HW-based?
		& % PRR-based?
		& % RNP-based?
		& % Score-based?
		& \cmark  % Image-based?
		& % topological
		& % Rx-side?
		& % Tx-side?
		& % pasive?
		& % active?
		& % Relation
		\\\hline
		
		Topological% feature
		& % HW-based?
		& % PRR-based?
		& % RNP-based?
		& % Score-based?
		& % Image-based?
		& \cmark  % topological
		& % Rx-side?
		& % Tx-side?
		& % pasive?
		& % active?
		& % Relation
		\\\hline
		
	\end{tabularx}
\end{table*}
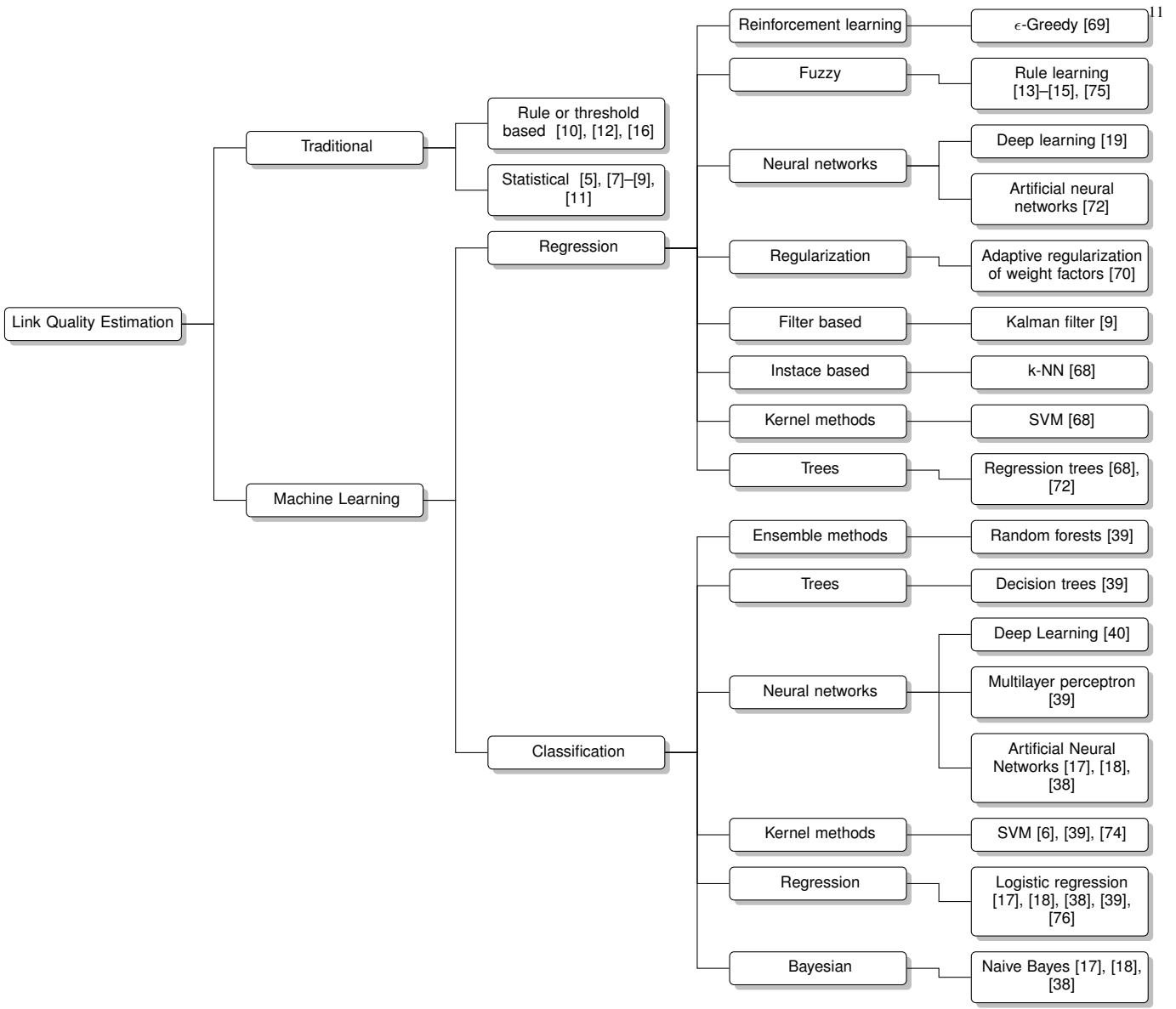
\begin{figure*}[]
	\vspace{-1cm}
	\newcolumntype{H}[1]{>{\centering}p{#1}}
	\begin{forest}
		forked edges,
		for tree={
			if level=0{align=center}{%
				align={@{}H{26mm}@{}},
			},
			grow=east,
			text width=2.53cm,
			draw,
			font=\sffamily\scriptsize,
			edge path={
				\noexpand\path [draw, \forestoption{edge}] (!u.parent anchor) -| +(0.5mm,0) -- +(5mm,0) |- (.child anchor)\forestoption{edge label};
			},
			parent anchor=east,
			child anchor=west,
			l sep=10mm,
			tier/.wrap pgfmath arg={tier #1}{level()},
			edge={thin},
			fill=white,
			rounded corners=2pt,
			drop shadow,
		}
		[Link Quality Estimation
		[Machine Learning
		[Classification
		[Bayesian 
		[Naive Bayes \cite{liu2011foresee,liu2012talent, liu2014temporal} ]
		]
		[Regression 
		[Logistic regression \cite{liu2011foresee,liu2012talent,liu2014temporal,cerar2020,rehan2016machine} ]
		]
		[Kernel methods 
		[SVM \cite{shu2017research,demetri2019automated,cerar2020}]
		]
		[Neural networks 
		[Artificial Neural Networks \cite{liu2011foresee,liu2012talent,liu2014temporal} ]
		[Multilayer perceptron \cite{cerar2020}]
		[Deep Learning \cite{luo2019link}]
		]
		[Trees
		[Decision trees \cite{cerar2020}]
		]
		[Ensemble methods
		[Random forests \cite{cerar2020}]
		]
		]
		[Regression
		[Trees
		[Regression trees \cite{millan2015time,bote2018online}]
		] 
		[Kernel methods
		[SVM \cite{millan2015time}]
		]
		[Instace based
		[k-NN \cite{millan2015time}]
		]
		[Filter based 
		[Kalman filter\cite{senel2007kalman}]
		]
		[Regularization 
		[Adaptive regularization of weight factors\cite{okamoto2017machine}]
		]
		[Neural networks 
		[Artificial neural networks \cite{bote2018online}]
		[Deep learning \cite{sun2017wnn}]
		]
		[Fuzzy 
		[Rule learning ~\cite{baccour2010fuzzy,rekik2015fli,guo2013fuzzy,baccour2011radiale}]
		]
		[Reinforcement learning 
		[$\epsilon$-Greedy \cite{ancillotti2017reinforcement} ]
		]
		]
		]
		[Traditional 
		[Statistical ~\cite{nguyen1996trace, balakrishnan1998explicit, woo2003taming, senel2007kalman,srinivasan2008prr}]
		[Rule or threshold \\ based ~\cite{fonseca2007four, boano2010triangle, audeoud2018quick}]
		]
		]
	\end{forest}
	\caption{Taxonomy of the LQE approaches using ML algorithms and traditional methods.}
	\label{fig:mllqetax}
\end{figure*}

\subsection{Input metrics for LQE models}
\label{subsec:input}
With respect to the input metrics used for estimating the quality of a link summarized in the fourth column of Tables~\ref{tab:lqe-part1} and~\ref{tab:lqe-part2}, we distinguish between the single and the multiple metric approaches. Single metric approaches use a one dimension vector while multiple metric approaches use a multidimensional vector as input for developing a model.

\textit{Single metric input} approaches have a number of advantages. The trace-set is smaller and thus often easier to collect, the model typically requires less computational power to compute, and as shown in~\cite{liu2011foresee} they can be more straightforward to implement, especially on constrained devices. However, by only analyzing and relying on a single measured variable, such as RSSI, important information might be left out. For this reason, it is better to collect traces with \textit{several, possibly uncorrelated metrics}, each of them being able to contribute meaningful information to the final model. A good example of the latter is using RSSI and spectral images for instance. 

The estimators surveyed based on single input metric appear in~\cite{woo2003taming, srinivasan2008prr, sun2017wnn, audeoud2018quick,cerar2020} whereas the estimators based on multiple metrics are considered in~\cite{nguyen1996trace, balakrishnan1998explicit, fonseca2007four, senel2007kalman, boano2010triangle, baccour2010fuzzy, liu2011foresee, liu2012talent, guo2013fuzzy, rekik2015fli, ancillotti2017reinforcement,shu2017research,okamoto2017machine,bote2018online,luo2019link,demetri2019automated}. 

One can readily observe from the fourth column of Tables~\ref{tab:lqe-part1} and~\ref{tab:lqe-part2} that the most widely used metric, either directly or indirectly, is the \gls{PRR}, which is used as model input in~\cite{nguyen1996trace, woo2003taming, fonseca2007four, senel2007kalman, baccour2010fuzzy, liu2011foresee, liu2012talent, srinivasan2008prr, guo2013fuzzy, rekik2015fli}. Other input metrics derived from \gls{PRR} values are also used as input metrics in~\cite{boano2010triangle, senel2007kalman}. Looking at the frequency of use, \gls{PRR} is followed by hardware metrics, i.e., \gls{RSSI}, \gls{LQI} and \gls{SNR} in~\cite{fonseca2007four, senel2007kalman, boano2010triangle, liu2011foresee, liu2012talent, sun2017wnn, audeoud2018quick}. Other features are less common and tend to appear scarcely in single papers.

Table~\ref{tab:feature-analysis} summarizes metrics that can be used for measuring the quality of the link. Every metric from the first column of the table can also be used as input for another new metric. The so-called hardware-based metrics~\cite{baccour2012radio}, such as \gls{RSSI}, \gls{LQI}, \gls{SNR} and \gls{BER} are directly produced by the transceivers, and they also depend on underlying metrics, such as \gls{RSS}, \gls{SNR}, noise floor, implementation artifacts and vendor. The so-called software-based metrics are usually computed based on a blend of hardware and software metrics. It is clear from the first and the last columns of Table~\ref{tab:feature-analysis} that the number of independent input variables is limited. However, recently, additional input has been taken into account in~\cite{millan2015time}. \textit{Topological features} assuming cross-layer information exchange, where LQE is informed of node degree, hop count, strength and distance is considered in~\cite{millan2015time}, while~\cite{okamoto2017machine} and~\cite{demetri2019automated} have exclusively shown that \textit{imaging data} can be used as input for LQE models as an alternative source of data, as outlined at the bottom of Table~\ref{tab:feature-analysis}.

In addition to finding other new sources of data, a challenging task would be to analyze a large set of measurements in various environments and settings, from a large number of manufacturers to understand how measurements vary across different technologies and differ across various implementations within the same technology, and derive the truly effective metrics for an efficient development of \gls{LQE} model.

\subsection{Models for LQE}
\label{sub:model-for-lqe}
Considering the models used for developing \gls{LQE} summarized in the fifth column of Tables~\ref{tab:lqe-part1} and~\ref{tab:lqe-part2}, the publications surveyed can be distinguished as those using statistical models~\cite{nguyen1996trace, balakrishnan1998explicit, woo2003taming, senel2007kalman,srinivasan2008prr}, rule and/or threshold based models~\cite{fonseca2007four, boano2010triangle, audeoud2018quick}, fuzzy \gls{ML} models~\cite{baccour2010fuzzy, baccour2011radiale, guo2013fuzzy, rekik2015fli}, statistical \gls{ML} models~\cite{liu2011foresee, liu2012talent,liu2014temporal, shu2017research,okamoto2017machine,bote2018online,demetri2019automated,cerar2020,rehan2016machine,millan2015time}, reinforcement learning models~\cite{ancillotti2017reinforcement} and deep learning models~\cite{sun2017wnn,luo2019link}. For readers' convenience, the corresponding taxonomy is portrayed in Fig.~\ref{fig:mllqetax}.

With regard to the \textit{statistical models}, the authors of~\cite{nguyen1996trace, balakrishnan1998explicit} manually derived error probability models from traces of data using statistical methods. Additionally, Woo~\textit{et al.}~\cite{woo2003taming} derived an exponentially weighted \gls{PRR} by fitting a curve to an empirical distribution, whereas Senel~\textit{et al.}~\cite{senel2007kalman} first used a Kalman filter to model the correct value of the \gls{RSS}, then they extracted the noise floor from it to obtain \gls{SNR}, and finally, they leveraged a pre-calibrated table to map the \gls{SNR} to a value for the \gls{PSR}. Srinivasan~\textit{et al.}~\cite{srinivasan2008prr} used the Gilbert-Elliot model, which is a two-state Markov process with \textit{good} and \textit{bad} states with four transition probabilities. The output of the model is the channel memory parameter that describes the ``burstiness'' of a link. 

Considering the \textit{rule based models}, \gls{4B}~\cite{fonseca2007four} constructs a largely rule based model of the channel that depends on the values of the four input metrics, whereas Boano~\textit{et al.}~\cite{boano2010triangle} formulate the metric using geometric rules. First, Boano~\textit{et al.}~\cite{boano2010triangle} computed the distance between the instant \gls{SNR} and \gls{LQI} vectors in a 2D space. Then, they used three empirically set thresholds to identify four different link quality grades: \textit{very good}, \textit{good}, \textit{average} or \textit{bad}. Finally, \cite{audeoud2018quick} manually rules out good and bad links based on LQI values and then, for the remaining links, computes additional statistics that are used to determine their quality with respect to some thresholds.  

The first \textit{fuzzy model}, \gls{F-LQE}~\cite{baccour2010fuzzy} uses four input metrics incorporating \gls{WMEWMA}, averaged \gls{PRR} value, stability factor of \gls{PRR}, asymmetry level of \gls{PRR} and average \gls{SNR}, and fuzzy logic to estimate the two-class link quality. Rekik~\textit{et al.}~\cite{rekik2015fli} adapts \gls{F-LQE} to smart grid environments with higher than normal values for electromagnetic radiation, in particular 50~Hz noise and acoustic noise. Finally, Guo~\textit{et al.}~\cite{guo2013fuzzy} proposed a different two-class fuzzy model based on the two input metrics, namely coefficient of variance of \gls{PRR} and quantitative description of packet loss burst, which are gathered independently, and are different from the ones used for \gls{F-LQE}.

One of the earliest \textit{statistical \gls{ML} model}, the so-called \gls{4C}, was proposed by Liu~\textit{et al.}~\cite{liu2011foresee}, where \gls{4C} amalgamated \gls{RSSI}, \gls{SNR}, \gls{LQI} and \gls{WMEWMA}, and smoothed \gls{PRR} to train three \gls{ML} models based on na\" ive Bayes, neural networks and logistic regression algorithms. Then, Liu~\textit{et al.}~\cite{liu2012talent,liu2014temporal} introduced TALENT, an online \gls{ML} approach, where the model built on each device adapts to each new data point as opposed to being precomputed on a server. TALENT yields a binary output (i.e., whether \gls{PRR} is above the predefined threshold), while \gls{4C} produces a multi-class output. TALENT also uses state-of-the-art models for \gls{LQE}, such as \gls{SGD}~\cite{bottou2010ml-sgd} and smoothed Almeida–Langlois–Amaral–Plakhov algorithm~\cite{almeida1999alap} for the adaptive learning rate and logistic regression. 

Other statistical models, such as Shu~\textit{et al.}~\cite{shu2017research} used \gls{SVM} algorithm along with \gls{RSSI}, \gls{LQI} and \gls{PRR} as input to develop a five-class model of the link. Besides, Okamoto~\textit{et al.}~\cite{okamoto2017machine} used an on-line learning algorithm called adaptive regularization of weight vectors to learn to estimate throughput from throughput and images. Then, Bote-Lorenzo~\textit{et al.}~\cite{bote2018online} trained online perceptrons, online regression trees, fast incremental model trees, and adaptive model rules with \gls{LQ}, \gls{NLQ} and \gls{ETX} metrics to estimate the quality of a link, whereas Demetri~\textit{et al.}~\cite{demetri2019automated} benefit from a seven-class \gls{SVM} classifier to estimate LoRa network coverage area by means of using 5 input metrics to train the classifier including multi-spectral aerial images. More recently, \cite{cerar2020} evaluated four different ML models, namely logistic regression, tree based, ensemble, multilayer percepron, against each other.

The only proposed \textit{reinforcement learning model} for link quality estimation appears in~\cite{ancillotti2017reinforcement}. The authors train a greedy algorithm with \gls{PER}, \gls{RSSI} and energy consumption input metrics to estimate \gls{PRR} in view of protocol improvement in mobility scenarios. 

The two \gls{LQE} models using \textit{deep learning algorithms} have also been proposed. For the first model, Sun~\textit{et al.}~\cite{sun2017wnn} introduce \gls{WNN-LQE}, a new \gls{LQE} metric for estimating link quality in smart grid environments, where they only rely on SNR to train a wavelet neural network estimator in view of accurately estimating confidence intervals for \gls{PRR}. In the latter model, Luo~\textit{et al.}~\cite{luo2019link} incorporate four input metrics, namely \gls{SNR}, \gls{LQI}, \gls{RSSI}, and \gls{PRR}, and trains neural networks to distinguish a five-class \gls{LQE} model.

\subsection{Output of link quality estimator}
\label{sub:lqe-output}

Regarding the output of link quality estimators summarized in the sixth column of Tables~\ref{tab:lqe-part1} and~\ref{tab:lqe-part2}, we can observe three distinct types of the output values.

The first type is a \textit{binary or a two-class output}, which is produced by the classification model. This type of output can be found in~\cite{woo2003taming, baccour2011radiale, guo2013fuzzy, liu2014temporal, rekik2015fli}. The applications noticed are mainly (binary) decision making~\cite{woo2003taming} and above/below threshold estimation~\cite{baccour2011radiale, guo2013fuzzy, liu2014temporal, rekik2015fli}.

The second type is \textit{multi-class output} value. Similar to the first type, it is also produced by the classification model. The multi-class output values are utilized in~\cite{boano2010triangle, rehan2016machine, shu2017research, audeoud2018quick, luo2019link, demetri2019automated}, where \cite{rehan2016machine,audeoud2018quick,cerar2020} use a three-class, \cite{boano2010triangle} utilizes a four-class, \cite{shu2017research, luo2019link} rely on a five-class, and \cite{demetri2019automated} leverages a seven-class output. The applications observed are the categorization and estimation of the future \gls{LQE} state, which is expressed through labels/classes.

It is not clear from the analyzed work how the authors selected the number of classes in the case of multi-class output \gls{LQE} models. The three class output models seem to be justified by the three regions of a wireless links~\cite{baccour2012radio}. The seven class output model \cite{demetri2019automated} justifies the 7 types of classes based on seven types of geographical tiles. For the rest or the work, it is not clear what is the justification and advantage of a four, or five class \gls{LQE} model. Generally, by adding more classes, the granularity of the estimation can be increased while the computing time, memory size and processing power increase.

The third type is the \textit{continuous-valued output}. In contrast to the first two types, it is produced by a regression model, which is considered by~\cite{nguyen1996trace, balakrishnan1998explicit, fonseca2007four, senel2007kalman, srinivasan2008prr, liu2011foresee, millan2015time, sun2017wnn, ancillotti2017reinforcement, okamoto2017machine, bote2018online}. The value is typically limited only by numerical precision. The applications observed are the direct estimation of a metric~\cite{nguyen1996trace, balakrishnan1998explicit, senel2007kalman, millan2015time, sun2017wnn, ancillotti2017reinforcement, okamoto2017machine, bote2018online}, probability value~\cite{srinivasan2008prr, liu2011foresee} and their proposed scoring metric~\cite{fonseca2007four}, which are later used for comparative analysis.

Some of the proposed or identified applications require continuous-valued \gls{LQE} estimation, for instance, network congestion controller (TCP Reno) \cite{balakrishnan1998explicit}, communication handover \cite{okamoto2017machine}, and routing table managers \cite{fonseca2007four, liu2011foresee, millan2015time, sun2017wnn, ancillotti2017reinforcement, bote2018online}. For other routing table managers and applications, a discrete valued \gls{LQE} suffices according to the surveyed work. Note that any continuous estimator can be subsequently converted to discrete valued one.

\begin{table*}[thb!]
	\centering
	\scriptsize
	\renewcommand{\arraystretch}{1.4}
	\caption{A survey of the comparison for LQE models and their respective evaluation metrics considering the research papers comprehensively surveyed in Tables~\ref{tab:lqe-part1} and~\ref{tab:lqe-part2}.}
	\label{tab:eval}
	\begin{tabularx}{\linewidth}{| C{0.2} | C{1.5} | C{.8} | C{1.5} |}
		\hline\rowcolor{gray}
		
		\bfseries ID
		& \bfseries Evaluation metrics
		& \bfseries The proposed LQE models
		& \bfseries Link quality estimators that the proposed LQE models are compared to
		\\\hline
		
		1
		& Confusion matrix % evaluation
		&\cite{boano2010triangle}, \cite{audeoud2018quick}% Research
		& % compares with
		\\\hline 
		
		2
		& Confusion matrix, accuracy, precision, recall, F1 % evaluation
		&\cite{cerar2020} % Research
		& % compares with
		\\\hline 
		
		3
		& Classification accuracy, confusion matrix % evaluation
		&\cite{liu2012talent, liu2014temporal} % Research
		& ETX~\cite{de2005high-ETX}, STLE~\cite{alizai2009bursty-STLE}, 4B~\cite{fonseca2007four}, 4C~\cite{liu2011foresee} % compares with
		\\\hline
		
		4
		& Confusion matrix, recall, classification accuracy % evaluation
		&\cite{luo2019link} % Research 
		& SVC, extreme learning (EML), WNN~\cite{sun2017wnn}
		\\\hline
		
		5
		& Classification accuracy % evaluation
		&\cite{shu2017research} % Research 
		& FLI~\cite{guo2013fuzzy}, LQI-PRR~\cite{luo2013new}
		\\\hline
		
		6
		& Classification accuracy, power estimation error % evaluation
		&\cite{demetri2019automated} % Research
		&
		\\\hline
		
		7
		& Avg. delivery cost, classification accuracy % evaluation
		&\cite{liu2011foresee} % Research
		& STLE~\cite{alizai2009bursty-STLE}, 4B~\cite{fonseca2007four} % compares with
		\\\hline
		
		8
		& RMSE, number of topology changes % evaluation
		& \cite{guo2013fuzzy} % Research
		& 4B~\cite{fonseca2007four} % compares with
		\\\hline

		9
		& (RMSE) Throughput estimation error % evaluation
		& \cite{okamoto2017machine} % Research 
		&
		\\\hline
		
		10
		& (RMSE) \gls{PRR} estimation error % evaluation
		& \cite{sun2017wnn} % Research 
		& SNR, back-propagation Neural Network, ARIMA, XCoPred
		\\\hline
		
		11
		& \gls{MAE} % evaluation
		& \cite{millan2015time} % Research 
		& SVM, regression trees, k-nearest neighbor, Gaussian process for regression
		\\\hline

		12
		& \gls{MAE}, computational load % evaluation
		& \cite{bote2018online} % Research 
		& Baseline routing performance, online perceptrons, online regression trees, fast incremental model trees vs. adaptive model rules
		\\\hline
		
		13
		& \gls{CDF}, LQE stability % evaluation
		& \cite{rekik2015fli} % Research
		& ETX~\cite{de2005high-ETX}, F-LQE~\cite{guo2013fuzzy}
		\\\hline
		
		14
		& Mean and stdev. of estimation error, CDF, R$^2$ % evaluation
		& \cite{nguyen1996trace} % Research
		& % compares with
		\\\hline

		15
		& LQE sensitivity, LQE stability, CDF % evaluation
		& \cite{baccour2010fuzzy, baccour2011radiale} % Research
		& ETX~\cite{de2005high-ETX}, WMEWMA~\cite{woo2003taming}, RNP~\cite{cerpa2005temporal-RNP}, 4B~\cite{fonseca2007four} % compares with
		\\\hline
		
		16
		& Number of downloads % evaluation
		& \cite{balakrishnan1998explicit} % Research
		& % compares with
		\\\hline
		
		17
		& \gls{PRR}, number of parent changes % evaluation
		& \cite{woo2003taming} % Research
		& % compares with
		\\\hline
		
		18
		& Total number of transmissions, average tree depth, delivery rate (PSR) % evaluation
		& \cite{fonseca2007four} % Research
		& ETX~\cite{de2005high-ETX}, Collection Tree Protocol (CTP)~\cite{gnawali2009collection}, MultiHopLQI % compares with
		\\\hline
		
		19
		& \gls{PSR} % evaluation
		& \cite{senel2007kalman} % Research 
		& ETX~\cite{de2005high-ETX}, RNP~\cite{cerpa2005temporal-RNP} % compares with
		\\\hline
		
		20
		& Throughput % evaluation
		& \cite{srinivasan2008prr} % Research 
		& % compares with
		\\\hline
		
		21
		& Channel rank estimation, energy consumption, channel switching delay, stability % evaluation
		& \cite{rehan2016machine} % Research 
		&
		\\\hline
		
		22
		& Average packet loss, num. of control packets, energy consumption % evaluation
		& \cite{ancillotti2017reinforcement} % Research 
		& %RPL, mRPL
		\\\hline

	\end{tabularx}
\end{table*}

\subsection{Evaluation of the proposed models}
\label{sub:evaluation-metrics}
We analyze the way Tables~\ref{tab:lqe-part1} and~\ref{tab:lqe-part2} evaluate the proposed \gls{LQE} models along several dimensions. The evaluation metric analysis of the surveyed literature is presented in Table~\ref{tab:eval}. The second column of the table lists the metrics used to evaluate the \gls{LQE} model by the research papers listed in the third column of the table. The fourth column of the table identifies what other existing link quality estimators were utilized and compared against the ones proposed in the papers outlined in the third column.

\subsubsection{Evaluation from the purpose of the LQE perspective}
Firstly, we analyze the evaluation of the models through the lens of the purpose of the \gls{LQE} as discussed in Section~\ref{subsec:purpose}. We identify direct evaluation, where the paper directly quantifies the performance of the proposed \gls{LQE} models vs. indirect evaluation, where the improvement of the protocol or the application as a result of the \gls{LQE} metric is quantified.

\textit{Direct evaluations} of \gls{LQE} models typically evaluate the predicted or estimated value against a measured or simulated ground truth. The metrics used for evaluation depend on the output of the proposed model for \gls{LQE} discussed in Section~\ref{sub:lqe-output}.

When the output are categorical values, then it is possible to use metrics based on predicted label count versus the label count of the ground truth. Confusion matrices are used by~\cite{boano2010triangle, liu2012talent, liu2014temporal, audeoud2018quick, luo2019link,cerar2020} as seen in rows 1, 2, 3 and 4 of Table~\ref{tab:eval}, classification accuracy is used by~\cite{liu2011foresee, liu2012talent, liu2014temporal, shu2017research,demetri2019automated, luo2019link} as observed in rows 3, 5, 6 and 7, and recall is used in combination with accuracy and confusion matrix by~\cite{luo2019link} as illustrated in the fourth row of the table. Only more recently, \cite{cerar2020} uses the combined confusion matrix, precision, recall and F1 to provide more detailed insights into the performance of their classifier. Well known evaluation metrics, such as classification precision, classification sensitivity, F1 and \gls{ROC} curve are used seldom or not at all among the evaluation metrics in the surveyed classification work. However, they can be computed for some of the metrics based on the provided confusion matrices. 

The \gls{LQE} metrics listed in rows 1-3 of Table~\ref{tab:eval} can be compared to each other in terms of performance by mapping the 5 and 7 class estimators to the 2 or 3 class estimator. This results in a number of comparable 2 or 3 dimension confusion matrices that can be analyzed. However, as the metrics are developed and evaluated under different datasets, the comparison would not be exactly fair and it would not be clear which design decision led one to be superior to another. The same discussion holds also for other rows of the table that share common evaluation metrics. High level comparisons that abstract such details are provided later in Sections \ref{sec:lqe-ML-analysis-app} and \ref{sec:lqe-ML-analysis} for selected ML works that reported their results in sufficient detail.

When the output is continuous, then each predicted value is compared against each measured or simulated value using a distance metric. For instance, the authors of~\cite{guo2013fuzzy, okamoto2017machine, sun2017wnn} use \gls{RMSE} as a distance metric as shown in rows 8-10 of Table~\ref{tab:eval}, whereas the authors of~\cite{millan2015time,bote2018online} use mean absolute error (\gls{MAE}) as in rows 11 and 12 of the table. Some other research papers as in~\cite{rekik2015fli, nguyen1996trace, baccour2010fuzzy,baccour2011radiale} use \gls{CDF} as illustrated in rows 13-15, while the authors of~\cite{nguyen1996trace} leverage R$^2$ in row 14 of Table~\ref{tab:eval}.

\textit{Indirect evaluations} of \gls{LQE} models evaluate against application specific metrics. The papers evaluate the performance of their objective functions based on the presence of link quality estimators. For example, the studies conducted in~\cite{nguyen1996trace, balakrishnan1998explicit, woo2003taming, srinivasan2008prr, boano2010triangle, rehan2016machine, ancillotti2017reinforcement, okamoto2017machine, audeoud2018quick, bote2018online, demetri2019automated} consider their respective objective functions for the particular applications and demonstrate to obtain better results by means of using estimators compared to the cases with the absence of estimators. While these research papers are likely to be leading on the respective use cases of \gls{LQE} models owing to their first attempts in their specific application domains, their results and design decisions are still difficult to compare against each other. Various application specific evaluation metrics, such as number of downloads ~\cite{balakrishnan1998explicit}, number of parent changes~\cite{woo2003taming}, throughput~\cite{srinivasan2008prr} can also be found as listed in the rows 16-22 of Table~\ref{tab:eval}.

\begin{figure}
	\centering
	\begin{tikzpicture}[
	node distance=30mm,
	auto,
	smer/.style={draw=black!50, fill=black!50, -stealth'},
	box/.style={font=\small, rectangle, rounded corners=1mm, draw=black!50}
	]
	\foreach \label/\name [count=\i, evaluate=\i as \x using 360/14*(\i-2)] in {
		woo2003taming/WMEWMA~\cite{woo2003taming},
		de2005high/ETX~\cite{de2005high-ETX},
		cerpa2005temporal/RNP~\cite{cerpa2005temporal-RNP},
		alizai2009bursty/STLE~\cite{alizai2009bursty-STLE},
		fonseca2007four/4B~\cite{fonseca2007four},
		senel2007kalman/Kalman~\cite{senel2007kalman},
		baccour2011radiale/F-LQE~\cite{baccour2011radiale},
		liu2011foresee/4C~\cite{liu2011foresee},
		guo2013fuzzy/FLI~\cite{guo2013fuzzy},
		liu2014temporal/TALENT~\cite{liu2014temporal},
		rekik2015fli/Opt-FLQE~\cite{rekik2015fli},
		sun2017wnn/WNN~\cite{sun2017wnn},
		shu2017research/SVM~\cite{shu2017research},
		luo2019link/SAE~\cite{luo2019link}}
	{
		\node[box, auto] (\label) at (\x:3.5cm) {\name};
	};
	
	\foreach \source/\targets in {
		fonseca2007four/{de2005high},
		senel2007kalman/{de2005high, cerpa2005temporal},
		baccour2011radiale/{de2005high, cerpa2005temporal, woo2003taming, fonseca2007four},
		liu2011foresee/{alizai2009bursty, fonseca2007four},
		guo2013fuzzy/{fonseca2007four},
		liu2014temporal/{de2005high, alizai2009bursty, woo2003taming, fonseca2007four, liu2011foresee},
		rekik2015fli/{de2005high, guo2013fuzzy},
		sun2017wnn/{senel2007kalman},
		shu2017research/{guo2013fuzzy},
		luo2019link/{sun2017wnn}}
	{
		\foreach \target in \targets {
			\path[smer] (\source) edge[bend right=10] node[font=\small,text=black!50,near end,above,sloped] {} (\target);
		};
	};
	\end{tikzpicture}
	\caption{Visualization of relationships for cross-comparison of the research papers with their corresponding evaluation metrics outlined in Table~\ref{tab:eval}.}
	\label{fig:estimators-cross-comparison}
\end{figure}

\subsubsection{Evaluation from cross-comparison perspective}
Secondly, we categorize papers that \textit{evaluate their outcomes against other estimators existing at the time of writing} versus papers that are somewhat \textit{stand alone}. For instance, in row 3 of Table~\ref{tab:eval}, TALENT~\cite{liu2012talent} is evaluated against \gls{ETX}, STLE, \gls{4B}, WMEWMA and \gls{4C}. For more clarity, this is represented visually in Fig.~\ref{fig:estimators-cross-comparison} with directed arrows exiting from TALENT and entering the boxes of the respective metrics, which explicitly depicts the relationship between the last two columns of Table~\ref{tab:eval}. Such comparisons are informative as demonstrated by~\cite{baccour2011radiale}. Among their findings, they showed that \gls{PRR}, \gls{WMEWMA}, and \gls{ETX}, which are \gls{PRR}-based link quality estimators, overestimate the link quality, while \gls{RNP} and \gls{4B} underestimate the link quality. \gls{F-LQE} performed better estimation than the other compared estimators.

However, metrics of the surveyed papers~\cite{rehan2016machine, ancillotti2017reinforcement, okamoto2017machine, audeoud2018quick, demetri2019automated} are not evaluated against other existing estimators, due to their unique approach (application) and/or being among the first to tackle certain aspect of estimation. For instance, the authors of~\cite{rehan2016machine} evaluate the estimated ranking/classification of subset of wireless channels and the authors of~\cite{ancillotti2017reinforcement} evaluate the impact of networking performance with estimator assisted routing algorithm against vanilla (m)RPL protocol, while the authors of~\cite{okamoto2017machine} evaluate estimated and real throughput degradation when line of sight is blocked by an object. Besides, the authors of~\cite{audeoud2018quick} evaluate data-driven bidirectional link properties, and~\cite{demetri2019automated} evaluates estimated vs. ground truth signal fading, which is influenced by \gls{ML} algorithm's ability to classify geographical tiles.

\subsubsection{Evaluation from infrastructure perspective}
Thirdly, we categorize papers to those that perform evaluation and validation on real testbeds~\cite{nguyen1996trace, balakrishnan1998explicit, woo2003taming, fonseca2007four, senel2007kalman, boano2010triangle, baccour2010fuzzy, baccour2011radiale, liu2011foresee, guo2013fuzzy, liu2012talent, liu2014temporal, ancillotti2017reinforcement, okamoto2017machine, demetri2019automated} shown as in rows 1, 3, 6, 8, 9, 14-19, 22, those that perform evaluation in simulation such as~\cite{rekik2015fli, rehan2016machine, ancillotti2017reinforcement} in rows 13, 21, 22, and the rest that perform only numerical evaluation. The papers in the first category, that perform evaluation and validation on testbeds, are better at presenting how the estimator will actually influence the network. The papers from second category performing simulation can provide good foundation for further examination and potential implementation. Finally, the papers in third category, that only do numerical evaluation, can unveil possible improvements through statistical relationships. 

\subsubsection{Evaluation from convergence perspective}
Fourthly, during our analysis it has emerged that a number of papers reflect on and quantify the convergence of their model. For instance, in~\cite{srinivasan2008prr}, they concluded that their model starts to converge at approximately 40,000 packets. In~\cite{senel2007kalman}, the authors demonstrated that the link degradation could be detected even with a single received packet. The metric proposed in~\cite{boano2010triangle} required approximately 10 packets to provide the estimation in either a static or mobile scenario. In~\cite{liu2011foresee}, they suggested that data gathered from 4-7 nodes for approximately 10 minutes should be sufficient to train their models offline. Although these papers indicate convergence rate/size, a community wide systematic investigation of \gls{LQE} model convergence is missing. 

At this point, we can conclude that research community in general have shown remarkable improvements, use cases, and skills toward better estimators. However, despite the aforementioned evaluation of proposed estimators, providing a completely fair comparison of \gls{LQE} models is not feasible considering the diverse evaluation metrics outlined in Table~\ref{tab:eval}.

\begin{figure*}[!thb]
	\centering
	\newcolumntype{H}[1]{>{\centering}p{#1}}
	\scalebox{0.9}{
	\begin{forest}
		forked edges,
		for tree={
			if level=0{align=center}{%
				align={@{}H{28mm}@{}},
			},
			grow=east,
			text width=3.4cm,
			draw,
			font=\sffamily\small,
			edge path={
				\noexpand\path [draw, \forestoption{edge}] (!u.parent anchor) -| +(0.5mm,0) -- +(5mm,0) |- (.child anchor)\forestoption{edge label};
			},
			parent anchor=east,
			child anchor=west,
			l sep=10mm,
			tier/.wrap pgfmath arg={tier #1}{level()},
			edge={thin},
			fill=white,
			rounded corners=2pt,
			drop shadow,
		}
		[Purpose of ML\\LQE models
		[New \& Improved LQE
		[Prediction / Estimation
		[Link quality \cite{millan2015time, bote2018online, rehan2016machine, sun2017wnn, luo2019link, demetri2019automated, cerar2020}]
		[Stability \cite{rehan2016machine}]
		[Throughput \cite{okamoto2017machine, demetri2019automated}]
		]
		]
		[Protocol Improvement
		[Maximize
		[Throughput~\cite{liu2011foresee, liu2012talent, liu2014temporal, shu2017research}]
		[Reliability \cite{guo2013fuzzy, rekik2015fli}]
		[Reactivity \cite{rekik2015fli, rehan2016machine}]
		]
		[Minimize
		[Topology changes \cite{guo2013fuzzy}]
		[Depth of routing tree  \cite{guo2013fuzzy}]
		[Probing overhead \cite{ancillotti2017reinforcement}]
		[Traffic congestion \cite{shu2017research}]
		]
		]
		]
	\end{forest}
	}
	\caption{Classification of the works by considering the purpose for which the ML LQE model was developed.}
	\label{fig:lqe-purpose}
\end{figure*}
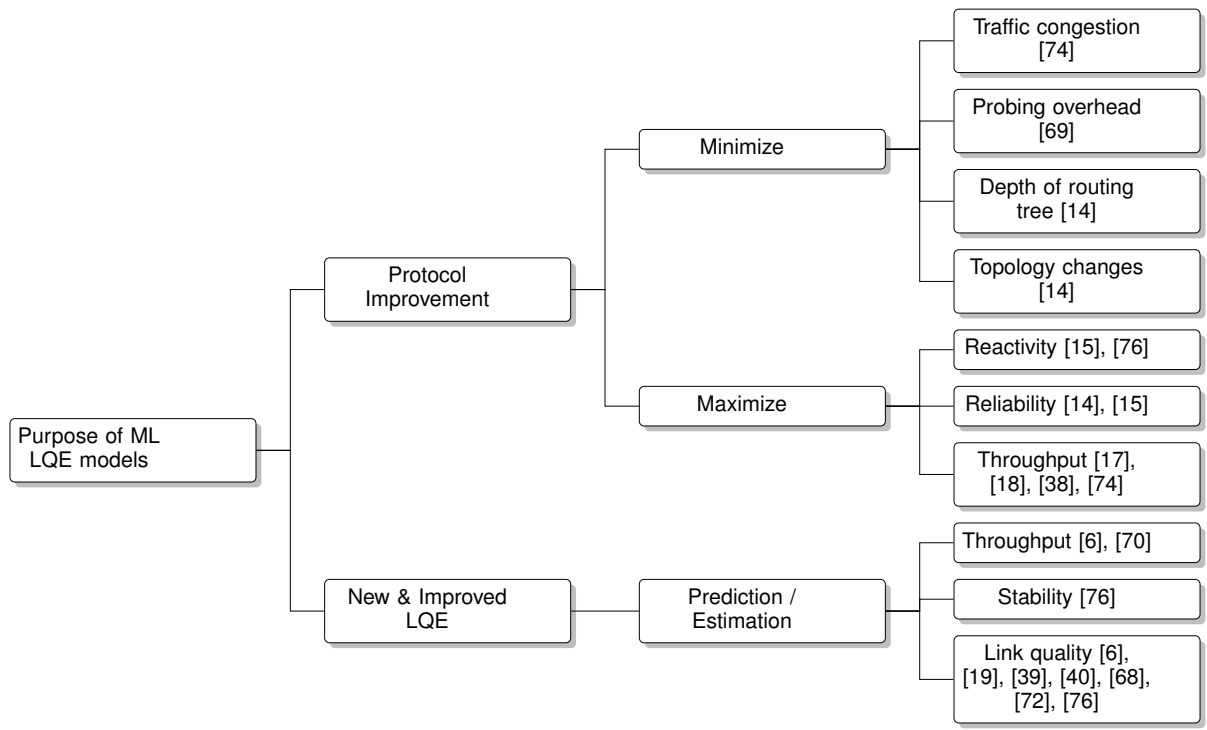

\begin{figure*}[htpb]
	\centering
	\scalebox{1.5}{\input{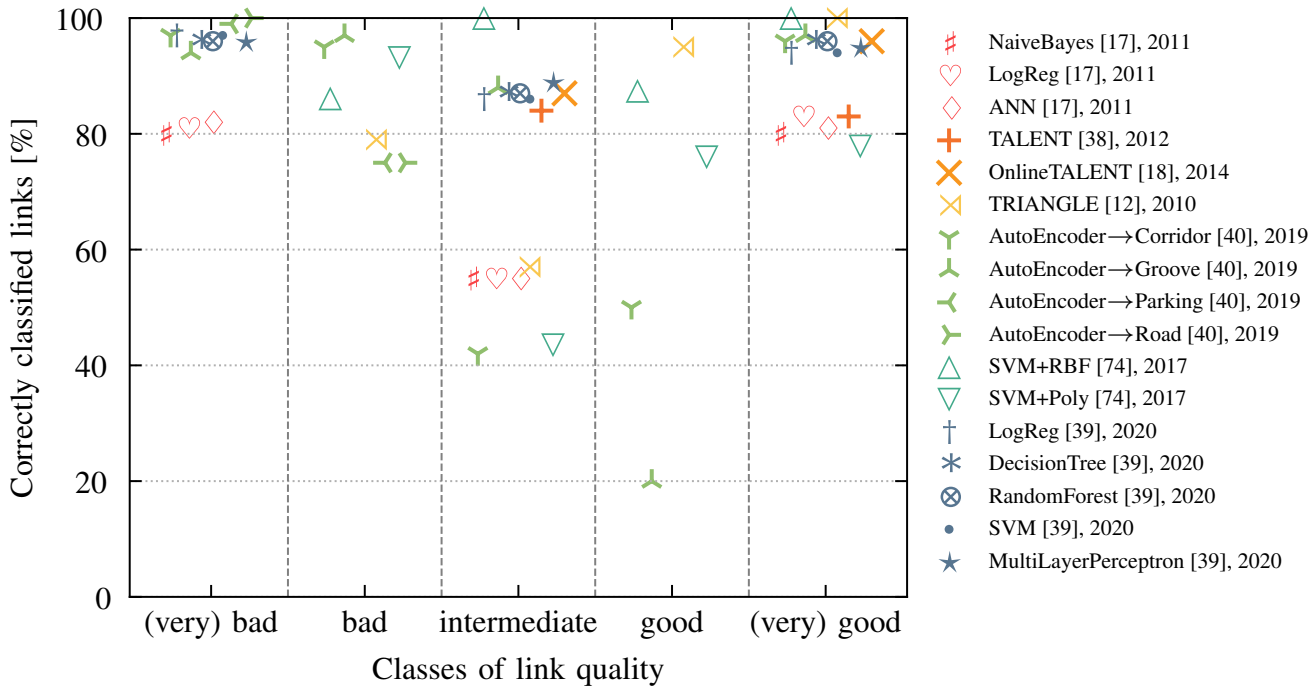}}
	\caption{Comparison of the wireless link quality classification performances throughout the surveyed papers.}
	\label{fig:quant-ML-comparison}
\end{figure*}

\subsection{Reproducibility}
\label{sub:reproducibility}
Reproducibility of the results is recognized as being an important step in the scientific process~\cite{gil2007examining, van2016contextual, baker20161} and is important for replication as well as for reporting explicit improvements over the baseline models. When researchers publicly share the data, simulation setups and their relevant codes it becomes easy for others to pick up, replicate and improve upon, thus speeding up the adoption and improvement. For instance,  when a new \gls{LQE} model is proposed, it can be ran on the same data or testbed as a set of existing models provided the data and models are publicly accessible to the community. The existing models can also be re-evaluated in the same setup, thus replicating the existing results or they can be used as baselines in new scenarios. With this approach, the performance of the new \gls{LQE} model can be directly compared to the existing models with relatively low effort.

With respect to the reproducibility of the results in the surveyed publications, we notice that only~\cite{liu2014temporal, srinivasan2008prr,cerar2020} are easily reproducible because they rely on publicly available trace-sets. Studies reported in~\cite{nguyen1996trace, balakrishnan1998explicit, woo2003taming, fonseca2007four, baccour2011radiale, liu2011foresee, audeoud2018quick} use open testbeds that, in principle, could be used to collect data and the results can be reproduced. However, it is not clear whether some of these testbeds are still operational given that 10-20 years have passed after the date of publication of the corresponding research. We were not able to find any evidence that the results in~\cite{senel2007kalman, boano2010triangle, guo2013fuzzy, rekik2015fli, sun2017wnn} could be reproduced as they strictly rely on an internal one-time deployment and data collection. 

\section{Application Perspective of ML-Based LQEs}
\label{sec:lqe-ML-analysis-app}

In this section, we provide an analysis of the ML-based LQEs from application perspectives. We identify what is important from an application perspective and how that affects ML methods utilized for the LQE modeling. We first focus on the purpose of the LQE model development followed by the analyses of the application quality aspects.

\subsection{LQE design purpose}
In Section~\ref{subsec:purpose}, we have reflected on the purpose for which an LQE model was developed, and as depicted in Fig.~\ref{fig:lqe-purpose}, we found that about half of the ML-based LQE studies developed an estimator with the goal of improving an existing protocol, while the other half aimed for a new and superior LQE model. Fig.~\ref{fig:lqe-purpose} presents that "protocol improvement" group attempts to minimize or maximize a particular objective, such as traffic congestion, probing overhead, topology changes, just to name a few. Most of the studies that fall into "new \& improved LQE" group only aim to improve the prediction or estimation of the quality of a link.

The body of the work considering "protocol improvements" is intricate to quantitatively compare against each other since numerical details of the LQE models are not explicitly provided in the respective works, as previously discussed in Section~\ref{sub:evaluation-metrics}. Similar difficulties also arise for a large part of the body of work related to "new \& improved LQE" models since they do not utilize consistent evaluation metrics. For instance, for LQE models formulated as a classification problem, only a subset of the works leverages accuracy as a metric, while other subsets use confusion matrix or specifically defined metrics, which indeed renders them impractical to quantitatively compare against each other, as outlined in Table~\ref{tab:eval} and discussed in Section~\ref{sub:evaluation-metrics}. Attaining a fair comparison is even more difficult for the works that formulate the LQE problem as a regression. Later in Section~\ref{sec:desguide}, we provide guidelines with regards to this aspect.

Fig.~\ref{fig:quant-ML-comparison} presents a high level comparison of the selected works that use ML for LQE model development~\cite{liu2011foresee,liu2012talent,liu2014temporal,luo2019link,cerar2020} and one that does not~\cite{boano2010triangle}. All the considered works formulated the LQE model as a classification problem and it is possible to extract the approximated per class performance from the reported performance results of those respective works. Notice that they are different in terms of; i) the input features used to train and evaluate the models (more details in Section~\ref{subsec:input}), ii) the number of classes used for the model (more details in Section~\ref{sub:lqe-output}), and iii) the considered ML algorithm (more details in Section~\ref{sub:model-for-lqe}).

On the x-axis, Fig.~\ref{fig:quant-ML-comparison} presents five different link quality classes, while on the y-axis it presents the percentage of correctly classified links. The comparison reveals that, autoencoder~\cite{luo2019link}, which is a type of deep learning method, on average performs best with above 95\% correctly classified \textit{very bad}, \textit{bad} and \textit{very good} links and about 87\% correctly classified \textit{intermediate} quality link classes. Autoencoders are outperformed by the non-ML baseline~\cite{boano2010triangle} and SVM with RBF kernel \cite{shu2017research} on the \textit{very good} link quality class by about 4 percentage points, by over 30 percentage points on the \textit{good} quality link class and by about 12 percentage points on the \textit{intermediate} quality link class. As autoencoders are known to be powerful methods, we speculate that such high performance difference on those three classes might be due to insufficient training data or other experimental artifacts.

Tree-based methods and SVM~\cite{cerar2020} as well as the customized online learning algorithm TALENT~\cite{liu2012talent} follow the performance of the autoencoders very closely with a tiny margin on \textit{very bad}, \textit{very good} and \textit{intermediate} link quality classes. Next, the offline version of TALENT~\cite{liu2014temporal} exhibits very similar performance to tree-based methods and SVM on the \textit{intermediate} class and about 17 percentage points worse on the \textit{very good} class. Moreover, traditional artificial neural networks, logistic regression and Naive Bayes~\cite{liu2011foresee} follow next with almost 20  percentage points difference compared to autoencoders on the \textit{very good} and \textit{very bad} link quality classes and almost 30 percentage points on the \textit{intermediate} link quality class. The relative performance difference of the work reported in~\cite{liu2011foresee} might be due to the poor data pre-processing practices, such as the lack of interpolation, which can significantly influence the final model performance that is discussed later in Section~\ref{sec:lqe-ML-analysis}.

To summarize, the analysis of Fig.~\ref{fig:quant-ML-comparison} reveals that autoencoders, tree based methods and SVM tend to consistently perform better than  logistic regression, naive Bayes and ANNs while the non-ML TRIANGLE estimator performs very well on two of the classes, namely \textit{very good} and \textit{good} link quality classes.

\subsubsection*{Discussion}

The observations from Fig.~\ref{fig:quant-ML-comparison} also conform to the general performance intuitions regarding ML approaches. Namely, fuzzy logic and Naive Bayes are generally comparable with the latter being far more practical and popular. Neither of the two are known to exhibit better relative performance against logistic or linear regression. As shown in~\cite{liu2011foresee,liu2012talent}, Naive Bayes tends to exhibit reduced performance compared to logistic regression, whereas ANNs are usually superior. Fuzzy logic, Naive Bayes, linear and logistic regression are relatively simple and require modest computational load and memory consumption. Therefore, these ML methods can be suitable for implementation in embedded devices, especially for small-dimensional feature spaces. Besides, ANNs can be designed to optimize computational load and memory consumption, particularly by simplifying their considered topologies, which in turn, comes with a cost to their performance.

For classification in constrained embedded devices, the authors of~\cite{liu2011foresee,liu2012talent} selected logistic regression for its simplicity among other three candidates. The selection was based on practical considerations, but their experiments proved that ANNs were superior compared to other LQE models. The reason behind this is because logistic and linear regressions are linear models that tend to be more suitable to approximate linear phenomena. Contrarily, LQE models do not follow linear models and therefore ANN-based model outperformed its counterpart LQE models in~\cite{liu2011foresee,liu2012talent}.

SVMs, part of the so-called Kernel Methods, were popular and frequently used at the beginning of the century before significant breakthroughs brought by deep learning (deep neural networks (DNN)). SVMs often exhibit at least similar performance to ANNs and also to decision/regression trees~\cite{millan2015time}. However, there are only a paucity of contributions on adapting them for embedded devices~\cite{pedersen2006embedded}. In~\cite{bote2018online}, the authors performed an in-depth comparison of ML algorithms including SVM, decision trees and k nearest neighbors (k-NN) from several perspectives, such as accuracy, computational load and training time. Their results showed that SVMs are constantly superior in terms of accuracy to k-NN and regression trees at the expense of significant resource consumption.

While many of the traditional ML methods including decision/regression trees and k-NN typically require an explicit, often manual feature engineering step, SVMs are able to automatically weight the features according to their importance automatizing part of the effort allocated for manual feature engineering. SVMs are known to be highly customizable through hyperparameter tuning, which is a dedicated research area within the ML community. Through appropriate selection of the kernel and parameter space~\cite{guo2008customizing}, they are able to perform very well on both linear and non-linear problems. Therefore, from this particular perspective, SVMs and the broader Kernel Methods are indeed favorable choices for developing LQE models.

Deep learning, represented by DNNs are a new class of ML algorithms that are currently under intense investigation in various research communities penetrating also wireless and LQE \cite{luo2019link}. These algorithms are very powerful and accurate for approximating both linear and non-linear problems, albeit requiring high memory and computational cost. Such models are prohibitive for embedding in constrained devices. However, there are a number of research efforts~\cite{ahmad2017embedded} invested in employing transfer learning approaches \cite{pan2009survey}. When an LQE~based data processing occurs on a non-constrained device, such as the case in~\cite{demetri2019automated}, DNNs can show an outstanding performance. While the authors of~\cite{demetri2019automated} proposed a novel and visionary approach for the development of an LQE model and accomplished robust results using SVMs, employing DNNs might assist in surpassing those existing results.

\begin{figure*}
	\centering
	\begin{tikzpicture}[
	scale=0.7,
	level0/.style={ultra thick, black, font=\bfseries\normalsize, align=center},
	level1/.style={ultra thick, black, font=\bfseries\small, align=center},
	level2/.style={ultra thick, black, font=\small, align=center},
	level3/.style={thick, black, font=\footnotesize, align=center},
	connect/.style = {thick, dashed, red},
	notice/.style  = { draw, rectangle callout, callout relative pointer={#1} },
	label/.style   = { text width=2cm }
	]
	\draw[level0]   (-2,0) -- node[above] (root) {Application\\Quality Aspects} (2,0);
	
	% Reliability part
	\draw[level1]   (3,5) -- node[above] (reliability) {Reliability} (5,5);
	\draw[level3]   (9,4) -- node[above] {Traffic\\\cite{guo2013fuzzy, shu2017research, okamoto2017machine}} (12,4);
	\draw[level3]   (9,5) -- node[above] {Link\\\cite{millan2015time, bote2018online, rehan2016machine, sun2017wnn, luo2019link, demetri2019automated, cerar2020}} (12,5);
	\draw[level3]   (9,6) -- node[above] {Topology, routing\\\cite{guo2013fuzzy, shu2017research, okamoto2017machine}} (12,6);
	\draw[connect]  (2,0) -- (3,5);
	\draw[connect] (5,5) -- (7,5) -- (9,4) (5,5) -- (7,5) -- (9,5) (5,5) -- (7,5) -- (9,6);
	
	% Adaptivity
	\draw[level1]   (3,0) -- node[above] (adaptivity) {Adaptivity} (5,0);
	\draw[level2]   (6,0) -- node[above] {Online\\learning} (8,0);
	\draw[level3]   (9,1)  -- node[above] {Naive Bayes \cite{liu2011foresee}} (12,1);
	\draw[level3]   (9,0) -- node[above] {Logistic Regression\\\cite{liu2012talent, liu2014temporal}} (12,0);
	\draw[level3]   (9,-1) -- node[above] {ANN\\\cite{liu2012talent, liu2014temporal, bote2018online}} (12,-1);
	\draw[connect] (2,0) -- (3,0);
	\draw[connect] (5,0) -- (6,0);
	\draw[connect] (8,0) -- (9,0) (8,0) -- (9,1) (8,0) -- (9,-1);
	
	% Stability
	\draw[level1]   (3,-3) -- node[above] (stability) {Stability} (5,-3);
	\draw[level2]   (6,-3) -- node[above] {Offline\\learning} (8,-3);
	\draw[level3]   (9,-2.5) -- node[above] {Fuzzy Methods \cite{guo2013fuzzy, rekik2015fli}} (12,-2.5);
	\draw[level3]   (9,-3.5) -- node[above] {Custom Algorithm \cite{rekik2015fli}} (12,-3.5);
	\draw[connect] (2,0) -- (3,-3);
	\draw[connect] (5,-3) -- (6,-3);
	\draw[connect] (8,-3) -- (9,-2.5) (8,-3) -- (9,-3.5);
	
	% Probing overhead
	\draw[level1]   (-3,-3) -- node[above] (probing) {Probing\\Overhead} (-5,-3);
	\draw[level2]   (-6,-3) -- node[above] {Trace-set\\Collection} (-8,-3);
	\draw[level3]   (-9,-2) -- node[above] {Adaptive probing rate \cite{ancillotti2017reinforcement}} (-12,-2);
	\draw[level3]   (-9,-3) -- node[above] {Async. \& sync. probing \cite{ancillotti2017reinforcement}} (-12,-3);
	\draw[level3]   (-9,-4) -- node[above] {Zero-overhead \cite{okamoto2017machine, demetri2019automated}} (-12,-4);
	\draw[connect] (-2,0) -- (-3,-3) (-5,-3) -- (-6,-3);
	\draw[connect] (-8,-3) -- (-9,-2) (-8,-3) -- (-9,-3) (-8,-3) -- (-9,-4);
	
	% Computation Cost
	\draw[level1]   (-3,5) -- node[above] (cost) {Computational\\Cost} (-5,5);
	\draw[connect] (-2,0) -- (-3,5);
	
	\draw[level2]   (-6,8) -- node[above] {High} (-8,8);
	\draw[level2]   (-6,5) -- node[above] {Medium} (-8,5);
	\draw[level2]   (-6,2) -- node[above] {Low} (-8,2);
	\draw[connect] (-5,5) -- (-6,8) (-5,5) -- (-6,5) (-5,5) -- (-6,2);
	
	% High
	\draw[level3]   (-9,9) -- node[above] {Regression Trees \cite{millan2015time}} (-12,9);
	\draw[level3]   (-9,8) -- node[above] {Online Perceptrons \cite{bote2018online}} (-12,8);
	\draw[level3]   (-9,7) -- node[above] {ANN \cite{liu2011foresee, liu2012talent, liu2014temporal}} (-12,7);
	\draw[connect]	(-8,8) -- (-9,9) (-8,8) -- (-9,8) (-8,8) -- (-9,7);
	
	% Medium
	\draw[level3]   (-9,5) -- node[above] {Online Logistic\\Regression \cite{liu2012talent, liu2014temporal}} (-12,5);
	\draw[connect] (-8,5) -- (-9,5);
	
	% Low
	\draw[level3]   (-9,1) -- node[above] {Logistic\\Regression \cite{liu2011foresee}} (-12,1);
	\draw[level3]   (-9,2) -- node[above] {Fuzzy Rules \cite{guo2013fuzzy}} (-12,2);
	\draw[level3]   (-9,3) -- node[above] {Naive Bayes \cite{liu2011foresee, liu2012talent, liu2014temporal}} (-12,3);
	\draw[connect] (-8,2) -- (-9,1) (-8,2) -- (-9,2) (-8,2) -- (-9,3);
	\end{tikzpicture}
	\caption{Classification of the surveyed LQE papers by taking into consideration the identified application quality aspects.}
	\label{fig:lqe-app-quality}
\end{figure*}
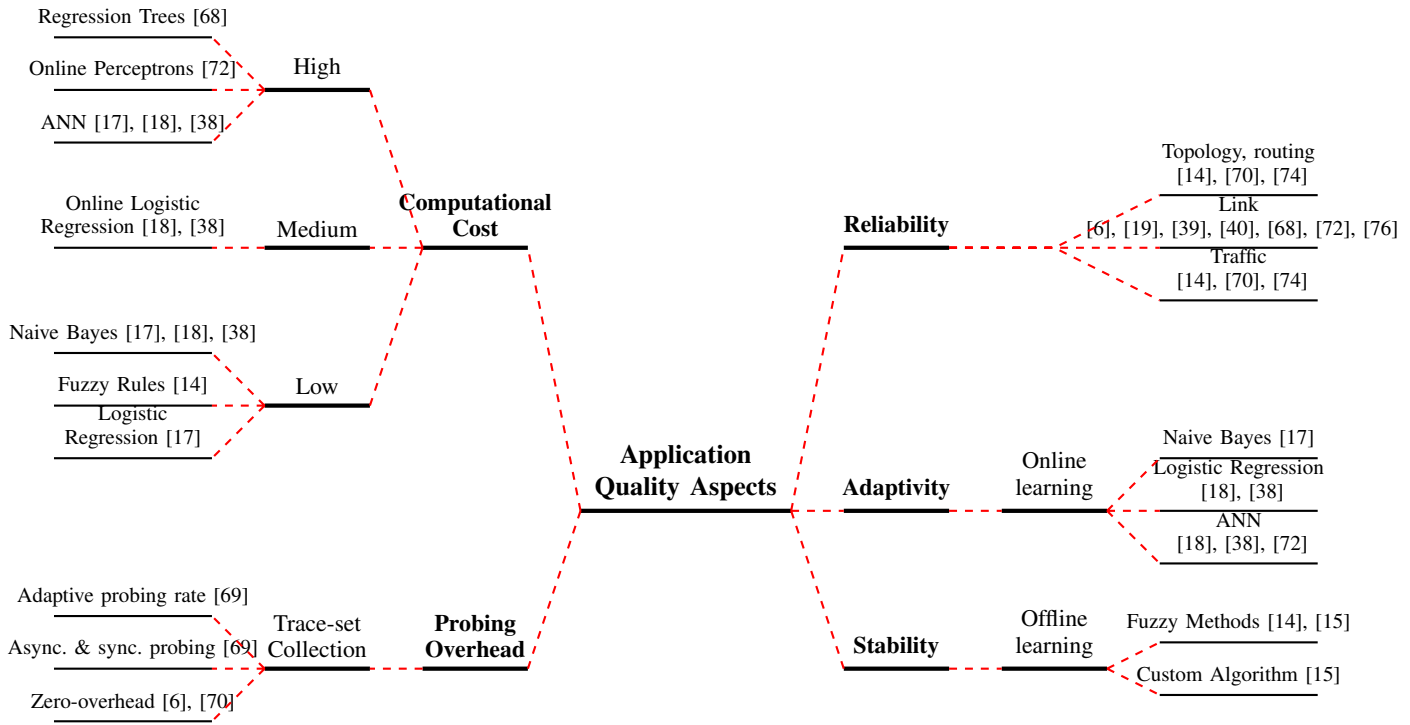

\subsection{Application quality aspects}
Following the analyses from Sections~\ref{subsec:purpose} and~\ref{sub:evaluation-metrics}, we have identified five important link quality aspects to consider when choosing or designing an LQE model (estimator). These aspects are often used to indirectly evaluate the performance of LQE models, by evaluating the behavior of the application that relies on LQE versus the one that does not rely on it.

\begin{enumerate}
	\item \textit{Reliability} - The LQE model should perform estimations that are as close as possible to the values observed. More explicitly, LQE models should maintain high accuracy.
	\item \textit{Adaptivity/Reactivity} - The LQE model should reach and adapt to persistent link quality changes. This indicates that when a link changes its quality for a longer period of time, the LQE model should be able to capture these changes and accordingly perform the estimations. Changes in estimation subsequently unveil routing topology changes.
	\item \textit{Stability} - The LQE model should be immune to transient link quality changes. This immunity ensures a relatively stable topology leading to reduced cost of routing overheads.
	\item \textit{Computational cost} - The computational complexity of LQE models should be considerate of the target devices, where computational load can be appropriately apportioned among constrained and powerful devices.
	\item \textit{Probing overhead} - LQE models consider a diverse set of metrics to estimate the link quality, as discussed in Section~\ref{subsec:input}, which are gathered through probing. LQE models should be designed in an optimal way so that the probing overhead is minimized.
\end{enumerate}

A comprehensive classification of the ML-based LQE studies according to the aforementioned five application quality aspects is exhibited in Fig.~\ref{fig:lqe-app-quality}, which reveals that most of the LQE studies explicitly consider \textit{computational cost} and \textit{reliability} aspects in their evaluations, whilst only a paucity of the studies considers \textit{probing overhead}, \textit{adaptability} and \textit{stability}. With respect to \textit{computational cost}, it can be readily observed from the figure that tree- and neural network-based methods tend to have higher computational cost, whereas online logistic regression has medium cost, and Naive Bayes, fuzzy logic and offline logistic regression have relatively low computational cost. With regards to the \textit{probing overhead} for trace-set collection, it is perceived from Fig.~\ref{fig:lqe-app-quality} that some LQE models are designed to incur zero-overhead, and one incurs both asynchronous and synchronous (async. \& sync.) probing, whereas the other is devised to use an adaptive probing rate. As far as \textit{reliability} is concerned, some LQE studies focus on the reliability of the routing tree topology, and on the link prediction/estimation, whereas others put emphasis on the traffic. \textit{Adaptability} is explicitly taken into consideration mostly in studies employing online learning algorithms, while \textit{stability} is considered for those studies focusing on offline learning algorithms.

\begin{table*}[!htp]
	\centering
	\scriptsize
	\renewcommand{\arraystretch}{1.2}
	\caption{Overview of the applications of the ML-based LQE models for the relevant papers surveyed in Tables~\ref{tab:lqe-part1} and~\ref{tab:lqe-part2}.}
	\label{tab:app-ml}
	\begin{tabularx}{\linewidth}{| C{.6} | C{1} | C{.4} | C{1} | C{0.35} | C{0.4} | C{0.25} | C{0.55} |  C{0.4} |}
		\hline\rowcolor{gray}
		\bfseries Purpose
		& \bfseries Specific Problems
		& \bfseries Research Papers
		& \bfseries ML Type and Method
		& \bfseries Reliability
		& \bfseries Adaptivity
		& \bfseries Stability
		& \bfseries Computational Cost
		& \bfseries Probing Overhead
		\\\hline
		
		\multirow{5}{*}{LQE for}
		\multirow{5}{*}{protocol}
		\multirow{5}{*}{performance}
		& \multirow{2}{*}{1. Reduce the cost of}
		\multirow{2}{*}{delivering a packet in}
		\multirow{2}{*}{multihop networks}
		\multirow{2}{*}{(CTP protocol)}
		& \cite{liu2011foresee} % Research 
		& \multirow{2}{*} {Classification:} 
		\multirow{2}{*} {Naive Bayes,} 
		\multirow{2}{*} {Logistic regression,} 
		\multirow{2}{*} {Artificial neural networks} 
		& 
		& No (offline)
		& 
		& Low
		&
		\\\cline{3-3}\cline{5-9}
		
		&
		& \cite{liu2012talent, liu2014temporal} % Research  
		&
		& Yes
		& Yes (online)  % method
		& 
		& Medium
		&
		\\\cline{2-9}
		
		& 2. Improve network reliability, reduce topology changes and routing depth
		& \cite{guo2013fuzzy} % Research 
		& Regression: Fuzzy logic (2 inference rules, defuzzification)% method
		& Yes
		&
		& Yes %topology changes, evaluated against 4BIT % evaluation
		& Low
		&
		\\\cline{2-9}
		
		& 3. Improve reliability and reactivity in an application specific network
		& \cite{rekik2015fli}
		& Classification: Custom algorithm based on fuzzy logic % method
		& Yes
		& Yes %Improves upon reactivity of F-LQE for smart grids.
		& Yes
		&
		&
		\\\cline{2-9}
		
		& 4. Minimize the overhead caused by active probing operations
		& \cite{ancillotti2017reinforcement} % Research 
		& Regression: Reinforcement learning % method
		&
		&
		&
		&
		& Yes
		\\\cline{2-9}
		
		& 5. Select links that maximize the delivery rate and minimize traffic congestion for routing.
		& \cite{shu2017research}  % Research 
		& Classification: SVM % method
		& Yes
		&
		&
		&
		&
		\\\hline
		
		\multirow{7}{*}{New or}
		\multirow{7}{*}{improved}
		\multirow{7}{*}{LQE}
		
		& \multirow{2}{*}{6. Prediction the quality}
		\multirow{2}{*}{of link in community }
		\multirow{2}{*}{network (WiFi)}
		& \cite{millan2015time}
		& Regression: SVM, regression trees, k-nearest neighbor, Gaussian process for regression
		& Yes
		& No (offline)
		& 
		& High
		&
		\\\cline{3-9}
		
		& 
		& \cite{bote2018online} % Research 
		& Regression: perceptron, regression trees, incremental model trees with drift detection and  adaptive model rules% method
		& Yes
		& Yes (online)
		&
		& High
		&
		\\\cline{2-9}
		
		&7. Link prediction quality, stability and reactivity
		&\cite{rehan2016machine} % Research 
		& Classification: custom algorithm + 2 extensions % method
		& %Channel rank estimation, energy consumption, channel switching delay, stability % evaluation
		& Yes
		& Yes
		&
		&
		\\\cline{2-9}
		
		&8. Reliable link quality estimation using probability-guaranteed estimation result
		&\cite{sun2017wnn} % Research 
		& Regression: Wavelet Neural networks
		& Yes
		&
		&
		&
		&
		\\\cline{2-9}
		
		&9. Improved LQE
		&\cite{luo2019link} % Research 
		& Classification: Deep learning (autoencoders) % method
		& Yes %Evaluated against \cite{sun2017wnn} in four environments: interor corridors, grove, park lots and road.  % evaluation
		&
		&
		&
		&
		\\\cline{2-9}
		
		&10. No overhead throughput estimation in mmWaves using RGB imaging
		&\cite{okamoto2017machine} % Research 
		& Regression: Adaptive regularization of weight vectors  % method
		& Yes
		& %No need for probing overhead % evaluation
		&
		&
		& Yes
		\\\cline{2-9}
		
		&11. Accurate estimation of LoRA transmissions using multispectral imaging
		&\cite{demetri2019automated} % Research 
		& Classification: SVMs with Radial Basis Function (RBF) kernel % method
		& Yes
		& %No need for pre-deployment measurements or probing overhead % evaluation
		&
		&
		& Yes
		\\\cline{2-9}
		
		&12. On Designing a Machine Learning Based WirelessLink Quality Classifier
		&\cite{cerar2020} % Research 
		& Classification: Logistic regression, decision trees, random forest, SVM, multi-layer perceptron % method
		& Yes
		& 
		&
		&
		&
		\\\hline
		
	\end{tabularx}
\end{table*}

\subsubsection*{Discussion}

To support a more in-depth understanding, Table~\ref{tab:app-ml} presents an aggregated and elaborated view of the papers that are systematically categorized in Figs.~\ref{fig:lqe-purpose} and~\ref{fig:lqe-app-quality}. The first column of the table shows the purpose for which LQEs have been developed, the second column of the table lists the problem that is being solved using ML-based LQE models, the third provides the relevant research papers solving those respective problems, column four includes the ML type and method, while the last five columns correspond to the link quality metrics previously enumerated in this section. The last five columns are filled in, if those quality aspects are given consideration in these respective research papers and left empty otherwise.
	
The first line of Table~\ref{tab:app-ml} indicates that the problem solved by \cite{liu2011foresee, liu2012talent, liu2014temporal} is to reduce the cost of packet delivery with a well-known multi-hop protocol, the so-called collection tree protocol (CTP). In their first approach, \cite{liu2011foresee} achieve this by developing three batch ML models that, according to their evaluation, perform better than 4BIT. However, ML models are trained in batch mode and remain static after training, therefore the estimator is not adaptive to persistent changes in the link. Batch or offline training of ML algorithms~\cite{banerjee2007topic} means that the model is trained, optimized and evaluated once on available training and testing sets, and has to be completely re-trained later in order to adapt the possible changes in the distribution of the updated data. In practice, this corresponds to sporadic updates, e.g., once in few hours and once per day depending on how the overall system is engineered. For the case of embedded devices, the device has to be fully or partially reprogrammed~\cite{ruckebusch2016gitar}. In the specific case of~\cite{liu2011foresee}, it is clear that the coefficients of the linear regression model learned during training are hard-coded on the target device and reprogramming is required for obtaining the updates.
	
When the behavior of the links changes significantly, especially for wireless networks having mobility, the offline model is expected to decrease in performance, since those link changes may not be recognized by the ML model residing on the devices. In~\cite{liu2012talent,liu2014temporal}, they improve their previously proposed offline modeling by introducing adaptivity to their models and thus developing online versions of the learning algorithms. Online ML algorithms are capable of updating their model~\cite{banerjee2007topic} as new data points arrive during regular operation. The authors of~\cite{liu2012talent,liu2014temporal} also address reliability and computational cost aspects in their evaluation, as can be readily seen in the respective columns of Table~\ref{tab:app-ml}.

Realizing the shortcomings of the offline-models~\cite{millan2015time} for estimating LQE in community networks and then developing on-line \cite{bote2018online} models can be also noticed in the sixth line of Table~\ref{tab:app-ml}. This research problem is formulated as a regression problem, while the previous one addressed in~\cite{liu2011foresee, liu2012talent, liu2014temporal} is formulated as a classification one. Both approaches are suitable for the purpose and both need to implement a threshold- or class-based decision making on whether to use the link or not. ML methods used in~\cite{millan2015time} and~\cite{bote2018online} target WiFi devices (routers) and are thus more expensive in terms of memory and computational cost than those that target constrained devices (sensors), as outlined at the first line of Table~\ref{tab:app-ml}. Generally speaking, ML algorithms, such as SVM and k-NN used in~\cite{millan2015time},~\cite{bote2018online} and outlined at line six of Table~\ref{tab:app-ml} are computationally more expensive than naive Bayes and logistic regression utilized in~\cite{liu2011foresee, liu2012talent, liu2014temporal} and outlined at the first line of Table~\ref{tab:app-ml}.

In addition to the adaptivity trade-offs noticed in research papers at the first and sixth rows of Table~\ref{tab:app-ml}, reactivity trade-offs can be perceived from research papers outlined in the second, third and seventh rows of Table~\ref{tab:app-ml}. More explicitly, in the second row, LQE model is used to improve network reliability by reducing topology changes and the depth of the routing tree~\cite{guo2013fuzzy}, while still maintaining high reliability, and in the third and seventh rows,~\cite{rekik2015fli} and~\cite{rehan2016machine} enhance reliability, stability and reactivity, respectively. The application requirements of these studies seem to favor reliable and cost effective routing with minimal routing topology changes. To sum up, the LQE model has to be as accurate as possible, update the model on significant link changes and remain immune to short-term variations for the sake of a stable topology. To achieve such goal, the right tuning of on-line learning algorithms that ensure a good stability vs adaptivity trade-offs has to be performed. 

The computation of LQE models involves probing overhead to collect relevant metrics, as discussed in Section~\ref{subsec:input} and Table~\ref{tab:feature-analysis}. Minimizing the probing overhead has also been a major concern for a number of research papers~\cite{demetri2019automated},~\cite{ancillotti2017reinforcement} and~\cite{okamoto2017machine}, as it can be readily observed from rows four, ten and eleven of Table~\ref{tab:app-ml}. In row four, probing overhead is reduced by using reinforcement learning to guide the probing process~\cite{ancillotti2017reinforcement}, while in~\cite{demetri2019automated} and~\cite{okamoto2017machine}, network related information obtained via probing is replaced with external non-networking sources based on imaging. Replacing the probing overhead with additional hardware components that involve learning from image data, image capturing and processing, consequently leads to increased computational complexity of the system.

The remaining research papers~\cite{sun2017wnn},~\cite{luo2019link} and~\cite{shu2017research} outlined at lines five, eight and nine of Table~\ref{tab:app-ml} address the aspects of developing more accurate estimators against predetermined baseline models. Additionally, the LQE model proposed by~\cite{sun2017wnn} provides probability-guaranteed estimation using packet reception ratio for satisfying reliability requirements of the smart grid communication standards.

\begin{figure*}
\centering
\begin{tikzpicture}[
scale=0.735,
level0/.style={ultra thick, black, font=\bfseries\normalsize, align=center},
level1/.style={ultra thick, black, font=\bfseries\small, align=center},
level2/.style={ultra thick, black, font=\footnotesize, align=center},
level3/.style={thick, black, font=\scriptsize, align=center},
connect/.style = {thick, dashed, red},
notice/.style  = { draw, rectangle callout, callout relative pointer={#1} },
label/.style   = { text width=2cm }
]
\draw[level0] (-2,0) -- node[above] {Design Decisions} (2,0);

% Type of ML
\draw[level1] (3,-2) -- node[above] {Type of\\ML} (5,-2);
\draw[connect] (2, 0) -- (3,-2);

\draw[level2] (6, 3.5) -- node[above] {Regression} (8, 3.5);
\draw[level2] (6,-4.5) -- node[above] {Classification} (8,-4.5);

\draw[connect] (5,-2) -- (6, 3.5) (5,-2) -- (6,-4.5);

% Regression
\draw[level3] (9, 8) -- node[above] {Fuzzy Logic \cite{guo2013fuzzy}} (12, 8);
\draw[level3] (9, 7) -- node[above] {SVM \cite{millan2015time}} (12, 7);
\draw[level3] (9, 6) -- node[above] {Regression Trees\\\cite{millan2015time, bote2018online}} (12, 6);
\draw[level3] (9, 5) -- node[above] {Gaussian Process \cite{millan2015time}} (12, 5);
\draw[level3] (9, 4) -- node[above] {kNN \cite{millan2015time}} (12, 4);
\draw[level3] (9, 3) -- node[above] {NN, WNN \cite{sun2017wnn}} (12, 3);
\draw[level3] (9, 2) -- node[above] {Reinforcement Learning\\\cite{ancillotti2017reinforcement}} (12, 2);
\draw[level3] (9, 1) -- node[above] {ARN \cite{okamoto2017machine}} (12, 1);
\draw[level3] (9, 0) -- node[above] {Online perceptron \cite{bote2018online}} (12, 0);

\foreach \x in {0,...,8} \draw[connect] (8, 3.5) -- (9, \x);

% Classification
\draw[level3] (9,-2) -- node[above] {Naive Bayes\\\cite{liu2011foresee, liu2012talent, liu2014temporal}} (12, -2);
\draw[level3] (9,-3) -- node[above] {Logistic Regression\\\cite{liu2011foresee, liu2012talent, liu2014temporal, cerar2020}} (12,-3);
\draw[level3] (9,-4) -- node[above] {Fuzzy \cite{rekik2015fli}} (12,-4);
\draw[level3] (9,-5) -- node[above] {SVM \cite{shu2017research, demetri2019automated, cerar2020}} (12,-5);
\draw[level3] (9,-6) -- node[above] {DL, ANN\\\cite{liu2011foresee, liu2012talent, liu2014temporal, luo2019link, cerar2020}} (12,-6);
\draw[level3] (9,-7) -- node[above] {Decision Trees \cite{cerar2020}} (12,-7);
\draw[level3] (9,-8) -- node[above] {Custom \cite{rehan2016machine}} (12,-8);

\foreach \x in {-8,...,-2} \draw[connect] (8, -4.5) -- (9, \x);

% Evaluation metrics
\draw[level1] (-3,-3) -- node[above] {Evaluation\\Metrics} (-5,-3);
\draw[connect] (-2,0) -- (-3, -3);

\draw[level2] (-6,0) -- node[above] {Standard\\Metrics} (-8,0);
\draw[level2] (-6,-5.5) -- node[above] {Application\\Specific} (-8,-5.5);
\draw[connect] (-5,-3) -- (-6,0) (-5,-3) -- (-6,-5.5);

\draw[level3] (-9, 3) -- node[above, text width=8em] {Accuracy\\\cite{liu2011foresee, liu2012talent, liu2014temporal, shu2017research, luo2019link, demetri2019automated, cerar2020}} (-12, 3);
\draw[level3] (-9, 2) -- node[above] {Recall \cite{luo2019link}} (-12, 2);
\draw[level3] (-9, 1) -- node[above] {Confusion Matrix\\\cite{liu2012talent, liu2014temporal, luo2019link, cerar2020}} (-12, 1);
\draw[level3] (-9, 0) -- node[above] {MSE, RMSE\\\cite{guo2013fuzzy, sun2017wnn, ancillotti2017reinforcement, okamoto2017machine, demetri2019automated}} (-12, 0);
\draw[level3] (-9,-1) -- node[above] {MAE \cite{millan2015time, bote2018online}} (-12, -1);
\draw[level3] (-9,-2) -- node[above] {CDF \cite{rekik2015fli}} (-12, -2);
\foreach \x in {-2,...,3} \draw[connect] (-8, 0) -- (-9, \x);

\draw[level3] (-9,-4) -- node[above] {Throughput \cite{okamoto2017machine}} (-12, -4);
\draw[level3] (-9,-5) -- node[above] {Topology Changes \cite{guo2013fuzzy}} (-12,-5);
\draw[level3] (-9,-6) -- node[above] {Stability \cite{rekik2015fli, rehan2016machine}} (-12,-6);
\draw[level3] (-9,-7) -- node[above] {Delivery Cost \cite{liu2011foresee}} (-12,-7);
\foreach \x in {-7,...,-4} \draw[connect] (-8, -5.5) -- (-9, \x);

% Cleaning & Interpolation
\draw[level1] (-3, 3) -- node[above] {Cleaning \&\\Interpolation} (-5, 3);
\draw[connect] (-2,0) -- (-3, 3);

\draw[level3] (-6, 4) -- node[above] {Missing Values\\\cite{liu2012talent, liu2014temporal, guo2013fuzzy, okamoto2017machine, luo2019link, cerar2020}} (-8, 4);
\draw[level3] (-6, 3) -- node[above] {Averaging \cite{okamoto2017machine, bote2018online, cerar2020}} (-8, 3);
\draw[level3] (-6, 2) -- node[above] {Scaling \cite{luo2019link}} (-8, 2);
\foreach \x in {2,3,4} \draw[connect] (-5, 3) -- (-6, \x);

% Resampling
\draw[level1] (3, 8) -- node[above] {Resampling} (5, 8);
\draw[connect] (2,0) -- (3, 8);

\draw[level3] (6, 8.5) -- node[above] {ML-based} (8, 8.5);
\draw[level3] (6, 7.5) -- node[above] {Random \cite{cerar2020}} (8, 7.5);
\foreach \y in {7.5, 8.5} \draw[connect] (5, 8) -- (6, \y);

% Feature Selection
\draw[level1] (-3, 8) -- node[above] {Feature\\Selection} (-5, 8);
\draw[connect] (-2,0) -- (-3, 8);

\draw[level3] (-6, 8.5) -- node[above, text width=12em] {Available \cite{liu2011foresee, liu2012talent, liu2014temporal, rekik2015fli, rehan2016machine, sun2017wnn, ancillotti2017reinforcement, shu2017research, okamoto2017machine, luo2019link, demetri2019automated, cerar2020}} (-11, 8.5);
\draw[level3] (-6, 7.5) -- node[above, text width=12em] {Synthetic \cite{liu2011foresee, liu2012talent, liu2014temporal, guo2013fuzzy, rekik2015fli,millan2015time, rehan2016machine, ancillotti2017reinforcement, shu2017research, bote2018online,luo2019link, demetri2019automated, cerar2020}} (-11, 7.5);
\foreach \x in {8.5,7.5} \draw[connect] (-5, 8) -- (-6, \x);
\end{tikzpicture}
\caption{Overview of the design decisions taken during the development of the ML-based LQE models for the relevant papers surveyed in Tables~\ref{tab:lqe-part1} and~\ref{tab:lqe-part2}.}
\label{fig:steps-desc-overview}
\end{figure*}
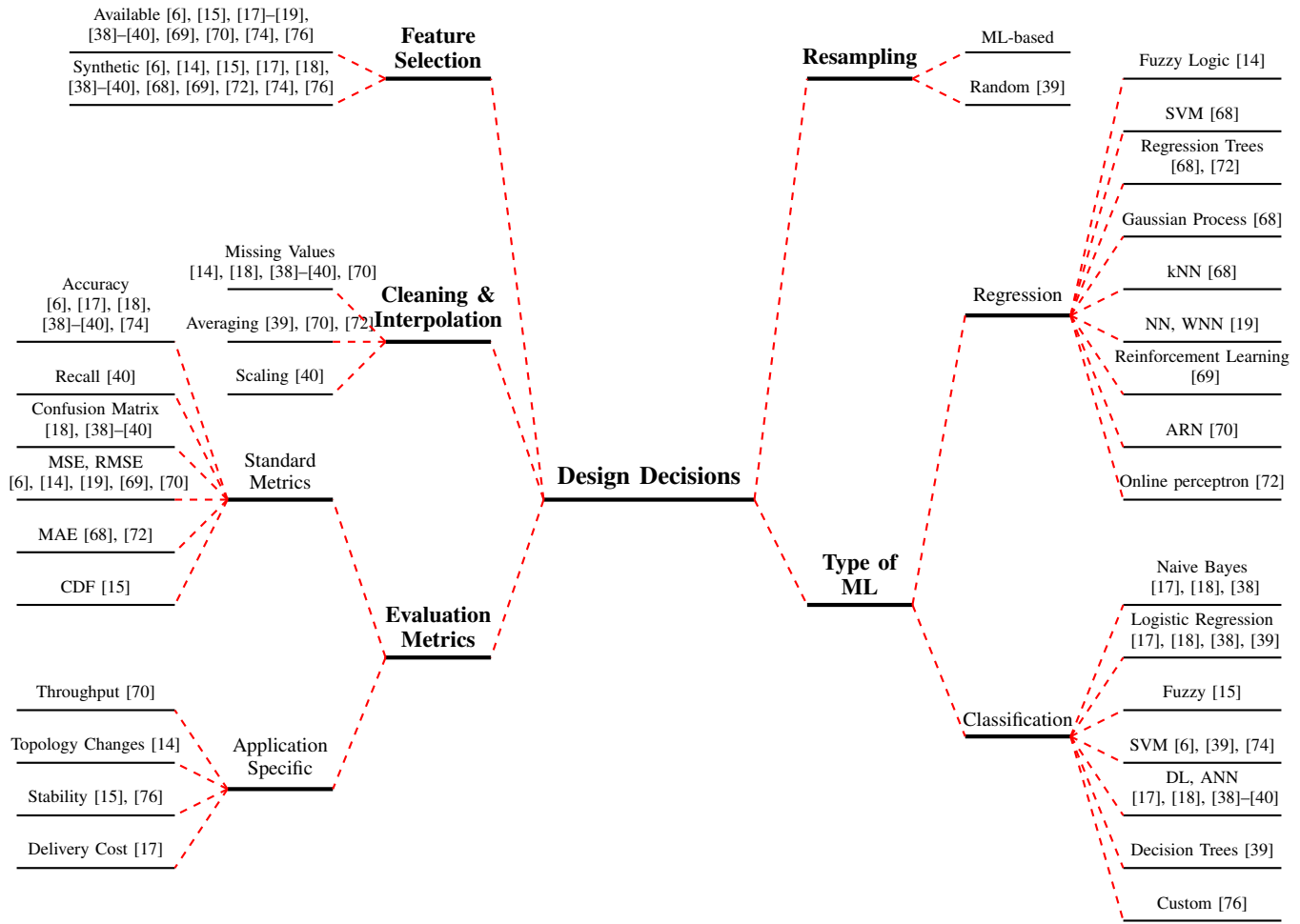

\section{Design Process Perspective of ML-Based LQEs}
\label{sec:lqe-ML-analysis}

For the development of any ML model, the researchers have to follow some very precise steps that are well established in the community, defined in the \gls{KDP}~\cite{fayyad1996data,kulin2016data}, namely data pre-processing, model building and model evaluation. The data pre-processing stage is known to be the most time-consuming process, tends to have a major influence on the final performance of the model and is applied on the training and evaluation data collected based on the input metrics discussed in Section~\ref{subsec:input}. This stage includes several steps, such as data cleaning and interpolation, feature selection and resampling. The model building and selection steps usually take a set of ML methods, train them using the available data and evaluate their results, as discussed in Section~\ref{sub:evaluation-metrics}.

\begin{figure*}[htpb]
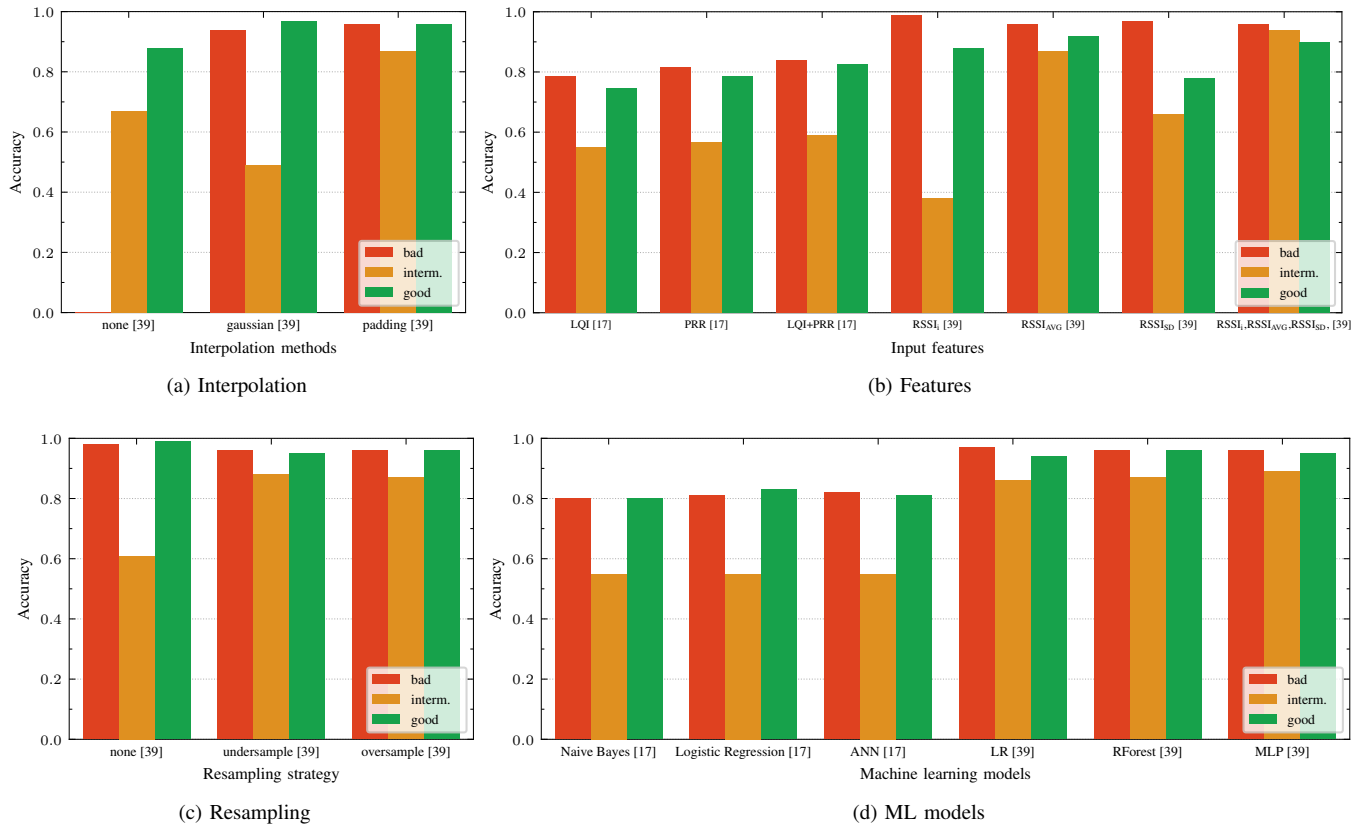

	\centering
	\subfloat[Interpolation\label{fig:eval:interpolation}]{\scalebox{0.78}{\input{./figures/benchmark/benchmark-interpolation.pgf}}}
	\subfloat[Features\label{fig:eval:features}]{\scalebox{0.78}{\input{./figures/benchmark/benchmark-features.pgf}}}
	
	\subfloat[Resampling\label{fig:eval:resampling}]{\scalebox{0.78}{\input{./figures/benchmark/benchmark-resampling.pgf}}}
	\subfloat[ML models\label{fig:eval:models}]{\scalebox{0.78}{\input{./figures/benchmark/benchmark-models.pgf}}}
	%\subfloat[Interpolation\label{fig:eval:interpolation}]{\includegraphics[width=0.35\linewidth]{figures/benchmark/benchmark-interpolation}}
	%\subfloat[Features\label{fig:eval:features}]{\includegraphics[width=0.65\linewidth]{figures/benchmark/benchmark-features}}
	%\subfloat[Resampling\label{fig:eval:resampling}]{\includegraphics[width=0.35\linewidth]{figures/benchmark/benchmark-resampling}}
	%\subfloat[Models\label{fig:eval:models}]{\includegraphics[width=0.65\linewidth]{figures/benchmark/benchmark-models}}
	\caption{Accuracy performance analyses for various steps of the design process as an exemplifying three-class LQE classification problem with unbalanced training data.}
	\label{fig:eval}
\end{figure*}

\begin{figure*}[htpb]
	\centering
	\includegraphics[width=0.95\linewidth]{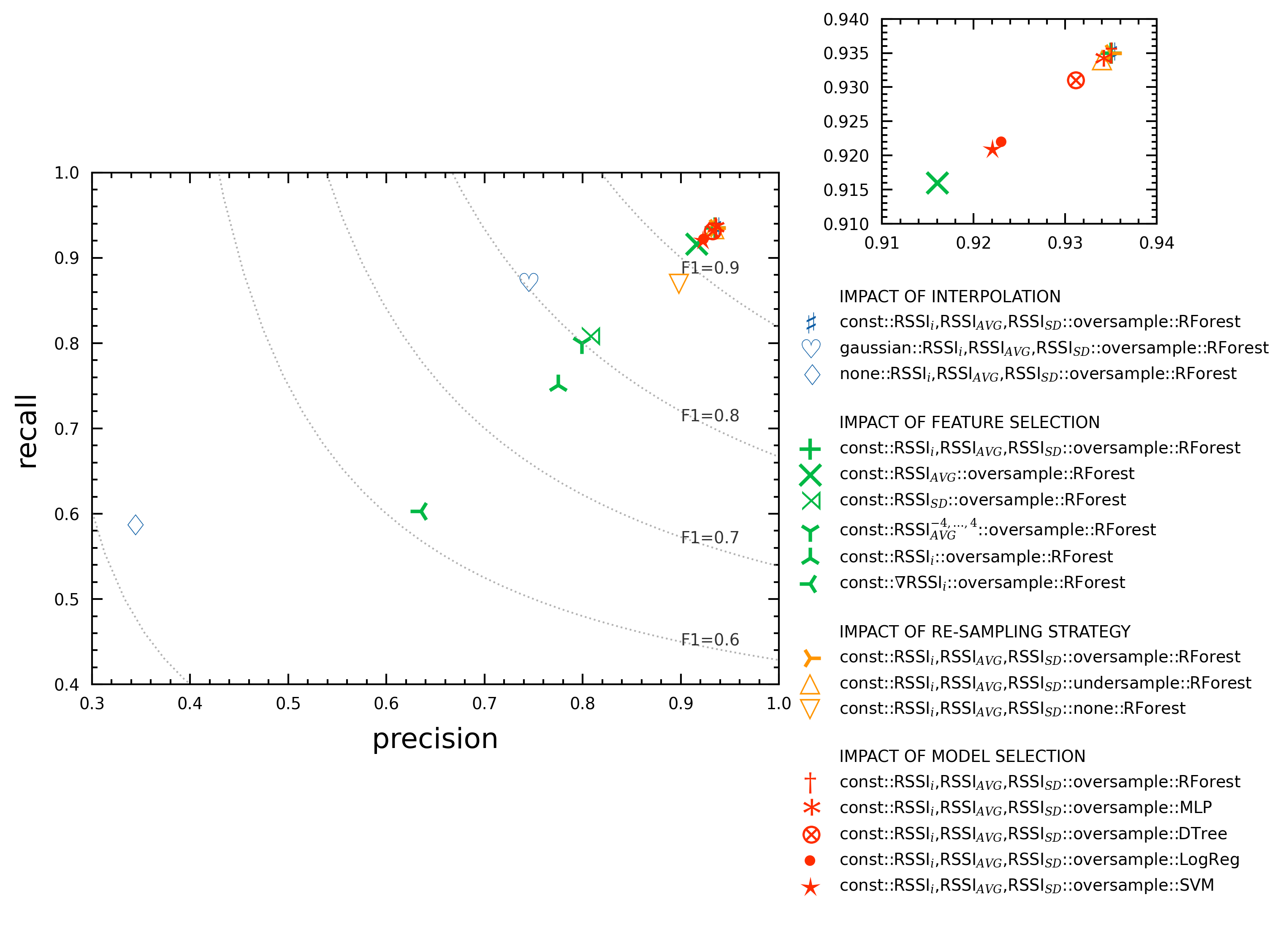}
	\caption{Precision vs. recall performance trade-off for various design decisions including interpolation, feature selection, resampling and model selection, where the figure situated at the top-right corner is a zoomed-in portion of the closest region to F1=1 of the main figure.}
	\label{fig:benchmark}
\end{figure*}

Analyzing the existing works from the perspective of the design process is equally important and complements the analysis from the application perspective performed in Section~\ref{sec:lqe-ML-analysis-app}. Fig.~\ref{fig:steps-desc-overview} classifies the studies based on the reported design decisions taken while developing ML-based \gls{LQE} models, namely cleaning and interpolation, feature selection, re-sampling strategy and ML model selection. Fig. \ref{fig:eval} compares the reported influence of the respective steps on the final model considering accuracy as the metric while Fig.~\ref{fig:benchmark} depicts the trade-off for the process considering the F1 score\footnote{$F1=2*precision*recall/(precision+recall)$} and the precision\footnote{$precision=true\ positives/(true\ positives\ +\ false\ positives)$}\ and recall\footnote{$recall=true\ positives/(true\ positives\ +\ false\ negatives)$} metrics.


\subsection{Cleaning \& interpolation steps}
\label{sub:cleaning-interpolating}
%motivation
From the Cleaning \& Interpolation branch of the mind map depicted in Fig.~\ref{fig:steps-desc-overview} it can be seen that only seven of the \gls{ML}-based LQE models provide explicit consideration of the cleaning and interpolation step. While in the general ML practice that use real world datasets, the cleaning step is very difficult to avoid and LQE-based research papers mostly leverage carefully collected datasets, often generated in-house from existing testbeds, as discussed in Section~\ref{sec:techstandards}. For instance, Okamoto~\textit{et al.}~\cite{okamoto2017machine} perform cleaning on the image data they selected to use as part of the model training.

With respect to interpolation, however, several works \cite{liu2012talent,liu2014temporal,guo2013fuzzy,luo2019link} fill in missing values with zeros. Their design decision with respect to this step of the process can also be referred to as interpolation using domain knowledge as they replace the missing \gls{RSSI} values with 0, which represents a poor quality link with no received signal, yielding \gls{PRR} equal to 0. It is not clear how~\cite{bote2018online} handle the missing data, however, they drop measurement data if there are not enough variations in their values. 

Explicitly mentioning the design decision with respect to cleaning and interpolation is important for reproducibility (discussed in Section~\ref{sub:reproducibility}) as well as for its potential influence on the final performance of the ML model. For instance, it can be readily seen from Fig.~\ref{fig:eval:interpolation} that, all the other settings kept the same, domain knowledge interpolation denoted by "padding" can increase the accuracy of a classifier on \textit{good} classes from 0.88 to 0.95, while also increasing the performance on the minority classes from 0.49 to 0.87 for \textit{intermediate} and nearly 0 to 0.98 for \textit{bad}, which can also be perceived from the findings of~\cite{cerar2020}. Going beyond accuracy as an evaluation metric, Fig.~\ref{fig:benchmark} shows significant performance increase, measured with F1 score which is the harmonic mean of the precision and recall, if the type of used interpolation is optimized for a particular scenario. More specifically, F1 score for no-interpolation is about 0.43 on the left lower part of the figure, then increases to 0.80 with Gaussian interpolation, and finally reaching 0.94 with constant interpolation (denoted by "padding" in Fig.~\ref{fig:eval:interpolation}) that utilizes domain knowledge.
 
\subsection{Feature selection}
\label{sub:feature-selection}
According to the feature selection branch of the mind map depicted in Fig.~\ref{fig:steps-desc-overview}, all research papers provide details on their feature selection. Often, all the features directly collected from testbed and simulator are used, as discussed in Section~\ref{sub:feature-selection}. Part of the literature, i.e.,~\cite{liu2011foresee},~\cite{liu2014temporal} and~\cite{liu2012talent} also considers the performance of the final model as a function of the input features as part of their analysis, while others only report a fixed set of features that are then used to develop and evaluate models. It may be because, the authors implicitly considered the feature selection step and solely reported the final features selected for their models to keep their paper concise. In such cases, the influence of other features or synthetic features~\cite{freitas2001understanding} cannot be readily assessed in the related works surveyed. 

Perceived from an extensive comparative evaluation in~\cite{cerar2020} and from another study that explicitly quantifies the impact of the feature selection on an LQE classification problem in~\cite{liu2011foresee}, we summarize the reported performances with respect to the feature selection step in Fig.~\ref{fig:eval:features}. While the works of the aforementioned figure leverage different datasets and distinct ML approaches, therefore they cannot be fairly benchmarked against each other, it is clear that the feature engineering can significantly increase the accuracy of a classifier within the same work by keeping all the other settings the same. Liu~\cite{liu2011foresee} reports up to 9 percentage points classification improvement in all classes by using LQI+PRR compared to the scenario using LQI only and PRR only, while Cerar~\cite{cerar2020} reports on average classification performance increases from 0.89 to 0.95, while also increasing the performance on the minority class from 0.38 to 0.87. Furthermore, according to Fig.~\ref{fig:benchmark}, classification performance of F1 score ranges from 0.61 to 0.93, of precision ranges from 0.62 to 0.93 and of recall ranges from 0.63 to 0.93.

\subsection{Resampling strategy}
\label{sub:resampling}
Resampling is used in ML communities when the available input data is imbalanced \cite{chawla2004special, yap2014application}. For instance, assume a classification problem where the aim is to classify links into \textit{good}, \textit{bad} and \textit{intermediate} classes, similar to the problem approached in \cite{rehan2016machine,audeoud2018quick}. If the \textit{good} class would represent 75\% of the examples in the training dataset, \textit{bad} would represent 20\% and \textit{intermediate} would represent the remaining 5\%, then a ML model would likely be well trained to recognize the \textit{good} classes as it has been exposed to many such instances. However, it might be difficult for the model to recognize the other two classes, as they are scarcely populated instances in the dataset. 

According to the resampling branch of the mind map in Fig.~\ref{fig:steps-desc-overview}, only one very recent  research papers elaborates on their resampling strategy. In other works it is often not clear whether they employed a resampling strategy in the case of imbalanced datasets. For instance, the performance of the predictor on two of the five classes is modest in~\cite{luo2019link}. It would be interesting to understand whether employing a resampling strategy would provide a better discrimination of the considered classes. Resampling could also improve other surveyed estimators in~\cite{liu2012talent, liu2014temporal, shu2017research, demetri2019automated}. 

From Fig.~\ref{fig:eval:resampling}, it can be seen that, all the other settings being the same, performing resampling can slightly decrease the accuracy of a classifier on the two majority classes from 0.97 to 0.95, albeit it can yield a dramatic increase in the classification performance of the minority \textit{intermediate} class with an accuracy raise from 0.61 to 0.88, which can also be worked out from the findings of~\cite{cerar2020}. Going beyond accuracy as an evaluation metric, Fig.~\ref{fig:benchmark} exhibits significant precision, recall and F1 score increase for the minority class, when a resampling strategy is leveraged. More specifically, an LQE model without resampling yields an F1 score of about 0.87, which then  increases to about 0.93 with undersampling and remains at 0.93 when oversampling is considered.

\subsection{Machine learning method}
\label{sub:ML-method}
According to the ML method branch of the mind map shown in Fig.~\ref{fig:steps-desc-overview} seven of the works estimate the link quality in terms of discrete values, therefore they perform classification, while the remaining seven estimate it as actual values, hence regression is employed. The preferred ML method is chosen according to the specific application considered. It can be seen from this branch that the same type of algorithm can be adopted for classification and regression, respectively. For example, SVMs are exploited for regression in~\cite{millan2015time} and for classification in~\cite{shu2017research,demetri2019automated}. Besides, every ML algorithm can be adapted to work in an on-line mode by means of retraining the model with every new incoming value during its operation. As discussed in Section~\ref{sec:lqe-ML-analysis-app}, online learning are particularly suitable for LQE models that also optimize the adaptivity in~\cite{liu2014temporal,okamoto2017machine,bote2018online}.

For classification, the most frequently used ML algorithms are naive Bayes, logistic regression, artificial neural networks (ANNs) and SVMs. The first three are used in~\cite{liu2011foresee,liu2012talent,liu2014temporal}, while SVMs are used in~\cite{shu2017research,demetri2019automated}. The ML algorithms used for regression are more diverse ranging from fuzzy logic to reinforcement learning. While the performance of the classification algorithms is often evaluated according to the precision/recall and F1 scores in ML communities, potentially via complementary confusion matrices, the performance of regression are evaluated using distance metrics, such as RMSE and MAE. 

Fig.~\ref{fig:eval:models} shows that, all the other settings being the same, the selection of the ML method for a selected classification problem has a relatively smaller impact on the accuracy of a classifier compared to the other steps of the design process. As reported in both ~\cite{liu2011foresee} and ~\cite{cerar2020}, the accuracy changes by up to 3 percentage points between the considered models. The zoomed portion of Fig.~\ref{fig:benchmark} exhibits the negligible impact of the model selection on the F1 score, which is up to around 0.02.

\section{Overview of Measurement Data Sources}
\label{sec:dataset-overview}

\begin{table*}[!htb]
	\centering
	\scriptsize
	\renewcommand{\arraystretch}{1.2}
	\caption{Publicly available trace-sets for the analysis of LQE.}
	\label{tab:datasets}
	\begin{tabularx}{\linewidth}{| C{1} | C{1} | C{1} | C{1} | C{.5} | L{1.5} |}
		
		\hline\rowcolor{gray}
		\bfseries Origin of Trace-sets 
		& \bfseries HW. \& Technology
		& \bfseries Measurements
		& \bfseries Data Points
		& \bfseries Type
		& \bfseries Additional Notes
		\\\hline
		
		% INFO: https://www.cs.princeton.edu/courses/archive/spring17/cos598A/lectures/roofnet.md
		% SRC: N/A, no origin, scattered around the internet usually as dataset for ML courses
		MIT, Roofnet, \cite{aguayo2004link, gokhale2008feasibility}, 2002
		& Cisco Aironet 350, IEEE~802.11b, mesh, custom Roofnet protocol
		& Source, destination, sequence, time, signal, noise and so on
		& 21\,258\,359 \newline (1725~links, 4~bitrates)
		& 1-to-N
		& Which packets were lost on a link is not provided.
		\\\hline

		% SRC: https://crawdad.org/rutgers/noise/20070420/
		Rutgers University, ORBIT testbed, \cite{kaul2006creating}, 2007
		& 29x~PC + Atheros~5212, IEEE~802.11abg
		& Seq.~number, RSSI
		& 611\,632 \newline(406~links, 300~packets/link, 1~packet/100 ms, 5 levels of noise)
		& 1-to-N
		& Minor preprocessing is involved.
		\\\hline
		
		% SRC: https://crawdad.org/due/packet-delivery/20150401/packet-metadata/
		``Packet-metadata'', \cite{fu2015experimental}, 2015
		& 2x~TelosB, IEEE~802.15.4
		& RSSI, LQI, noise floor, packet size, no. retries, energy, Tx power, ACK, queue size and so on
		& 14\,515\,200 \newline(300~packets per 80646~runs per 6~distances)
		& 1-to-1
		& It requires minor preprocessing.
		\\\hline
		
		% SRC: https://crawdad.org/cu/rssi/20090528/
		Colorado, \cite{bauer2009physical}, 2009
		& 5x~listeners, IEEE~802.11
		& Signal strength, data rate, channel, time-stamp and so on
		& 29\,000 \newline(500 packets per 58 locations)
		& 1-to-1
		& It requires preprocessing.
		\\\hline
		
		% SRC: https://crawdad.org/umich/rss/20110810/
		University of Michigan, \cite{chen2011robust}, 2006
		& 14x~Mica2, proprietary protocol, sub-GHz ISM
		& RSSI
		& 580\,762 \newline(1~packet/0.5s, 30~min/device, 3191~records/link)
		& 1-to-N
		& MATLAB's binary format is considered and inconsistent data is observed (leading zeros and no units). Source and destination nodes are not clearly identified.
		\\\hline
		
		% SRC: https://archive.ics.uci.edu/ml/machine-learning-databases/00321/
		EVARILOS, UGent, \cite{van2015platform}, 2015
		& 6 nodes, Bluetooth
		& RSSI, time-stamp
		& 5\,938 \newline(\textless2\,000~records/link)
		& N-to-1
		& Hospital environment is considered in the absence of interference.
		\\\hline
		
		EVARILOS, UGent, \cite{van2015platform}, 2015
		& 5 nodes, IEEE~802.15.4
		& RSSI, time-of-arrival, time-stamp
		& 110\,126 \newline(\textless35\,000~records/link)
		& 1-to-N
		& Hospital environment is considered in the absence of interference.
		\\\hline
		
		% SRC: https://crawdad.org/cu/antenna/20090508/
		University of Colorado, \cite{anderson2009impact, anderson2011modeling}, 2009
		& 6x~PC with omni-directional antennas, 1x distinctly configured omni-directional antenna for transmitter, IEEE~802.11
		& Seq.~number, coordinates, direction, TX power, 5x~RSSI values per log
		& 5x~623\,207 \newline(500~packets per 180~positions per 4~directions per 11~Tx~levels per 5~nodes)
		& 1-to-N
		& Experiment is composed of nodes equipped with antennas that are capable of serving 4 different directions. Tx power is variable and extensive documentation is available.
		\\\hline
		
		% SRC: http://di.ulb.ac.be/labo/datasets.html
		Brussels University, \cite{le2007principal}, 2007
		& 19x~Tmote~Sky, IEEE~802.15.4
		& Seq.~number, RSSI, LQI, time-stamp
		& 112\,793 \newline(\textless1\,600~packet/link) % \gcerar{Timi got (547\,200 overall), did we look same dataset?}
		& 1-to-N
		& It requires advanced preprocessing. Sequence numbers are rarely inconsistent. There are three more trace-sets available from this experiment that is intended for localization.
		\\\hline
	\end{tabularx}
\end{table*}

\begin{table*}[!htb]
	\centering
	\scriptsize
	\renewcommand{\arraystretch}{1.2}
	\caption{Available features of the trace-sets surveyed in Table~\ref{tab:datasets} for the sake of LQE.}
	\label{tab:datasets-features}
	\begin{tabularx}{\linewidth}{| X | c | c | c | c | c | c | c | c | c |}
		
		\hline\rowcolor{gray}
		\bfseries Trace-set
		& \bfseries Seq.~Numbers
		& \bfseries Time-stamp
		& \bfseries RSSI
		& \bfseries LQI
		& \bfseries SNR (Signal/Noise)
		& \bfseries Location
		& \bfseries Queue (Size/Length)
		& \bfseries Frame Size
		& \bfseries HW. Specs.
		\\\hline

		Roofnet~\cite{aguayo2004link, gokhale2008feasibility} % Name
		&  % Seq
		& \cmark (implicit) % timestamp
		& % RSSI
		& % LQI
		& \cmark / \cmark % SNR (signal/noise)
		& % Location
		& % queue size / length
		& % frame length
		& \cmark % HW specs
		\\\hline

		Rutgers~\cite{kaul2006creating} % Name
		& \cmark % Seq
		& \cmark % timestamp
		& \cmark % RSSI
		& % LQI
		& \xmark / \cmark % SNR (signal/noise)
		& \cmark % Location
		& % queue size / length
		& % frame length
		& \cmark % HW specs
		\\\hline

		``Packet-metadata''~\cite{fu2015experimental} % Name
		& \cmark % Seq
		& \cmark % timestamp
		& \cmark % RSSI
		& \cmark % LQI
		& \cmark / \cmark % SNR (signal/noise)
		& \cmark % Location
		& \cmark / \cmark % queue size / length
		& \cmark % frame length
		& \cmark % HW specs
		\\\hline

		Colorado~\cite{bauer2009physical} % Name
		& \cmark % Seq
		& \cmark % timestamp
		& \cmark % RSSI
		& % LQI
		& % SNR (signal/noise)
		& \cmark % Location
		& % queue size / length
		& \cmark % frame length
		& \cmark % HW specs
		\\\hline

		University of Michigan~\cite{chen2011robust} % name
		& \cmark % Seq
		& % timestamp
		& \cmark % RSSI
		& % LQI
		& % SNR (signal/noise)
		& % Location
		& % queue size / length
		& % frame length
		& \cmark % HW specs
		\\\hline

		% SRC: https://archive.ics.uci.edu/ml/machine-learning-databases/00321/
		EVARILOS~\cite{van2015platform} % name
		& \cmark % Seq
		& \cmark % timestamp
		& \cmark % RSSI
		& % LQI
		& % SNR (signal/noise)
		& \cmark % Location
		& % queue size / length
		& % frame length
		& \cmark % HW specs
		\\\hline

		Colorado~\cite{anderson2009impact, anderson2011modeling} % name
		& \cmark % Seq
		& \cmark % timestamp
		& \cmark % RSSI
		& % LQI
		& % SNR (signal/noise)
		& \cmark % Location
		& % queue size / length
		& % frame length
		& \cmark % HW specs
		\\\hline

		Brussels~\cite{le2007principal} % name
		& \cmark % Seq
		& \cmark % timestamp
		& \cmark % RSSI
		& \cmark % LQI
		& % SNR (signal/noise)
		& \cmark % Location
		& % queue size / length
		& % frame length
		& % HW specs
		\\\hline
	\end{tabularx}
\end{table*}

To complement the survey of the \gls{LQE} models developed using data, we perform a survey of the publicly available trace-sets that have already been used or could be used for \gls{LQE}. The data collected for a limited period of time on a given radio link, is referred to as \textit{traces} in this section. When a set of these traces is recorded using more links and/or periods in several rounds of tests for a given testbed, we refer for it as a \textit{trace-set}. Traces and trace-sets, in general, are prone to have irregularities and missing values that need to be preprocessed, especially when ported into \gls{ML} algorithms. In this paper, we refer to a trace-set that has been preprocessed as \textit{dataset}. Ideally, a trace-set should include all the information available that is directly or indirectly related to the packets' trip.

To support our analysis, Tables~\ref{tab:datasets} and \ref{tab:datasets-features} summarize the publicly available trace-sets and the available features in each trace-set respectively. Our survey only analyzes publicly available trace-sets for \gls{LQE} research that we were able to look into, however we mention other applicable trace-sets that are not publicly available. Table~\ref{tab:datasets} reviews the source of the trace-sets and the estimated year of creation along with the hardware and technology used for the trace-set gathering. Additionally, data that each trace relies on, the size of the trace/trace-set, the type of communication used in the measurement campaign, and additional notes on the specification and characteristic of the trace-sets can also be found in Table~\ref{tab:datasets}. Table~\ref{tab:datasets-features} lists the trace-sets in the first column while the remaining columns refer to various metrics contained within the trace-set. This table maps the available metrics, also referred to as features, to the analyzed trace-sets.

To summarize the important points of these trace-sets, they were collected by the research teams at various universities worldwide using their own testbeds~\cite{aguayo2004link, kaul2006creating, van2015platform} or via conducting one-time deployments~\cite{fu2015experimental, bauer2009physical, chen2011robust, anderson2009impact, le2007principal}. This confirms that the trace-sets were likely generated on testbeds developed and maintained in universities, which is consistent also with our findings in Section~\ref{sec:techstandards}. According to the second column of Table~\ref{tab:datasets}, four of the trace-sets are based on IEEE~802.11, three utilize IEEE~802.15.4, one is based on IEEE~802.15.1, and one operates on a proprietary radio technology. According to the fourth column of the table, the number of entries, i.e. data points, ranges from only 6 thousand up to 21 million, whereas the number of measured data per entry ranges from one to about fifteen. The third column of the table lists the measurements available in each trace-set. For more clarity, the measurements are summarized in Table~\ref{tab:datasets-features} for each trace-set and their meaning and importance for \gls{LQE} is summarized as follows:

\begin{itemize}
 \item A sequence number holds key information on the consecutive orders of the received packets and/or frames. With the aid of the sequence number, reconstruction of time series is enabled and thus it inherently provides information on packet loss and duplicated packets. It is already part of the frame headers owing to the standardization efforts. Sequence numbers can be processed to provide \gls{PRR} and its counterpart \gls{PER} that are useful input for \gls{LQE} model.
 \item A time-stamp, which can be relative or absolute, is a suitable addition to the aforementioned sequence number. It reveals the amount of elapsed time between measurements. Therefore, it can help for deciding on whether a previous data point is still relevant and thus improving \gls{LQE} in a dynamic environment. If a high precision timer and dedicated radio hardware are available, time-stamps can also empower localization. 
 \item Measurement points indicating the quality of received signal on the links are mainly described by SNR, RSSI and LQI. \gls{SNR} represents the ratio between the signal strength and the background noise strength. Compared to all other features, it allows the most clear-cut observation of the radio environment. However, some hardware, especially constrained devices, might not support direct \gls{SNR} observation. In contrast to \gls{SNR}, \gls{RSSI} is the most widely-used measurement data and it can be accessed on the majority of radio hardware. It shows high correlation with \gls{SNR}, since it is obtained in a similar way. Researchers may argue on its inaccuracy due to the low precision, i.e., quantization is around 3dB on most hardware. As opposed to the \gls{SNR} and \gls{RSSI}, \gls{LQI} is a score-based measurement data and mostly found in radios of ZigBee-like (IEEE 802.15.4) technologies, which provides an indication of the quality of a communication channel for the transmission and the flawless reception of signals. However, the drawback of \gls{LQI} is the lack of strict definition, leaning it to the vendor to decide its way of implementation and it may lead to the difficulty of cross-hardware comparison across vendors.
 \item For a more dynamic environment of wireless networks, where nodes are mainly mobile, information regarding the physical (geographical) locations can be beneficial.
 \item Additionally, there are other software related measurements data including queue size, queue length and frame length just to name few. If we refer to domain knowledge\footnote{Domain knowledge is the knowledge relating to the associated environment in which the target system performs, where the knowledge concerning the environment of a particular application plays a significant role in facilitating the process of learning in the context of \gls{ML} algorithms.}, shorter frames tend to be more prone to errors, while queuing statistics can reveal information concerning buffer congestions.
 \item For the interpretation of the technical research outcome, revealing which hardwares were utilized during data collection is important to help diagnosing potential erratic behaviors of some hardware, including sensitivity degradation with time.
\end{itemize}

As can be seen from Table~\ref{tab:datasets-features}, no single metrics appears in all trace-sets, however, sequence numbers, time stamps, \gls{RSSI}, location and hardware specifications are available in the majority.

% Roofnet
The Roofnet~\cite{aguayo2004link} is a well known WiFi-based trace-set built by MIT. It contains the largest number of data points among the trace-sets listed in Table~\ref{tab:datasets}. However, it is difficult to obtain the exact Roofnet setup/configuration used during the collection of the measurement data, since it has evolved with other contributions. One particular drawback of Roofnet is that \gls{PRR}, as a potential \gls{LQE} candidate, can only be computed as an aggregate value per link without the knowledge of how the link quality varied over time. Table~\ref{tab:datasets-features} shows that this particular trace-set strictly depends on \gls{SNR} values for the analysis of \gls{LQE}.

% Rutgers
The Rutgers trace-set~\cite{kaul2006creating} was gathered in the ORBIT testbed. It is large enough for \gls{ML} models, requires only moderate preprocessing and is appropriately formed for data-driven \gls{LQE}. It contains the overall packet loss of 36.5\%. The meta-data contains information regarding physical positions, timestamps and hardware used. The trace-set for each node contains raw \gls{RSSI} value along with the sequence number, as depicted in Table~\ref{tab:datasets-features}. From the surveyed papers, \cite{liu2014temporal} relies on both Rutgers and Colorado, while~\cite{srinivasan2008prr} considers only Rutgers.

% packet-metadata
The ``packet-metadata''~\cite{fu2015experimental} comes with a plethora of features convenient for \gls{LQE} research, as indicated in Table~\ref{tab:datasets-features}. In addition to the typical \gls{LQI} and \gls{RSSI}, it provides information about the noise floor, transmission power, dissipated energy as well as several network stacks and buffer related parameters. One of the major characteristic of this trace-set is to enable the observation of packet queue. Packet loss can only be observed in rare cases with very small packet queue length.

Upon closer investigation for the remaining six trace-sets listed in Table~\ref{tab:datasets}, they are not primarily targeted for data-driven \gls{LQE} research. The trace-set from the University of Michigan~\cite{chen2011robust} is somewhat incomplete and suffers from an inconsistent data format containing lack of units, missing sequence numbers and inadequate documentation. The two EVARILOS trace-sets~\cite{van2015platform} are mainly well formated, whereas each contains fewer than 2,000 entries, and thus both are not well suited for data-driven \gls{LQE} research. In Colorado trace-set~\cite{anderson2009impact}, the diversity of the link performance is missing as all links seem to exhibit less than 1\% packet loss. Finally, the trace-set of Brussels University~\cite{le2007principal}, at the time of writing, is inadequate for data-driven \gls{LQE} analysis, and suffers from an inconsistent data structure and deficient documentation.

% Which is the most suitable then?
After careful evaluation of the candidate trace-sets, we can conclude that the most suitable candidate for data-driven analysis of \gls{LQE} is the Rutgers trace-set. Roughly speaking, all the other candidates lack sufficient size, are structured in improper format, contain negligible packet loss hindering from practical \gls{LQE} investigation and/or rely on deficient documentation. However, these are the main characteristics required for \gls{ML}-based \gls{LQE} investigation, where it's classification primarily depends on \gls{PRR}. Even though we concluded that the Rutgers trace-set is the most suitable one for data-driven \gls{LQE} research, it also lacks some critical aspects for near-perfect data-driven \gls{LQE} research including explicit time-stamps and non-artificial noise sources just to name a few. We take this conclusion in account later in Section~\ref{sec:desguide} where we suggest industry and research community a design guideline on how a good trace-set should be collected.

\section{Findings}
\label{sec:findings}
In this section, we present our findings as a result of the comprehensive survey of data-driven \gls{LQE} models, publicly available trace-sets and the design of \gls{ML}-based \gls{LQE} models. First, we elaborate on the lessons learned from the aforementioned survey of the literature, then we suggest design guidelines for developing ML-based LQE models based on application quality aspects and for generic trace-set collection to the industry and research community.

\subsection{Lessons Learned}\label{sec:learnedles}
Having surveyed the comprehensive literature for LQE models using ML algorithms in Section~\ref{sec:overview-of-lqe}, we now outline the lessons we have learned throughout this section.

\begin{itemize}	
	\item While traditionally, most LQE models were developed to be eventually used by a routing protocol, recently researchers have also identified their potential application in single hop networks, particularly with the intention of reducing network planning costs via automation~\cite{demetri2019automated}.
			
	\item Recently, new sources of information or input metrics, such as topological- and imaging-based are considered for the development of LQE models, as noted in Section \ref{subsec:input}.
	
	\item From Sections~\ref{sub:model-for-lqe} and~\ref{sec:lqe-ML-analysis-app}, it can be concluded that reinforcement learning is a relatively less popular ML method for LQE research.
	
	\item A number of LQE models provide categorization (grade) for link quality rather than continuous values. The analysis in Section~\ref{sub:lqe-output} shows that the number of categories or classes (link quality grades) varies between 2 and 7.
	
	\item There is no standardized and easy way of evaluating and benchmarking LQE models against each other, as it is evident from the analysis in Section~\ref{sub:evaluation-metrics}.
	
	\item Only a small number of research papers provide all the details and datasets so that the results can be readily reproduced by the research community to improve upon and to be utilized as a baseline/benchmarking model for the sake of comparative analysis, as discussed in Section~\ref{sub:reproducibility}.
	
\end{itemize}

We highlight the following lessons learned from the application perspective analysis of the ML-based LQE models performed in Section~\ref{sec:lqe-ML-analysis-app}.
	
\begin{itemize}	
	\item From the application that uses LQE, such as a multi-hop routing protocol, we were able to identify five application quality metrics that are indispensable for the development of an ML-based LQE model: reliability, adaptivity/reactivity, stability, computational cost and probing overhead. These application quality metrics are outlined and explained in Section~\ref{sec:lqe-ML-analysis-app} and distilled from the extensive survey in Section~\ref{sec:overview-of-lqe}. These metrics are sometimes used to evaluate the performance of the application with/without using LQE.
	
	\item Only a paucity of contributions explicitly considers adaptivity, stability, computational cost and probing overhead in their evaluation for the performance of an LQE model, as perceived from the analysis in Section~\ref{sec:lqe-ML-analysis-app}. No research paper considers all five aspects together.
	
	\item To develop LQE models for wireless networks with dynamic topology, adaptivity can be enabled with the aid of online learning algorithms. Important link changes are difficult to capture with offline models, resulting in a degradation of the performance of the LQE model, as the up-to-date link state is unknown to the intended devices.
\end{itemize}

The lessons learned from design decisions taken for developing existing ML-based LQE models as analyzed in Section~\ref{sec:lqe-ML-analysis} can be summarized as follows.

\begin{itemize}	
	
	\item Training data for ML models often miss data points, for example no records for the lost packets can be found. The approach adopted for compensating the missing data, such as interpolation, may have significant impact on the final performance of the LQE model and explicitly describing the process is important for enabling reproducibility.
	
	\item The feature sets that are utilized for LQE research are not always explicitly reported nor identical among different LQE models, which hinders fair comparative analysis for diverse parameter settings.
	
	\item Training data for ML models can be highly imbalanced. Classification-wise, for example, the training dataset can be dominated by one type of link quality class (grade), which consequently leads to a highly biased LQE model that is unable to recognize minority classes. To counter this artifact, resampling has to be employed for highly imbalanced datasets. No research papers explicitly state their resampling strategy, as readily observed in Fig.~\ref{fig:steps-desc-overview} of Section~\ref{sub:resampling}.
	
	\item Logistic and linear regressions are linear models that tend to be more suitable to approximate linear phenomena. In practical scenarios, LQE models do not obey linearity and therefore ANN-based models outperform linear models. However, ANN- and DNN-based models usually require high memory and computational resources, which is unfavorable for constrained devices, albeit they may be tuned to necessitate less resources but at the expense of proportional performance.
		
\end{itemize}

From the overview of measurement data sources in Section~\ref{sec:dataset-overview}, we have learned the following lessons.

\begin{itemize}	
	
	\item Only a limited number of publicly available datasets record overlapping/identical metrics, which can indeed empower fair comparative analyses between diverse LQE models.
	
	\item Measurement points indicating the quality of the received	signal on links are commonly defined by SNR, RSSI and LQI.	
\end{itemize}

\subsection{Design Guidelines for ML-based LQE Model}\label{sec:desguideML}

\begin{figure*}
	\centering
	\begin{tikzpicture}[
	scale=0.68,
	level0/.style={ultra thick, black, font=\bfseries\normalsize, align=center},
	level1/.style={ultra thick, black, font=\bfseries\small, align=center},
	level2/.style={ultra thick, black, font=\footnotesize, align=center},
	level3/.style={thick, black, font=\footnotesize, align=center},
	connect/.style = {thick, dashed, red},
	notice/.style  = { draw, rectangle callout, callout relative pointer={#1} },
	label/.style   = { text width=2cm }
	]
	\draw[level0] (-2,0) -- node[above] {Design Guidelines\\for Developing\\LQEs} (2,0);

	% RELIABILITY
	\draw[level1] (3, 2) -- node[above] {Reliability} (5, 2);
	\draw[connect] (2, 0) -- (3, 2);
	
	%% Reliability -- Traceset collection
	\draw[level2] (6, 7) -- node[above] {Traceset\\collection} (8, 7);
	\draw[connect] (5, 2) -- (6,7) (5,2) -- (6,1) (5,2) -- (6,-5);
	
	\foreach \x [count=\xi from 0] in {
		Consider longer\\observation time,
		Consider more metrics\\(see Tab.~\ref{tab:datasets-features}),
		Consider external metrics\\from reliable sources,
		Consider active\\probing mechanisms,
		Consider increasing\\number of datapoints
	} {
		\def\y{10-1.3*\xi};
		\draw[level3] (8.5, \y) -- node[above] {\x} (13, \y);
		\draw[connect] (8, 7) -- (8.5, \y);
	}
	
	%% Reliability -- Preprocessing
	\draw[level2] (6,1) -- node[above] {Preprocessing} (8,1);
	\foreach \x [count=\xi from 0] in {
		Consider producing sythetic\\features (see Sec.~\ref{sub:feature-selection}),
		High dimensionality\\feature vector,
		Consider adding alternative\\data representations
	} {
		\def\y{2-1.3*\xi};
		\draw[level3] (8.5, \y) -- node[above] {\x} (13, \y);
		\draw[connect] (8, 1) -- (8.5, \y);
	}

	\draw[level2] (6,-5) -- node[above] {Method\\Selection} (8,-5);
	\foreach \x [count=\xi from 0] in {
		Consider DNN for\\highdimensional data,
		Consider SVM with\\non-linear kernel,
		Consider ensemble\\methods
	} {
		\def\y{-3-1.3*\xi};
		\draw[level3] (8.5, \y) -- node[above] {\x} (13, \y);
		\draw[connect] (8, -5) -- (8.5, \y);
	}

	%% Computation Cost
	\draw[level1] (-3, 5) -- node[above] {Computational\\Cost} (-5, 5);
	\draw[connect] (-2, 0) -- (-3, 5) (-5,5) -- (-6,8) (-5,5) -- (-6, 2.5);
	
	\draw[level2] (-6, 8) -- node[above] {Preprocessing} (-8, 8);
	\foreach \x [count=\xi from 0] in {
		Minimize\\feature set,
		Consider reducing data\\dimensionality,
		Consider reducing data\\precision,
		Consider smaller\\window size
	} {
		\def\y{10-1.3*\xi};
		\draw[level3] (-8.5, \y) -- node[above] {\x} (-13, \y);
		\draw[connect] (-8, 8) -- (-8.5, \y);
	}

	\draw[level2] (-6, 2.5) -- node[above] {Method\\selection} (-8, 2.5);
	\foreach \x [count=\xi from 0] in {
		Less expensive approaches\\e.g. NaiveBayes;\\Linear algorithms;\\pretrained \& pruned DNN,
		Consider online\\ML approaches
	} {
		\def\y{3-1.3*\xi};
		\draw[level3] (-8.5, \y) -- node[above] {\x} (-13, \y);
		\draw[connect] (-8, 2.5) -- (-8.5, \y);
	}

	% STABILITY
	\draw[level1] (-3,-2) -- node[above] {Stability} (-5,-2);
	\draw[connect] (-2, 0) -- (-3, -2) (-5,-2) -- (-6,-0.5) (-5,-2) -- (-6,-5);
	
	%% STABILITY -- preprocessing
	\draw[level2] (-6,-0.5) -- node[above] {Preprocessing} (-8,-0.5);
	\foreach \x [count=\xi from 0] in {
		Consider reducing influence\\of transient effects,
		Consider larger\\window size
	} {
		\def\y{0-1.3*\xi};
		\draw[level3] (-8.5, \y) -- node[above] {\x} (-13, \y);
		\draw[connect] (-8, -0.5) -- (-8.5, \y);
	}
	
	%% STABILITY -- methods selection
	\draw[level2] (-6,-5) -- node[above] {Methods\\selection} (-8,-5);
	\foreach \x [count=\xi from 0] in {
		Optimize to detect\\persisting changes,
		Minimize influence\\of transient effects
	} {
		\def\y{-4-1.3*\xi};
		\draw[level3] (-8.5, \y) -- node[above] {\x} (-13, \y);
		\draw[connect] (-8, -5) -- (-8.5, \y);
	}

	% ADAPTABILITY
	\draw[level1] (-3,-10) -- node[above] {Adaptability} (-5,-10);
	\draw[connect] (-2, 0) -- (-3, -10) (-5,-10) -- (-6,-8) (-5,-10) -- (-6,-11) (-5,-10) -- (-6,-14);

	%% ADAPTABILITY -- traceset collection
	\draw[level2] (-6,-8) -- node[above] {Traceset\\collection} (-8,-8);
	\foreach \x [count=\xi from 0] in {
		Fast enough to pick\\transient effects,
		Consider smaller\\window
	} {
		\def\y{-7-1.3*\xi};
		\draw[level3] (-8.5, \y) -- node[above] {\x} (-13, \y);
		\draw[connect] (-8, -8) -- (-8.5, \y);
	}
	
	%% ADAPTABILITY -- preprocessing
	\draw[level2] (-6,-11) -- node[above] {Preprocessing} (-8,-11);
	\foreach \x [count=\xi from 0] in {
		Consider operating on\\time window; batches\\or streams,
		Consider smaller\\window size
	} {
		\def\y{-10.2-1.3*\xi};
		\draw[level3] (-8.5, \y) -- node[above] {\x} (-13, \y);
		\draw[connect] (-8, -11) -- (-8.5, \y);
	}
	
	%% ADAPTABILITY -- methods selection
	\draw[level2] (-6,-14) -- node[above] {Method\\selection} (-8,-14);
	\foreach \x [count=\xi from 0] in {
		Consider online\\version of algorithms,
		Consider fast\\offline methods,
		Reinforcement\\learning
	} {
		\def\y{-13-1.2*\xi};
		\draw[level3] (-8.5, \y) -- node[above] {\x} (-13, \y);
		\draw[connect] (-8, -14) -- (-8.5, \y);
	}

	% Probing overhead
	\draw[level1] (3,-9) -- node[above] {Probing\\Overhead} (5, -9);
	\draw[connect] (2, 0) -- (3, -9) (5,-9) -- (6, -11);

	\draw[level2] (6, -11) -- node[above] {Traceset\\collection} (8, -11);
	\foreach \x [count=\xi from 0] in {
		Consider reducing data\\collection,
		Pick directly\\available data,
		Minimize feature set\\(see Tab.~\ref{tab:datasets-features}),
		Consider passive probing,
		Consider Alternative\\learning datasets
	} {
		\def\y{-9-1.3*\xi};
		\draw[level3] (8.5, \y) -- node[above] {\x} (13, \y);
		\draw[connect] (8, -11) -- (8.5, \y);
	}
\end{tikzpicture}
\caption{Mind map representation of design guidelines for LQE model development.}
\label{fig:design-guideline-ML-App}
\end{figure*}
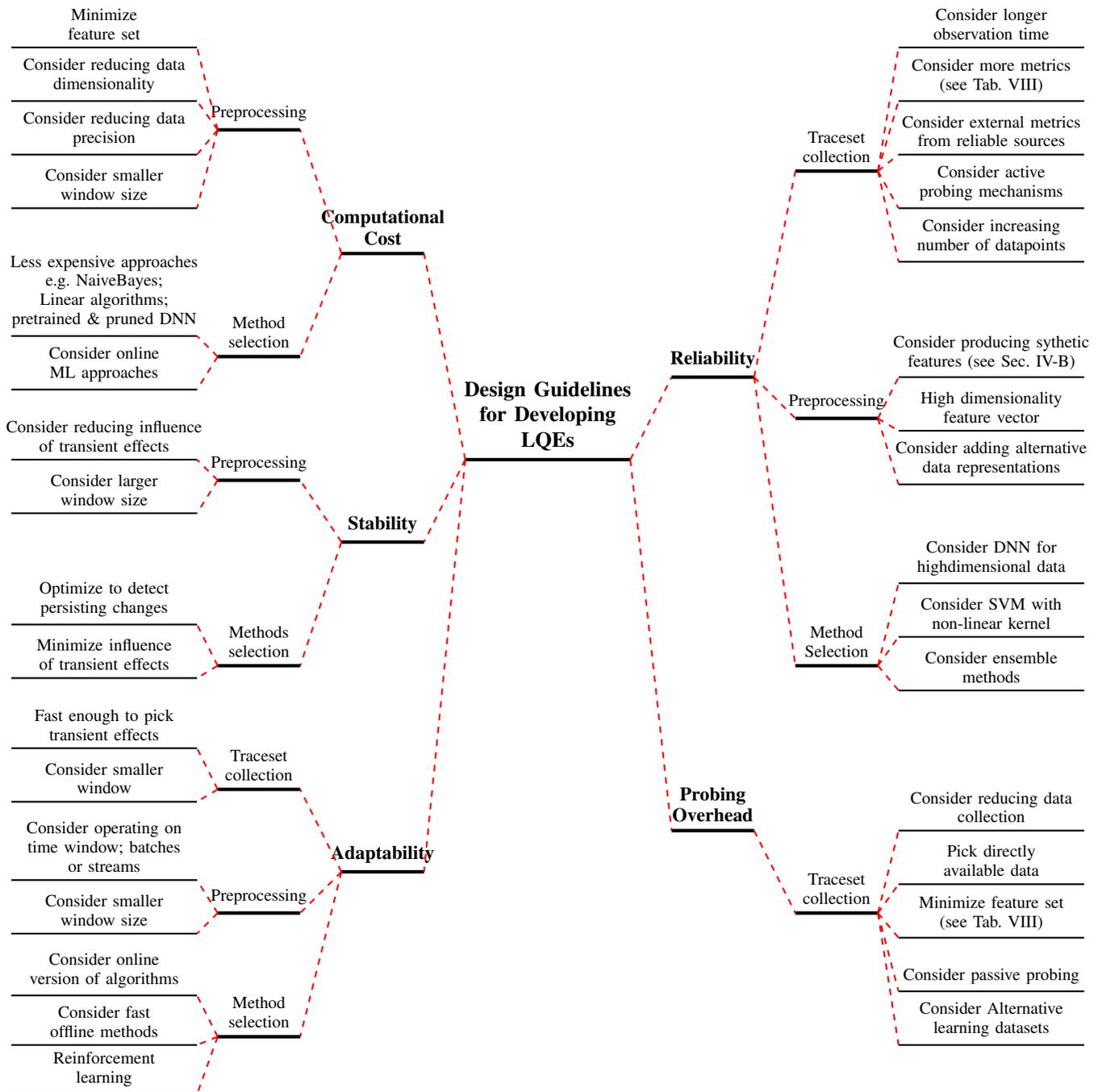

Due to a very large decision space for developing a ML-based LQE model, it can be challenging to provide a universal decision diagram or methodology. However, showing how application requirements affect design decisions, and by reflexivity, how certain design decisions can favor some application requirements can be invaluable for the development of ML-based LQE models. In this section, we provide design guidelines on developing a ML-based LQE model starting from the five application quality aspects identified in Section~\ref{sec:lqe-ML-analysis-app} and their implications on decisions during the design steps of the ML process discussed in Section~\ref{sec:lqe-ML-analysis}. The visual relationship of how the application quality aspects influence the design decisions for developing LQE models is illustrated in Fig.~\ref{fig:design-guideline-ML-App}.

\subsubsection{Reliability}When \textit{reliability} is \textit{the only} application quality aspect to be optimized for developing a ML-based LQE model, trace-set collection, data pre-processing and ML method selection should be carefully considered, as depicted in the Reliability branch of the mind map in Fig.~\ref{fig:design-guideline-ML-App}. 
	
\paragraph*{Trace-set collection} The trace-set collection and subsequent probing mechanism utilized during the actual operation of an LQE model, can collect all the input metrics listed in Table~\ref{tab:datasets-features} and perhaps even other inventive metrics that have not been used up-to-date in the existing literature.

\paragraph*{Data pre-processing} During data pre-processing, high dimensional feature vectors using recorded input metrics as well as synthetically generated ones (see Section~\ref{sub:feature-selection}) can be used as there are no constraints on the memory use or computational power of the machine used to train the subsequent model.
	
\paragraph*{ML method selection} During ML method selection, more computationally expensive methods, such as DNN, SVMs with non-linear kernel as well as ensemble methods, such as random forests can be considered. For accurate models that provide very good \textit{reliability}, these methods are able to train on high dimensional feature vectors. However, they will also require many training data-points, possibly hours or days of measurements. While DNNs are known to be very powerful, they are also excessively data hungry. Their performance can be significantly diminished if the data-points are not sufficient.

\subsubsection{Adaptivity}When \textit{adaptivity} is \textit{the only} application quality aspect to be optimized for developing an ML-based LQE model, data pre-processing and ML method selection are the two aspects to be examined, as illustrated in the Adaptivity branch of the mind map in Fig.~\ref{fig:design-guideline-ML-App}. 

\paragraph*{Data pre-processing} Adaptivity requires LQE model to capture non-transient link fluctuations, therefore it has to monitor temporal aspects of the link. This is usually realized by introducing time windows on which the pre-processing is done. As opposed to pre-processing all available data in a bulk mode for subsequent offline development as employed for \textit{reliability} aspect, each window is pre-processed separately for the \textit{adaptivity}. The size of the window then influences the \textit{adaptivity} of the model, where a smaller window size yields a more adaptive model.

\paragraph*{ML method selection} During the ML method selection, online versions of ML methods or reinforcement learning are more suitable for capturing the changes in time. Generally, the online version of an offline ML method may be slightly more expensive computationally and its performance may be slightly reduced. Reinforcement learning is a class of ML algorithms that learn from experience and these are inherently designed to adapt to changes. The higher the required \textit{adaptivity}, the faster the model has to change, leading to a more reactive ML (method) parameter tuning.

\subsubsection{Stability}When \textit{stability} is \textit{the only} application quality aspect to be optimized for developing an ML-based LQE model, the same ML design steps are affected as outlined in the \textit{Adaptivity} aspect, namely data pre-processing and ML method selection, as portrayed in the Stability branch of the mind map in Fig.~\ref{fig:design-guideline-ML-App}. However, they are reversely affected when compared to the \textit{adaptivity} aspect.

\paragraph*{Data pre-processing} Stability requires LQE model to be immune to transient link behavior. While it may assume changes over time, it encourages only relevant changes. The size of the window chosen in this case typically represents a compromise between the batch approach mentioned for \textit{reliability} and the relatively small reactive window that maximizes \textit{adaptivity}.

\paragraph*{ML method selection} During the ML method selection, online versions of ML methods or reinforcement learning are more suitable for capturing changes in time, however, they need to be optimized to detect persistent link changes, while being immune to transient ones.

\subsubsection{Computational Cost}When \textit{computational cost} is \textit{the only} application quality aspect to be optimized for developing an ML-based LQE model, data pre-processing and ML method selection should be carefully contemplated, as outlined in the Computational Cost branch of the mind map in Fig. ~\ref{fig:design-guideline-ML-App}.
	
\paragraph*{Data pre-processing} Computational cost optimization requires reducing memory and energy consumption as well as processor performance  aspects required for the LQE model development. For offline or batch processing, the size of the feature vectors should be kept to a minimum, therefore it has to include only the most relevant real or synthetic features. Alternatively, projecting large feature vectors to a lower dimensional space might help for training. Additionally, for online processing, smaller time windows that minimize RAM consumption are favored.

\paragraph*{ML method selection} During ML method selection, less intensive methods, such as naive Bayes or linear/logistic regression are preferred. When online versions of the ML methods are utilized, their configurations should be appropriately adjusted so that the resource usage is kept at minimum. For instance, transfer learning~\cite{pan2009survey} approaches enable stripped down versions of a complete model that was previously learned on a powerful machine, which is then deployed to the production environment. Transfer learning is becoming a relatively popular way of deploying DNN-based models on flying drones for instance~\cite{pan2009survey}.

\subsubsection{Probing overhead}When \textit{Probing overhead} is \textit{the only} application quality aspect to be optimized for developing an ML-based LQE model, trace-set collection is the only design process that requires careful attention, as illustrated in the probing overhead branch of Fig.~\ref{fig:design-guideline-ML-App}.

\paragraph*{Trace-set collection} Trace-set collection and subsequent probing mechanism utilized during actual operation of the LQE model should only collect few and most important metrics from the ones listed in Table~\ref{tab:datasets-features}. Ideally, LQE model can be engineered to work on passive probing so that it can only use the metrics that the transmitter captures.

\subsubsection{Practical scenarios}A practical application using LQE will likely request optimizing more than one of the five identified application quality aspects. As a result, the guideline and its illustrations for such cases would be more sophisticated and interconnected than in  Fig.~\ref{fig:design-guideline-ML-App}. However, the proposed guideline provides an overview of the measures to be taken and presents an invaluable trade-off between these application quality aspects that require careful attention for the development of an ML-based LQE model.

For example, when the application requires high \textit{reliability} and \textit{adaptivity}, large feature spaces can be used with powerful online algorithms on appropriately identified time windows. However, if \textit{computational cost} is appended to the requirements, the feature space should be limited and the algorithm parameters should be optimized. If the LQE model is still computationally expensive, transfer learning or other out-of-the-box ML methods should be employed. When \textit{probing overhead} is also appended to the previously-mentioned application quality aspects, then the feature set should only include locally available data (passive probing) and limited number of metrics (possibly none) involving active probing, as discussed in Section~\ref{subsec:input}. In brief, this guideline can be used as a reference for the development of an ML-based LQE model depending on the combination or quality aspects relevant for the application.

%\subsection{Guidelines for evaluating the performance of ML based LQE}\label{sec:evalGuideline}
%
%Should have a two fold evaluation:
%1) as per the standard 5 quality features of the applications
%2) as per standard ML community: MSE (in addition to others) for regression and confusion matrix (in addition to others) for classification

\subsection{Design Guidelines for Trace-Set Collection}\label{sec:desguide}
% In Section~\ref{sec:overview-of-lqe} we analyzed existing research work on \glspl{LQE} in general (with or without ML). We went through on what trace-sets these researches used, which are enlisted on Tables~\ref{tab:lqe-part1} and~\ref{tab:lqe-part2}. Their steps in data processing are examined on Table~\ref{tab:steps--desc-overview}. What we discoverd was that only minority of researches are reproducible, as we already discussed in Section~\ref{sub:reproducibility}. In most cases, trace-sets are not publicly available.
% 
% Next, we examined publicly available trace-sets in Section~\ref{sec:dataset-overview} and listed them on Table~\ref{tab:datasets} and~\ref{tab:datasets-features} (feature examination is in Table~\ref{tab:feature-analysis}). Each of these trace-sets were collected with different goal in mind, where some of them were built with specific purpose.

\begin{figure*}[!htb]
	\centering
	\includegraphics[scale=0.75]{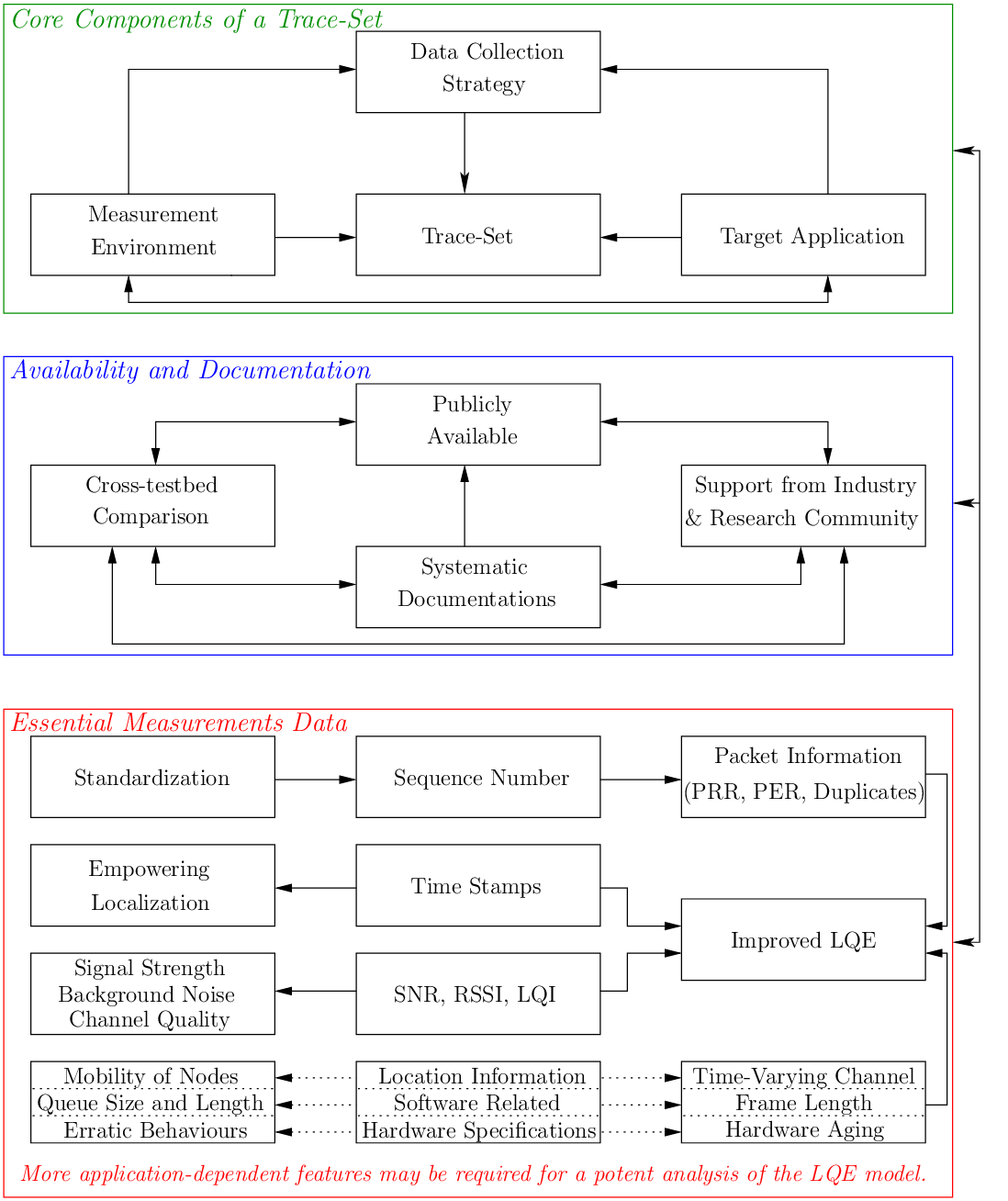}
	\caption{Design guidelines recommended for the industry and research community to follow in order to design and collect trace-sets for the sake of LQE research.}
	\label{fig:design-guideline}
\end{figure*}

We now attempt to provide a generic guideline on how to design and collect an \gls{LQE} trace-set, as portrayed in Fig.~\ref{fig:design-guideline}. It is worth noting that this design guideline comprises of plausible and reasonable observations gleaned from this survey of \gls{LQE} and trace-sets, and from the analysis of \gls{ML} methods reviewed for the sake of \gls{LQE} models. Our plausible recommendations on how to design and collect an \gls{LQE} trace-set can be summarized as follows, which can also be followed as in Fig.~\ref{fig:design-guideline}.

\textbf{\textit{1. Core components of a trace-set}}:~Deciding on the data collection strategy, the application and the environment is a crucial stage, since the development of an \gls{LQE} model is strictly dependent on the trace-set environment including industrial, outdoor, indoor and ``clean'' laboratory environments.  State of the radio spectrum and interference level are important metrics to be taken into account before collecting a trace-set. For example, for an \gls{LQE} model to work efficiently in a particular environment that is exposed to interference, then the \gls{LQE} model has to be developed and trained over this kind of trace-set. More explicitly, one cannot expect an ML-based \gls{LQE} model to perform well in an interference-exposed environment without having it implemented and tested on a trace-set containing interference measurement data, which leads us to data collection strategy and the application.

\textbf{\textit{2. Availability and documentation}}:~Making trace-set publicly available is also another important stage, which can indeed empower better cross-testbed comparisons and provide good support/foundation from research community to conduct and disseminate research on \gls{LQE} models. There are numerous ways to make trace-sets publicly available. One well known repository for wireless trace-sets is CRAWDAD\footnote{A repository for archiving wireless data at Dartmouth: \url{https://crawdad.org}.}, although researchers can also take advantage of other methods like public version control systems, e.g., GitHub, GitLab and BitBucket just to name a few. Moreover, a systematic description on how the trace-set was collected is also required for research community to understand, test and improve upon. This will indeed help in capacity building between research groups.

\textbf{\textit{3. Essential measurements data}}:~Plausible logic dictates that a generic trace-set that can be utilized for any kind of \gls{LQE} research is infeasible considering numerous features induced by the wireless communication parameters. By interpreting our overall observations gleaned from this survey paper, some of the most important measurements data or features that are recommended for an effective \gls{LQE} research are already included in the design guideline of Fig.~\ref{fig:design-guideline} with a notice that other application-dependent features may be required for a strong analysis of the \gls{LQE} model. The elaborated details of these essential measurements data can be found in Section~\ref{sec:dataset-overview}. 

There may be other application-dependent metrics and features (measurements data) related to the set of parameters of wireless communication that could be taken into account for a healthy investigation of a particular \gls{LQE} model. We observe from the outcomes of this survey paper that each application can have unique characteristics and requirements for maintaining reliability, for satisfying a certain QoS and more generally for accomplishing a target objective, such as in smart grid, wireless sensor network, mobile cellular communication, air-to-air communication, air-to-ground communication, traditional terrestrial communication, underwater communication and other wirelessly communicating networks. Explicitly, for each application of these networks, determining a suitable evaluation metric is vitally important for the sake of maintaining a reliable and adequate communication. Therefore, trace-sets have to be designed and collected based on not only applications but also on evaluation metrics considering diverse environments, settings and technologies in order to be able to derive the properly effective metrics for an efficient development of the link quality estimation models.

Nonetheless, from the perspective of innovative data sources, a trace-set can be built without on-site measurements and before embarking on hardware deployments in order to provide a good estimate for the link quality for the sake of maintaining reliable communications. To achieve such goal, Demetri~\textit{et al.}~\cite{demetri2019automated} exploited readily available multi-spectral images from remote sensing, which are then utilized to quantify the attenuation of the deployment environment based on the classification of landscape characteristics. This particular research demonstrates that the quantification and classification of links can be conducted via solely relying on the image-based data source rather than the traditional on-site measurements data.

For urban area applications, the aforementioned technique can also be leveraged for maintaining up to a certain degree of the link quality, but only considering the stationarity of the deployment environment. This is mainly because the spectral images obtained via remote sensing represent a stationary instance of the landscape and thus this technique would dramatically fail, since the \gls{LQE} model developed using remote sensing would not be able to cope with the high mobility in such a scenario with moving vehicles, slowly-fading pedestrian channels, mobile UAVs and so on.

Besides, 3D model of large buildings can also be leveraged for the optimal indoor deployment of access points and wireless devices in order to supply with the adequate connectivity and coverage. The trace-set built from this indoor deployment can be utilized for other large and similar indoor buildings along with an indoor-generic \gls{LQE} model to understand the characteristics of indoor links and to provide high quality link performance. Similarly, the same strategy can be implemented for a particular city to understand the link behavior in different weather conditions. One study for such scenario is conducted using high frequency~\cite{Fenicia2012,kelmendi2017rain}, where the impact of rainfall on wireless links was researched. They utilized rain gauges and their models are demonstrated to contain large bias, and rainfall predictions were underestimated, which indicates that a long-lasting and realistic measurement conditions are required along with a plethora of measurements data before developing a healthy \gls{LQE} model.

Finally, recording hardware related metrics on a trace-set could also help in diagnosing potential problems during the model development. This would indeed require commercial radio chips that are capable of reporting the chip errors or chip related issues in order to pinpoint problems that may be encountered at the time of measurements data collection~\cite{Spuhler2013}.

\section{Summary}
\label{sec:summary}
Having outlined the lessons learned along with a comprehensive design guideline derived for ML-based LQE model development and trace-set collection, we now provide our concluding remarks and future research directions along with challenging open problems.

\subsection{Conclusions}
The data-driven approaches have been long ago adopted in the study of \gls{LQE}. However, with the adoption of \gls{ML} algorithms, it has recently gained new momentum stimulating for a broader and deeper understanding of the impact of communication parameters on the overall link quality. In this treatise, we first provide an in-depth survey of the existing literature on LQE models built from data traces, which reveals the expanding use of \gls{ML} algorithms. We then analyze \gls{ML}-based LQE models using performance data with the perspective of application requirements as well as with the ML-based design process that is commonly utilized in the ML research community. We complement our survey with the review of publicly available datasets relevant for LQE research. The findings from the analyses are summarized and design guidelines are provided to further consolidate this area of research.

\subsection{Future Research Directions}
Finally, we conclude the paper with a discussion on the open challenges, followed by several directions for future research, regarding (i) data sources utilized for developing \gls{LQE} models, (ii) applicability of \gls{LQE} models to heterogeneous networks incorporating multi-technology nodes, and (iii) a broader and deeper understanding of the link quality in various environments.

It is highly likely that commercial markets will leverage either pre-built \gls{LQE} models for a particular application or entire training data to develop models from scratch. The potential opportunity of "model stores" and "dataset stores" can follow a similar way to conventional application stores/markets, distributing models for diverse applications. The competition will gradually become ripe as time elapsed. However, data-driven models are still in their infancy and several critical open challenges await concerning \gls{LQE} models, which are outlined as follows.

\begin{enumerate}
	\item % LQE comparison
	A significant challenge is to directly compare different wireless link quality estimators. As discussed in Section~\ref{sub:evaluation-metrics}, there is no standardized approach to evaluate the performance of the estimators, and only a very small subset of estimators are compared directly in existing works. Establishing a uniform way of benchmarking new LQE models against existing ones using standard datasets and standard ML evaluation metrics, such as practiced in various ML communities, would greatly contribute to the ability to reproduce and compare innovative ML-based LQE models.
	\item % LQE comparison metrics
	The performance of the existing LQE models using classifiers are solely evaluated based on the $accuracy$ metric, possibly in addition to another application-specific metric, as discussed in Section~\ref{sub:evaluation-metrics}. However, it is well-known in the ML communities that $accuracy$ is a misleading performance evaluation metric, especially for imbalanced datasets~\cite{jeni2013facing}. Adopting standardized metrics for classification, e.g., $precision$, $recall$, $F1$ and, where necessary, the detailed $confusion$ $matrix$ would lead to a more in-depth understanding of the actual performance and behavior of the LQE models for all the target classes. The same challenge applies to LQE models solving a regression problem.
	\item % 1) Share trace-sets
	Another challenge is to encourage researchers and industry to share trace-sets collected from real networks. More suitable public trace-sets would allow algorithms and machine learning models to be properly evaluated across different networks and scenarios considering the important metrics discussed in Section~\ref{sec:dataset-overview}. Indeed, trace-sets collected in an industrial environment could better represent a realistic communication network potentially with a broad number of parameters.
	\item % 2) multi-hop & dynamic
	The other challenge is to go beyond one-to-one trace-sets. Research community is required to extend the scope to a more realistic measurement setup, e.g., considering multi-hop, non-static networks representing several wireless technologies. Such instances of trace-sets are scarce due to the necessity of exhausting efforts to monitor and record a packet's travel through a particular communication network.
	\item % 3) generating data
	Another challenge is that certain types of trace-sets are very expensive and time-consuming to gather. One way to overcome this is to conduct a synthesis of artificial data using generative adversarial neural networks as pointed out in~\cite{goodfellow2014generative}. Roughly speaking, this open challenge is a formidable task, since conducting such synthesis could potentially introduce unwanted bias to existing data, even though for specific applications a number of suitable examples of this method can be found in the literature, such as wireless channel modeling~\cite{Ye86442502018}, ~\cite{Yang86639872019}.
	\item % 4) Awareness and analysis of interference of heterogeneous networks.
	The traditional approach to measure interference is mainly conducted through \gls{SNR} or \gls{RSSI} measurement data, which strictly relies on the data collection at certain intervals, and communication established from other nodes is mainly treated as a background noise for the sake of simplicity. The aim of interference measurement as part of this challenge is to develop \gls{LQE} models that are aware of the on-going communication within a heterogeneous communication environment. None of the trace-set layouts surveyed in Section~\ref{sec:dataset-overview} is designed for such asynchronous information. Therefore, research community and industry have to pay attention to collecting such realistic trace-sets in order to be able to develop robust, agile and flexible \gls{LQE} models that can readily adapt in dynamic and realistic communication environments.
	\item % 4) Aw
	The wireless link abstraction comprised of channel, physical layer and link layer represents a complex system affected by a multitude of parameters, but most of the LQE datasets and research only leverages a small number of observed parameters. While recently additional image-based and topological-based contextual information has been incorporated in LQE models, it would be necessary in future large scale multi-parameter measurement campaigns to also capture the type of antenna, modulation and coding utilized, producer of the transceiver, firmware versions, to name a few. Such efforts would lead to a more in-depth understanding of the real-world operational networks and potential use of the findings to make well-informed decisions for the design of next-generation wireless systems, even beyond ML-based LQE model development.
\end{enumerate}

In order to realize beyond simple decision making, i.e., channel and radio behavior modeling, \textit{hand-tuning} of communication parameters within transceivers must be avoided. It is anticipated that the transceivers' internal components will be gradually replaced by software-based counterparts. Therefore, an inevitable incorporation of software-defined radio (SDR), FPGAs and link quality estimators is expected for intelligently handling parameters and operations through self-contained smart components. These joint \gls{LQE} models can be designed in a similar manner to~\cite{Haggui8726401}, particularly for heterogeneous networks involving the 5G and beyond communications.

The recent advancements in data-driven approaches in the form of machine learning and deep learning have already proven to be successful for the applications of communication networks. For example, attempts to use neural network-based autoencoders for channel decoding provide promising solutions~\cite{gruber2017deep}, which can also be adopted for data-driven \gls{LQE} investigation as it is discussed in~\cite{luo2019link}.

The performance of link quality estimator is constrained by the dynamic network topology and one can keep track of the network topology changes considering replay-buffer-based deep Q-learning algorithm developed in~\cite{Koushik8674587}, where authors control the position of UAVs, acting as relays, to compensate for the deteriorated communication links.

Additionally, \gls{LQE} models involved in the optimization problems may become very large in size, and thus algorithms that can reduce complexity have to be developed to tackle with the scale of the problem. For example, a similar deep learning approach to~\cite{Liu7996587} can be adopted for improving the performance of the proposed \gls{LQE} model by means of eliminating the links from optimization problem that are not utilized for transmission.

Referring back to Section~\ref{sub:evaluation-metrics}, we discussed the convergence rate of \gls{LQE} models. While some contributions~\cite{srinivasan2008prr, senel2007kalman, boano2010triangle, liu2011foresee} focus their attention on the convergence of their LQE model, majority of the papers tend to neglect it. Motivated by this premise, we suggest the research community to pay particular attention on the \gls{LQE} model convergence in order to prove the validity of their proposed models.

In addition to finding other new sources of data, a challenging task would be to analyze a large set of measurements in various environments and settings, from a large number of manufacturers to understand how measurements vary across different technologies and differ for various implementations within the same technology, and derive truly effective metrics for an efficient development of the link quality estimation model.

\section*{Acknowledgment}
This work was funded in part by the Slovenian Research Agency under Grants P2-0016 and J2-9232, and in part by the European Community H2020 NRG-5 project under Grant 762013. The authors would also like to thank Timotej Gale and Matja\v z Depolli for their valuable insights.

%Print the glossary
%\newpage
\printglossary[type=acronym,style=super,nogroupskip]

\bibliographystyle{IEEEtran}
\bibliography{ftr_eng}

%\newpage
%\input{Bio/bios.tex}

\end{document}